\documentstyle[12pt,fleqn]{article}
\newcommand{\be}{\begin{equation}}
\newcommand{\ee}{\end{equation}}
\newcommand{\bea}{\begin{eqnarray}}
\newcommand{\eea}{\end{eqnarray}}

\begin{document}
\title{Convergence properties of the\\
equal-time connected Green function approach\\
for temporal gauge $\rm{SU(2)}_{2+1}$ Yang-Mills theory
\footnote{supported by DFG, Forschungszentrum J\"ulich and GSI Darmstadt}
\footnote{part of the dissertation of J\"orn M. H\"auser}}
\author{J.M. H\"auser, W. Cassing, S. Leupold, and M.H. Thoma
\thanks{Heisenberg fellow}\\
Institut f\"ur Theoretische Physik, Universit\"at Gie\ss en\\
35392 Gie\ss en, Germany}
\maketitle
\date{\today}
\begin{abstract}
The hierarchy of equations of motion for equal-time Green
functions in temporal gauge SU(N) Yang-Mills theory is truncated using
an expansion in terms of connected Green functions. A second hierarchy
of constraint equations arises from Gauss law and can be truncated
in a similar way.
Within this approximation scheme we investigate
SU(2) Yang-Mills theory on a torus in $2+1$ spacetime dimensions
in a finite basis of plane wave states
and focus on infrared and ultraviolet properties of the approach.
We study the consequences of restoring the hierarchy
of Gauss law constraints and of different momentum cutoffs
for the 2- and the 3-point functions.
In all truncation schemes considered up to the 4-point level the
connected Green function approach is found to be UV divergent and
either violating gauge invariance and/or energy conservation.
The problems associated with adiabatically generating a perturbed
ground state are discussed as well.
\end{abstract}
\bigskip
\begin{center}
\end{center}
\newpage
\section{Introduction}
\label{introduction}
Nonabelian gauge theories in general require the use of nonperturbative
techniques as soon as one leaves the region of asymptotic freedom.
Important examples of corresponding problems are the mechanism of
confinement, probably related to the vacuum structure \cite{shu84},
the investigation of hadron spectra and hadronic matter, the
understanding of the deconfinement transition, the chiral
symmetry restoration, the mechanism of hadronization in
relativistic heavy-ion collisions, following the possible preformation
of a quark-gluon plasma, and non-perturbative effects
in a quark-gluon plasma above the critical temperature \cite{hat92}.
Besides numerically highly involved lattice calculations
\cite{pet91} there are only preliminary attempts to achieve
a nonperturbative description of nonabelian theories within the
framework of Dyson-Schwinger equations
\cite{tho89}-\cite{hab90},
discretized light front quantization \cite{pau96},
variational calculations \cite{sch85}-\cite{mis96},
and transport theories based on gauge-covariant
Wigner functions with or without the restriction to an
equal-time limit \cite{elz186}-\cite{bia93}.

In this article we follow a different approach along the line of
$n$-point correlation dynamics \cite{wan85}-\cite{wan294},
which consists in a truncation of the hierarchy of equations of
motion for equal-time Green functions in terms of connected Green
functions up to order $n$.
In relativistic field theory the method has previously been applied
successfully to $\Phi^4$ theory in $0+1$, $1+1$, and $2+1$ spacetime
dimensions \cite{hae295, pet96, pet97}.
The novel phenomenon in the application of $n$-point correlation
dynamics to Yang-Mills theory in temporal gauge is the requirement
of Gauss law constraints on physical states.
We apply 2-, 3-, and 4-point correlation dynamics to the simplest
nontrivial Yang-Mills theory, i.e. SU(2) in $2+1$ dimensions.
The theory has the advantage of being not only superrenormalizable
but finite \cite{fey81}.

This article is organized as follows: In Section \ref{quantization}
we discuss our procedure of quantization in temporal gauge, which
involves fixing the residual gauge freedom.
In Section \ref{eomcgf} we present the derivation of the correlation
dynamical equations of motion while in Section \ref{gausshierarchy}
we discuss the hierarchy of constraints for equal-time Green
functions that arises from Gauss law.
In Section \ref{propagate} we present numerical results for the
propagation of the system with given initial conditions for a variety
of approximation schemes and investigate scaling, infrared and
ultraviolet properties of the corresponding approximations.
In Section \ref{adiabaticswitching} we discuss the method of dynamically
generating an approximate ground state configuration by the method
of adiabatic switching. Section \ref{summary} contains a summary
of our results.
Appendix \ref{simpleex} contains a simple quantum mechanical example
for our approach to fixing the residual gauge freedom,
while in Appendix \ref{fourpointcolourstructure} we list the
colour structure of the connected four-point functions within our
approach.
\section{Quantization in temporal gauge}
\label{quantization}
In this section we discuss the quantization of SU(N) Yang-Mills
theories in temporal gauge ($A^a_0=0$) and the corresponding free
propagator.

We first turn to the discussion of canonical quantization.
The Lagrangian density of $d+1$ dimensional SU(N) Yang-Mills theory is
\begin{eqnarray}
{\cal L}=-\frac{1}{4} F^a_{\mu \nu} F^{a \; \mu \nu} \; ,
F^a_{\mu \nu}=\partial_\mu A^a_\nu - \partial_\nu A^a_\mu
+ g f^{abc} A^b_\mu A^c_\nu \; ,
\label{lagrangian}
\end{eqnarray}
where $a$, $b$, $c$ are the colour indices in the adjoint
representation of SU(N) and the greek
indices run over the $d+1$ spacetime components of the gauge field.
Throughout the article, repeated indices of any kind will be summed
over.
The corresponding $A^a_0=0$ gauge quantum Hamiltonian is \cite{chr80}
\begin{eqnarray}
H=\int d^d x \left( \frac{1}{2} \Pi^a_i(\vec{x}) \Pi^a_i(\vec{x})
+ \frac{1}{4} F^a_{ij}(\vec{x}) F^a_{ij}(\vec{x}) \right)
\label{hamiltonian}
\end{eqnarray}
and the canonical equal-time commutation relations are
\begin{eqnarray}
[A^a_i(\vec{x},t),\Pi^b_j(\vec{y},t)]
= i \delta_{ij} \delta^{ab} \delta(\vec{x}-\vec{y}) \; ,
\label{commutator}
\end{eqnarray}
where now the latin indices $i$, $j$ only run over the $d$ spatial
components of the gauge field and $\Pi^a_i$ denote the conjugate
momenta.
The Euler-Lagrange equations following from (\ref{lagrangian})
in the $A^a_0=0$ gauge for the spatial components of the gauge
field $A^a_i$ can be reproduced with (\ref{hamiltonian}) and
(\ref{commutator}) as the Heisenberg equations
for the operators $A^a_i$ and $\Pi^a_i$. In contrast to that,
the Euler-Lagrange equation for $A^a_0$, the Gauss law,
is not contained in the dynamics generated by (\ref{hamiltonian}).
It can, furthermore, not be realized as an operator identity,
since this would contradict the commutation relations (\ref{commutator}).
The Gauss law thus has to be imposed as a constraint on physical states:
\begin{eqnarray}
G^a(\vec{x}) |\Psi_{ph}\rangle=0 \; ,
\; \; G^a(\vec{x})=\left( \partial_{x_i} \Pi^a_i(\vec{x})
+ g f^{abc} A^b_i(\vec{x}) \Pi^c_i(\vec{x}) \right) \; .
\label{gausslaw}
\end{eqnarray}
One possibility to arrive at the above formulation of temporal gauge
Yang-Mills theory is to directly deduce it from the Lagrangian
(\ref{lagrangian}), as we have just outlined, which involves right
from the start to introduce a Hilbert space of wave functionals
that depend on the spatial components of the gauge field, only.

It is worthwhile to point out that, alternatively,
one can start from a Lagrangian containing an additional
(covariant gauge) gauge fixing term, and first introduce a
Hilbert space of wave functionals depending on all $d+1$ spacetime
components of the gauge field. The wave functionals then, in addition
to Gauss law, have to fulfill the Lorentz condition,
which can be used to obtain the above temporal gauge formulation
by a separation of variables \cite{che87}.

The nontrivial aspects of the temporal gauge formulation
now arise by the fact that the condition $A^a_0=0$ does not
fix the gauge completely, and there is a residual gauge symmetry
consisting in the subgroup of all time independent gauge
transformations (that tend to 1 at spatial infinity in the continuum
or are periodic on a torus).
The Gauss law operators $G^a(\vec{x})$ defined in (\ref{gausslaw})
are the generators of this residual gauge symmetry, i.e. we have
\begin{eqnarray}
[G^a(\vec{x}),H]=0
\label{gaussconserve}
\end{eqnarray}
and for infinitesimal $\Theta^a(\vec{x})$ (tending to zero at
spatial infinity in the continuum or being periodic on a torus)
\begin{eqnarray}
&& U[\Theta]=e^{-i\int d^d x \Theta^b(\vec{x}) G^b(\vec{x})} \; ,
\nonumber \\
&& U^{-1}[\Theta]A^a_i(\vec{x})U[\Theta]
=g f^{abc} \Theta^b(\vec{x}) A^c_i(\vec{x})
- \partial_{x_i} \Theta^a(\vec{x}) \; ,
\nonumber \\
&& U^{-1}[\Theta]\Pi^a_i(\vec{x})U[\Theta]
=g f^{abc} \Theta^b(\vec{x}) \Pi^c_i(\vec{x}) \; .
\label{resgaugetraf}
\end{eqnarray}
Condition (\ref{gausslaw}) thus requires the physical
states to be gauge invariant, i.e.
\begin{eqnarray}
U[\Theta] |\Psi_{ph} \rangle = |\Psi_{ph} \rangle \; ,
\label{stateinv}
\end{eqnarray}
which also implies that the physical states are non-normalizable.

The residual gauge freedom can now be dealt with in different ways.
One possibility is to eliminate the remaining unphysical degrees of
freedom and to formulate the theory entirely in terms of unconstrained
variables. This can either be achieved by a change of variables
\cite{chr80, che87} or by applying unitary gauge fixing
transformations \cite{len94}. A formulation in terms of unconstrained
degrees of freedom as a starting point for nonperturbative calculations
has the advantage, that the Gauss law (\ref{gausslaw}) does not
have to be observed as an additional constraint in an approximation
scheme, since it is already implemented in the choice of variables.
Furthermore, the wave functionals defined with respect to the
unconstrained variables are then normalizable.
However, such formulations have the disadvantage of leading to highly
complicated (in general non-polynomial) expressions for the
corresponding quantum Hamiltonians \cite{chr80}.
The application of correlation dynamics is, however, restricted to
simple forms of the Hamiltonian containing only low order polynomials
in the fields.

We will thus employ a different method for dealing with the
residual gauge freedom (\ref{gaussconserve}), (\ref{resgaugetraf}), which
consists in introducing a nontrivial integral measure for
matrix elements between physical states obeying (\ref{gausslaw})
by means of the usual Fadeev-Popov method
(cf. \cite{ros84} and for a simple example Appendix \ref{simpleex}).
We start by considering the matrix element of a gauge
invariant operator $O_{inv}$ (i.e., $[O_{inv},G^a(\vec{x})]=0$)
between two physical states $|\Phi_{ph}\rangle$ and
$|\Psi_{ph}\rangle$:
\begin{eqnarray}
\langle \Phi_{ph}| O_{inv} |\Psi_{ph} \rangle
=\int {\cal D}\mu[A] \; \Phi^*_{ph}[A] O^A_{inv} \Psi_{ph}[A]
\; , \; \; O^A_{inv} \Psi_{ph}[A]
\equiv \langle A | O_{inv} |\Psi_{ph}\rangle \; .
\label{startmatel}
\end{eqnarray}
Since the integrand in (\ref{startmatel})
is gauge invariant, in order to obtain a
well defined result it is sufficient to choose the integral measure
${\cal D}\mu[A]$ to contain a factor of
the inverse of the (infinite) volume of the residual gauge group, i.e.
\begin{eqnarray}
{\cal D}\mu[A]=\left( \int {\cal D}h[U] \right)^{-1} {\cal D}A \; ,
\label{invariantintmeasure}
\end{eqnarray}
so that the physical states can be chosen to be normalized in
the following way:
\begin{eqnarray}
\langle \Psi_{ph}|\Psi_{ph}\rangle=\int {\cal D}\mu[A] \;
\Psi^*_{ph}[A]\Psi_{ph}[A]
=(\int {\cal D}h[U])^{-1} \int {\cal D}A \;
\Psi^*_{ph}[A]\Psi_{ph}[A] =1 \; .
\label{normalization}
\end{eqnarray}
Here ${\cal D}h[U]$ is the Haar measure for the integration
over the residual gauge group and
${\cal D}A=\prod_{\vec{x},i,a} dA^a_i(\vec{x})$.
The integral measure ${\cal D}\mu[A]$, however, still leads to
ill defined matrix elements for operators that are not gauge invariant.
In order to retain the matrix elements of gauge invariant operators
defined by ${\cal D}\mu[A]$
and at the same time to have well defined matrix elements of gauge
dependent operators, we now apply the Fadeev-Popov method
to the integral (\ref{startmatel}). We first define a Fadeev-Popov
determinant $\Delta[A]$ by
\begin{eqnarray}
1=\Delta[A] \int {\cal D}h[U] \; \delta[F[A^U]] \; ,
\label{fadpopdet}
\end{eqnarray}
where $F[A]=0$ is some purely spatial gauge fixing condition and $A^U$
denotes the field $A$ transformed by the gauge transformation $U$.
We then get in the usual manner
\begin{eqnarray}
&& \langle \Phi_{ph}| O_{inv} |\Psi_{ph} \rangle
=\int {\cal D}\mu[A] \; \Delta[A] \int {\cal D}h[U] \;
\delta[F[A^U]] \Phi^*_{ph}[A] O^A_{inv} \Psi_{ph}[A]
\nonumber \\
&& = \int {\cal D}h[U] \int {\cal D}\mu[A^U] \; \Delta[A^U]
\delta[F[A^U]] \Phi^*_{ph}[A^U] O^{A^U}_{inv} \Psi_{ph}[A^U]
\nonumber \\
&& = \int {\cal D}A \; \Delta[A] \delta[F[A]]
\Phi^*_{ph}[A] O^A_{inv} \Psi_{ph}[A] \; .
\label{fadpoptrick}
\end{eqnarray}
One could at this stage proceed to
smear out the functional $\delta$-distribution
in the usual way, but this is not required for our further analysis.
Eq. (\ref{fadpoptrick}) defines an integral measure
\begin{eqnarray}
{\cal D}\nu[A]={\cal D}A \; \Delta[A] \delta[F[A]] \; ,
\label{intmeasure}
\end{eqnarray}
that yields the same results as ${\cal D}\mu[A]$ for the matrix elements
of gauge invariant operators and at the same time gives well defined
finite results for the matrix elements of gauge dependent operators.

The use of (\ref{intmeasure}) for the definition of matrix
elements in the space of physical states now has important
consequences for the hermiticity properties of gauge invariant
operators such as the Hamiltonian (\ref{hamiltonian}).
It is hermitean in the space of normalizable states,
and retains this property also in the space of
physical states \cite{che87, ros84}.
However, if it stands to the left of a gauge
dependent operator (which leads out of the space of physical
states into the space of gauge dependent non-normalizable states),
hermiticity will be lost (cf. Appendix \ref{simpleex}).

The standard example for the non-hermiticity of an operator,
that is hermitean in the space of normalizable states,
is the expectation value of the commutator
$[G^a(\vec{x}),A^b_i(\vec{y})]$, where the (gauge-covariant)
Gauss law operator loses its hermiticity
\cite{ros84, hos84, bia84}.

In order to demonstrate the above mentioned non-hermiticity
of the (gauge-invariant) Hamiltonian we consider the time evolution
of the equal-time expectation value\\
$\langle \Psi_{ph}|\partial_{x_i} A^a_i(\vec{x})
\partial_{y_j} A^b_j(\vec{y})|\Psi_{ph} \rangle$
in the abelian limit ($g=0$). If we assume the system to be
in an eigenstate of the Hamiltonian,
\begin{eqnarray}
H|\Psi_{ph}\rangle=E|\Psi_{ph}\rangle
\Leftrightarrow H^A \Psi_{ph}[A]=E \Psi_{ph}[A] \; ,
\label{eigenstate}
\end{eqnarray}
and work in the Schr\"odinger picture, we have
\begin{eqnarray}
\lefteqn{
\partial_t \langle \Psi_{ph} |\partial_{x_i} A^a_i(\vec{x})
\partial_{y_j} A^b_j(\vec{y}) |\Psi_{ph} \rangle |_{S.p.} }
\nonumber \\
&& =\int {\cal D}\nu[A] \left( -i H^A \Psi_{ph}[A] \right)^*
\partial_{x_i} A^a_i(\vec{x}) \partial_{y_j} A^b_j(\vec{y})
\Psi_{ph}[A]
\nonumber \\
&&+ \int {\cal D}\nu[A] \Psi^*_{ph}[A]
\partial_{x_i} A^a_i(\vec{x}) \partial_{y_j} A^b_j(\vec{y})
\left( -i H^A \Psi_{ph}[A] \right) = 0
\label{schrodinger}
\end{eqnarray}
(for simplicity, we use the same notation for classical fields
and field operators).
In contrast to that, the Heisenberg picture yields
\begin{eqnarray}
&&\partial_t \langle \Psi_{ph} |\partial_{x_i} A^a_i(\vec{x})
\partial_{y_j} A^b_j(\vec{y}) |\Psi_{ph} \rangle
|_{H.p.}=\langle \Psi_{ph} | \frac{1}{i}
[\partial_{x_i} A^a_i(\vec{x}) \partial_{y_j} A^b_j(\vec{y}), H]
| \Psi_{ph} \rangle
\nonumber \\
&& = \langle \Psi_{ph} | \partial_{x_i} \Pi^a_i(\vec{x})
\partial_{y_j} A^b_j(\vec{y}) | \Psi_{ph} \rangle
+ \langle \Psi_{ph} | \partial_{x_i} A^a_i(\vec{x})
\partial_{y_j} \Pi^b_j(\vec{y}) | \Psi_{ph} \rangle
\nonumber \\
&& = -i \partial_{x_i} \partial_{y_j} \delta(\vec{x}-\vec{y})
\delta^{ab} \delta_{ij} = i \vec{\partial_{x}}^2
\delta(\vec{x}-\vec{y}) \delta^{ab}\; ,
\label{heisenberg}
\end{eqnarray}
where we have used (\ref{commutator}) and
the abelian limit of (\ref{gausslaw}).
The discrepancy between (\ref{schrodinger})
and (\ref{heisenberg}) can be understood
in terms of wave functionals, if one looks at the transformation
from the Schr\"odinger picture to the Heisenberg picture at the time
where the two pictures coincide: in order to
get from one picture to the other, one has to partially integrate
the functional derivatives appearing in the electric
field energy $\frac{1}{2} \int d^d x \left[ \Pi^a_i(\vec{x})
\Pi^a_i(\vec{x}) \right]^A = -\frac{1}{2} \int d^d x
\frac{\delta}{\delta A^a_i(\vec{x})}
\frac{\delta}{\delta A^a_i(\vec{x})}$,
which gives additional nontrivial
terms when the derivatives act on the $A$-dependent terms appearing
in the integral measure (\ref{intmeasure}):
\begin{eqnarray}
\lefteqn{
\int {\cal D}\nu[A] \left( \int d^d z
\frac{\delta}{\delta A^c_k(\vec{z})}
\frac{\delta}{\delta A^c_k(\vec{z})}
\Psi_{ph}[A] \right)^*
\partial_{x_i} A^a_i(\vec{x}) \partial_{y_j} A^b_j(\vec{y})
\Psi_{ph}[A]}
\nonumber \\
&&= \int d^d z \int {\cal D}A \; \Psi_{ph}[A]^* \times
\nonumber \\
&& \times \frac{\delta}{\delta A^c_k(\vec{z})}
\frac{\delta}{\delta A^c_k(\vec{z})}
\left( \Delta[A] \delta[F[A]]
\partial_{x_i} A^a_i(\vec{x}) \partial_{y_j} A^b_j(\vec{y})
\Psi_{ph}[A] \right)
\nonumber \\
&&= \int d^d z \int {\cal D}A \; \Psi_{ph}[A]^* \times
\nonumber \\
&& \times \left\{
\left( \frac{\delta}{\delta A^c_k(\vec{z})}
\frac{\delta}{\delta A^c_k(\vec{z})}  \Delta[A] \delta[F[A]] \right)
\partial_{x_i} A^a_i(\vec{x}) \partial_{y_j} A^b_j(\vec{y}) \Psi_{ph}[A]
\right.
\nonumber \\
&& \left. +2\left(\frac{\delta}{\delta A^c_k(\vec{z})}
\Delta[A] \delta[F[A]] \right)
\left( \frac{\delta}{\delta A^c_k(\vec{z})}
\partial_{x_i} A^a_i(\vec{x}) \partial_{y_j} A^b_j(\vec{y}) \Psi_{ph}[A]
\right) \right\}
\nonumber \\
&& + \int {\cal D}\nu[A] \Psi_{ph}[A]^* \left(
\int d^d z \frac{\delta}{\delta A^c_k(\vec{z})}
\frac{\delta}{\delta A^c_k(\vec{z})}
\partial_{x_i} A^a_i(\vec{x}) \partial_{y_j} A^b_j(\vec{y}) \Psi_{ph}[A]
\right) \; .
\label{partialintegrate}
\end{eqnarray}
A different point of
view leading to the same result is the following \cite{gir86}:
suppose instead of (\ref{intmeasure})
we would use (\ref{invariantintmeasure}) and regularize the matrix
elements of gauge dependent operators by only integrating over
a finite range of field amplitudes. The above mentioned additional
nontrivial terms then arise as hermiticity-destroying
surface contributions in the partial integration.

In general, we note that the Heisenberg time evolution of gauge
dependent quantities does not give the same result as the
Schr\"odinger time evolution and
yields time dependencies even if the system is in an equilibrium situation
(e.g., an eigenstate of (\ref{hamiltonian})). Also, in the
Schr\"odinger picture the gauge fixing condition $F[A]=0$ is
fulfilled at all times, while in the Heisenberg picture it is
only fulfilled at the time of coincidence with the Schr\"odinger
picture. However, if one
considers the time evolution of gauge invariant quantities,
both pictures give identical results (e.g., if one reconstructs
an equilibrium expectation value of a gauge invariant quantity from
gauge dependent quantities, all time dependencies in the Heisenberg
picture must cancel out). Thus, since all physical obervables are gauge
invariant, the Schr\"odinger and the Heisenberg picture
are of course still equivalent.
From a technical point of view, in order to generate equations of
motion for Green functions from the Schr\"odinger picture, one would
have to explicitly carry out the partial integration outlined above,
which leads to nontrivial additional terms (containing the
Fadeev-Popov (ghost) and gauge fixing contributions which arise due to
the fixing of the residual spatial gauge freedom).

One immediate consequence of the above considerations is the appearance
of a term explicitly breaking time translation invariance in the
free propagator as determined by the abelian limit of the
Heisenberg equations of motion.
If we choose the condition that fixes the residual
gauge freedom to be the Coulomb gauge condition
\begin{eqnarray}
F[A]=\partial_{x_i} A^a_i(\vec{x})=0 \; ,
\label{coulombfix}
\end{eqnarray}
we obtain (in the continuum), if the Heisenberg and the
Schr\"odinger picture coincide at time $t_0$,
\begin{eqnarray}
\lefteqn{
\langle T A^a_i(\vec{x},t) A^b_j(\vec{x}',t') \rangle
=\frac{\delta^{ab}}{2}\int \frac{d^d k}{(2\pi)^d}
e^{i\vec{k}(\vec{x}-\vec{x}')}
\left[\frac{1}{|\vec{k}|}e^{-i|\vec{k}||t-t'|}
\left(\delta_{ij}-\frac{k_i k_j}{\vec{k}^2} \right) \right.}
\nonumber \\
&& \left. -\frac{i}{2}
\left( |t-t'|+(t+t')-2 t_0 \right) \frac{k_i k_j}{\vec{k}^2}
\right] \; ,
\label{mypropagator}
\end{eqnarray}
where $T$ denotes the time ordering operator.
A derivation of this result within the framework of canonical
quantization can e.g. be found in \cite{gir86, yam87}.
We note in passing, that (\ref{coulombfix}) at an intermediate time
$t_0 \neq \pm \infty$ might still not be a sufficient gauge fixing
condition due to Gribov ambiguities. A detailed discussion of this
issue, however, is beyond the scope of this article.

The first authors to come up with a longitudinal propagator as
contained in (\ref{mypropagator}) were Carraciolo, Curci and Menotti
\cite{car82}, who obtained their result in a more heuristic way
by first writing down the most general solution of the equation
of motion. This solution, which is in addition required
to be symmetric in $t$ and $t'$ and to fulfill the condition
$\partial_{t}^2 \langle T A^a_{L,i}(\vec{x},t)
A^b_{L,j}(\vec{x}',t') \rangle = -\partial_{t} \partial_{t'}
\langle T A^a_{L,i}(\vec{x},t) A^b_{L,j}(\vec{x}',t') \rangle$,
reads
\begin{eqnarray}
\langle T A^a_{L,i}(\vec{x},t) A^b_{L,j}(\vec{x}',t') \rangle
=-\frac{i}{2} \left( |t-t'|+\alpha (t+t')+\gamma \right)
\frac{\delta^{ab}}{2}\int \frac{d^d k}{(2\pi)^d}
e^{i\vec{k}(\vec{x}-\vec{x}')} \frac{k_i k_j}{\vec{k}^2} \; .
\label{ccmgeneral}
\end{eqnarray}
In (\ref{ccmgeneral}) the subscript $L$ stands for the longitudinal
part of the gauge field. Now perturbatively evaluating the Wilson
loop up to order $g^4$ and comparing the final result to that obtained
in Feynman and Coulomb gauge, the requirement of gauge invariance
leads to the condition $\alpha=\pm 1$, while $\gamma$ gives
no contribution to the Wilson loop.
The result $\alpha=-1$ is obtained in the above mentioned canonical
derivation if instead of (\ref{gausslaw}) one uses the condition
\begin{eqnarray}
\langle \Psi_{ph}|G^a(\vec{x})=0 \Leftrightarrow
G^a(\vec{x})^\dagger|\Psi_{ph}\rangle=0 \; ,
\label{gaussdagger}
\end{eqnarray}
which also leads to a consistent formulation of the theory.
Note that $G^a(\vec{x}) \neq G^a(\vec{x})^\dagger$ and that
both (\ref{gausslaw}) and (\ref{gaussdagger}) cannot be imposed
simultaneously, since this would contradict (\ref{commutator}).
The authors in \cite{car82} then conclude that contact with the
Coulomb gauge can only be made at $t_0=\pm \infty$
(or equivalently at one of the endpoints of a finite time strip).
This is true if the Coulomb gauge condition requires every Green
function containing a longitudinal gauge field $A^a_{L,i}(\vec{x},t_0)$
to vanish. This, however, would imply
that $|\gamma|$ has to be taken to infinity at some stage of the
calculation.

In order to shed some light on the last point we will conclude
this section by briefly discussing path integral quantization
in temporal gauge. After carrying out the $A^a_0=0$ gauge fixing,
one is left with a ghost-free path integral, that however
still contains an integration over the residual gauge degrees
of freedom. If one now fixes this residual gauge freedom via the
Fadeev-Popov method with the Coulomb gauge condition at time $t_0$, i.e.
\begin{eqnarray}
\partial_{x_i} A^a_i(\vec{x},t_0)=0 \; ,
\label{pathcoulombfix}
\end{eqnarray}
one arrives at \cite{ler87}
\begin{eqnarray}
\langle T A^a_{L,i}(\vec{x},t) A^b_{L,j}(\vec{x}',t') \rangle
=-\frac{i}{2} \left( |t-t'|-|t-t_0|-|t'-t_0| \right)
\frac{\delta^{ab}}{2}\int \frac{d^d k}{(2\pi)^d}
e^{i\vec{k}(\vec{x}-\vec{x}')} \frac{k_i k_j}{\vec{k}^2}
\label{leroypropagator}
\end{eqnarray}
for the longitudinal part of the free propagator.
Moreover, in the nonabelian theory the gauge fixing condition
(\ref{pathcoulombfix}) leads to ghosts that live
on a spacetime hypersurface of fixed time $t_0$.
Eq. (\ref{leroypropagator}) gives zero if $t_0$ lies in between $t$ and
$t'$ and it agrees with (\ref{ccmgeneral}) for $\alpha=-1$ if
$t,t'>t_0$ and for $\alpha=+1$ if $t,t'<t_0$. Agreement between
(\ref{leroypropagator}) and (\ref{mypropagator}) can thus be achieved
only in the limit $|t_0| \rightarrow \infty$. The disagreement
between both results in case of a finite value of $t_0$, however,
can easily be understood to arise from the difference in the
way the Coulomb gauge condition is implemented in our canonical
approach and in the path integral approach.
The use of the integral measure (\ref{intmeasure}) induced by
(\ref{coulombfix}) for the definition of matrix elements
between physical states only leads to vanishing Green functions
if $A^a_{L,i}(\vec{x},t_0)$ stands to the left of all other operators
(for a simple example we refer the reader to Appendix \ref{simpleex}),
while the implementation of (\ref{coulombfix}) in the path integral
formulation forces all Green functions containing
$A^a_{L,i}(\vec{x},t_0)$ to vanish irrespective of
the operator ordering
(that arises from the time ordering being carried out).

We therefore conclude that there is no a priori contradiction
between (\ref{mypropagator}) and (\ref{leroypropagator}),
since both are in fact derived from inequivalent gauge
fixing conditions.

On one hand, we are now in a position to formulate simple Heisenberg
dynamics employing the Hamiltonian (\ref{hamiltonian}), on the other
hand we have to deal with explicit time dependencies and the Gauss law
(\ref{gausslaw}) as an additional constraint.
\section{Equations of motion for connected equal-time Green functions}
\label{eomcgf}
We now turn to the derivation of a closed set of
equations of motion for connected equal-time Green functions
within the approximation scheme of correlation dynamics.
The derivation of these equations up to the 4-point level
has already been presented in \cite{wan95} for the full SU(N) QCD
Lagrangian containing gauge bosons and massless fermions,
however, in order to make this article self-contained
we will repeat the main steps
for the pure gauge boson sector in this section. We also
point out that in the present article we no longer adopt the
quantization scheme outlined in \cite{wan95, wan194},
which is based on the ''weak'' version of Gauss law, i.e. requiring
only $\langle G^a(\vec{x}) \rangle=0$.

The starting point for our derivation are the operator equations
of motion that arise from the Heisenberg equation
(if not explicitly stated otherwise, all time derivatives will
from now on refer to the Heisenberg picture)
\begin{eqnarray}
i\partial_t O(t)=[O(t),H] \; .
\label{heisenbergequation}
\end{eqnarray}
With (\ref{heisenbergequation}), (\ref{hamiltonian})
and (\ref{commutator}) one obtains
\begin{eqnarray}
\partial_t A^a_i(\vec{x})=\Pi^a_i(\vec{x})
\label{aeqom}
\end{eqnarray}
and
\begin{eqnarray}
\lefteqn{
\partial_t \Pi^a_i(\vec{x})= \left( \vec{\nabla}^2 \delta_{ik}
- \partial_{x_i} \partial_{x_k} \right) A^a_k(\vec{x})}
\nonumber \\
&&+ 2 g f^{abc} A^b_k(\vec{x}) \partial_{x_k} A^c_i(\vec{x})
+   g f^{abc} A^c_i(\vec{x}) \partial_{x_k} A^b_k(\vec{x})
-   g f^{abc} A^b_k(\vec{x}) \partial_{x_i} A^c_k(\vec{x})
\nonumber \\
&&+ g^2 f^{abc} f^{cde} A^b_k(\vec{x}) A^d_k(\vec{x}) A^e_i(\vec{x})
\; ,
\label{peqom}
\end{eqnarray}
where all operators are taken at the same time $t$, which has been
suppressed in our notation. In the same way one obtains the equations of
motion for higher order equal-time operator products, e.g.
\begin{eqnarray}
\lefteqn{
\partial_t \left( A^a_i(\vec{x}) A^b_j(\vec{y}) \right)
=(\partial_t A^a_i(\vec{x}) ) A^b_j(\vec{y})
+ A^a_i(\vec{x}) ( \partial_t A^b_j(\vec{y}) ) }
\nonumber \\
&&=\Pi^a_i(\vec{x}) A^b_j(\vec{y}) + A^a_i(\vec{x}) \Pi^b_j(\vec{y})
\; ,
\label{aaeqom}
\end{eqnarray}
\begin{eqnarray}
\lefteqn{
\partial_t \left( \Pi^a_i(\vec{x}) A^b_j(\vec{y}) \right)
=(\partial_t \Pi^a_i(\vec{x}) ) A^b_j(\vec{y})
+ \Pi^a_i(\vec{x}) ( \partial_t A^b_j(\vec{y}) )}
\nonumber \\
&&= \left\{ \left( \vec{\nabla}^2 \delta_{ik}
- \partial_{x_i} \partial_{x_k} \right) A^a_k(\vec{x}) \right.
\nonumber \\
&&+ 2 g f^{acd} A^c_k(\vec{x}) \partial_{x_k} A^d_i(\vec{x})
+   g f^{acd} A^d_i(\vec{x}) \partial_{x_k} A^c_k(\vec{x})
-   g f^{acd} A^c_k(\vec{x}) \partial_{x_i} A^d_k(\vec{x})
\nonumber \\
&& \left.
+ g^2 f^{acd} f^{def} A^c_k(\vec{x}) A^e_k(\vec{x}) A^f_i(\vec{x})
\right\} A^b_j(\vec{y})
\nonumber \\
&& + \Pi^a_i(\vec{x}) \Pi^b_j(\vec{y}) \; ,
\label{paeqom}
\end{eqnarray}
\begin{eqnarray}
\lefteqn{
\partial_t \left( \Pi^a_i(\vec{x}) \Pi^b_j(\vec{y}) \right)
=(\partial_t \Pi^a_i(\vec{x}) ) \Pi^b_j(\vec{y})
+ \Pi^a_i(\vec{x}) ( \partial_t \Pi^b_j(\vec{y}) )}
\nonumber \\
&&= \left\{ \left( \vec{\nabla}^2 \delta_{ik}
- \partial_{x_i} \partial_{x_k} \right) A^a_k(\vec{x}) \right.
\nonumber \\
&&+ 2 g f^{acd} A^c_k(\vec{x}) \partial_{x_k} A^d_i(\vec{x})
+   g f^{acd} A^d_i(\vec{x}) \partial_{x_k} A^c_k(\vec{x})
-   g f^{acd} A^c_k(\vec{x}) \partial_{x_i} A^d_k(\vec{x})
\nonumber \\
&& \left.
+ g^2 f^{acd} f^{def} A^c_k(\vec{x}) A^e_k(\vec{x}) A^f_i(\vec{x})
\right\} \Pi^b_j(\vec{y})
\nonumber \\
&& + \Pi^a_i(\vec{x}) \left\{ \left( \vec{\nabla}^2 \delta_{ik}
- \partial_{y_i} \partial_{y_k} \right) A^b_k(\vec{y}) \right.
\nonumber \\
&&+ 2 g f^{bcd} A^c_k(\vec{y}) \partial_{y_k} A^d_i(\vec{y})
+   g f^{bcd} A^d_i(\vec{y}) \partial_{y_k} A^c_k(\vec{y})
-   g f^{bcd} A^c_k(\vec{y}) \partial_{y_i} A^d_k(\vec{y})
\nonumber \\
&& \left.
+ g^2 f^{bcd} f^{def} A^c_k(\vec{y}) A^e_k(\vec{y}) A^f_i(\vec{y})
\right\}
\; , \; \; {\rm etc.}
\label{ppeqom}
\end{eqnarray}
We thus arrive at an infinite coupled hierarchy
of operator equations of motion, where the time derivative
of each product of $n$ field operators containing at least one conjugate
field momentum couples to products of $n$, $n+1$, and $n+2$ field
operators; the latter correspond to the free propagation
and the gauge boson selfinteraction via the 3 point vertex
and the 4 point vertex, respectively.

Eqs. (\ref{aeqom})-(\ref{ppeqom})
and the corresponding equations for higher order operator products
now straightforwardly induce an infinite hierarchy
of equations of motion for full equal-time Green functions,
the latter being defined as the expectation values
of the above equal-time operator products.

The most general expression for the expectation value
of an operator $O$ is
\begin{eqnarray}
&&\langle O \rangle = Tr \left( O \rho \right) \; ,
\nonumber \\
&&\rho=\sum_{\alpha} \rho_{\alpha} P_{\Psi_{ph,\alpha}} \; , \; \;
\sum_{\alpha} \rho_{\alpha}=1 \; , \; \; \rho_{\alpha} \geq 0
\; \forall \alpha
\label{traceo}
\end{eqnarray}
where $P_{\Psi_{ph,\alpha}}$ is the projection operator onto
the physical state $|\Psi_{ph,\alpha}\rangle$ and the
set of states $\left\{ |\Psi_{ph,\alpha}\rangle \right\}$ is
(without loss of generality) assumed to be an ortho-normalized basis
system for the space of physical states, that diagonalizes the
(non-equilibrium) statistical density operator $\rho$.
The trace is only taken with respect to the space of physical states
and each matrix element appearing in the trace is defined
via the integral measure (\ref{intmeasure}).
The explicit form of (\ref{traceo}) in terms of wave functionals
then reads
\begin{eqnarray}
\langle O \rangle = \sum_{\alpha} \rho_\alpha \int {\cal D}\nu[A]
\Psi^*_{ph,\alpha}[A] O^A \Psi_{ph,\alpha}[A] \; .
\label{explicittraceo}
\end{eqnarray}

Since we work in the Heisenberg picture, $\rho$ is time
independent and we have
\begin{eqnarray}
\partial_t Tr \left( O(t) \rho \right)
=Tr \left( \partial_t O(t) \rho \right)
=Tr \left( \frac{1}{i} [O(t),H] \rho \right)
\label{dtoutoftrace}
\end{eqnarray}
for any operator $O(t)$.
The equations of motion for full equal-time Green functions
corresponding to (\ref{aeqom}), (\ref{peqom}), etc.,
are therefore given by
\begin{eqnarray}
\partial_t \langle A^a_i(\vec{x}) \rangle
=\langle \Pi^a_i(\vec{x}) \rangle \; ,
\label{greenaeqom}
\end{eqnarray}
\begin{eqnarray}
\lefteqn{
\partial_t \langle \Pi^a_i(\vec{x}) \rangle
= \left( \vec{\nabla}^2 \delta_{ik}
- \partial_{x_i} \partial_{x_k} \right) \langle A^a_k(\vec{x}) \rangle}
\nonumber \\
&&+2 g f^{abc}\langle A^b_k(\vec{x})\partial_{x_k}A^c_i(\vec{x})\rangle
+ g f^{abc}\langle A^c_i(\vec{x}) \partial_{x_k} A^b_k(\vec{x})\rangle
- g f^{abc}\langle A^b_k(\vec{x}) \partial_{x_i} A^c_k(\vec{x})\rangle
\nonumber \\
&&+ g^2 f^{abc}f^{cde}
\langle A^b_k(\vec{x}) A^d_k(\vec{x}) A^e_i(\vec{x}) \rangle
\; , \; \; {\rm etc.}
\label{greenpeqom}
\end{eqnarray}

The hierarchy consisting of (\ref{greenaeqom}), (\ref{greenpeqom})
and the corresponding higher order equations can be seen as
the field theoretical analogue to the BBGKY (Born Bogoliubov
Green Kirkwood Yvon) density matrix hierarchy known from
nonrelativistic many body theory. The only new feature is the
difference between the hierarchies arising from the Heisenberg and
the Schr\"odinger picture, as it has already been discussed
in Section \ref{quantization}.

We note in passing that, in the Schr\"odinger picture,
the von Neumann equation remains valid also for the
description of the time evolution of gauge dependent operators,
since for its derivation only the hermiticity of $H$ in the
space of physical states is needed (all manipulations in the
derivation are carried out to the right of $O$). One could
then be tempted to recover the Heisenberg picture result by
using invariance of the trace under cyclic permutations.
This, however, is incorrect,
since the trace cannot be extended to run over a large enough
space of states in order to allow for an application of the
completeness relation as required for the proof
of cyclic permutation invariance.

For practical calculations, the infinite coupled hierarchy of
Heisenberg equations of motion for full equal-time Green
functions has to be truncated. In our approach, this is done
using the cluster expansion for $n$-point Green functions
\cite{wan294}, i.e. their decomposition into sums of products
of connected Green functions. The cluster expansion is expected
to exhibit good convergence properties with respect to the order
of connected Green functions taken into account, if the system
is in a single phase (localized wave functional) configuration.
On the other hand, if the system is in a configuration with
coexistence of different phases (e.g. in a groundstate wave
functional that is highly delocalized due to tunneling),
the higher order connected $n$-point functions
are expected to contribute to the full $n$-point functions in the same
order of magnitude as the corresponding disconnected parts \cite{pet97}.

The explicit form of the cluster expansion can be derived from the
generating functionals of full and connected Green functions,
$Z[J,\sigma]$ and $W[J,\sigma]$, given by
\begin{eqnarray}
&& Z[J,\sigma]=Tr\left\{ \left[ T e^{i \int d^{d+1}x
\left(J^a_i(x) A^a_i(x)
+ \sigma^a_i(x) \Pi^a_i(x) \right)} \right]
\rho \right\} \nonumber \\
&& {\rm and} \; \; Z[J,\sigma]=e^{W[J,\sigma]} \; ,
\label{genfunctionals}
\end{eqnarray}
where $x=(\vec{x},t)$.
We start with the cluster expansions of time ordered Green functions
with different time arguments:
\begin{eqnarray}
\langle A^a_i(x) \rangle =\langle A^a_i(x) \rangle_c \; , \; \;
\langle \Pi^a_i(x) \rangle=\langle \Pi^a_i(x) \rangle_c \; ,
\label{onepointclusterex}
\end{eqnarray}
\begin{eqnarray}
\lefteqn{
\langle T A^{a_1}_{i_1}(x_1) A^{a_2}_{i_2}(x_2) \rangle
=\lim_{J,\sigma \rightarrow 0}
\frac{\delta}{i\delta J^{a_1}_{i_1} (x_1)}
\frac{\delta}{i\delta J^{a_2}_{i_2} (x_2)} e^{W[J,\sigma]} }
\nonumber \\
&& = \lim_{J,\sigma \rightarrow 0} e^{W[J,\sigma]} \left\{ \left(
\frac{\delta}{i\delta J^{a_1}_{i_1} (x_1)}
\frac{\delta}{i\delta J^{a_2}_{i_2} (x_2)} W[J,\sigma] \right) \right.
\nonumber \\
&&+ \left. \left( \frac{\delta}{i\delta J^{a_1}_{i_1} (x_1)}
W[J,\sigma] \right)
\left( \frac{\delta}{i\delta J^{a_2}_{i_2} (x_2)} W[J,\sigma] \right)
\right\}
\nonumber \\
&& = \langle T A^{a_1}_{i_1}(x_1) A^{a_2}_{i_2}(x_2) \rangle_c
+ \langle A^{a_1}_{i_1}(x_1) \rangle
\langle A^{a_2}_{i_2}(x_2) \rangle \; ,
\label{aaclusterex}
\end{eqnarray}
where $\langle \cdot \rangle_c$ denotes the connected part of the
expectation value. Analogously we obtain
\begin{eqnarray*}
\langle T \Pi^{a_1}_{i_1}(x_1) A^{a_2}_{i_2}(x_2) \rangle
= \langle T \Pi^{a_1}_{i_1}(x_1) A^{a_2}_{i_2}(x_2) \rangle_c
+ \langle \Pi^{a_1}_{i_1}(x_1) \rangle
\langle A^{a_2}_{i_2}(x_2) \rangle \; ,
\end{eqnarray*}
\begin{eqnarray*}
\langle T \Pi^{a_1}_{i_1}(x_1) \Pi^{a_2}_{i_2}(x_2) \rangle
= \langle T \Pi^{a_1}_{i_1}(x_1) \Pi^{a_2}_{i_2}(x_2) \rangle_c
+ \langle \Pi^{a_1}_{i_1}(x_1) \rangle
\langle \Pi^{a_2}_{i_2}(x_2) \rangle \; ,
\end{eqnarray*}
\begin{eqnarray}
\lefteqn{
\langle T A^{a_1}_{i_1}(x_1) A^{a_2}_{i_2}(x_2)
A^{a_3}_{i_3}(x_3) \rangle =
\langle T A^{a_1}_{i_1}(x_1) A^{a_2}_{i_2}(x_2)
A^{a_3}_{i_3}(x_3) \rangle_c }
\nonumber \\
&&+ \langle T A^{a_1}_{i_1}(x_1) A^{a_2}_{i_2}(x_2) \rangle_c
\langle A^{a_3}_{i_3}(x_3) \rangle
+ \langle T A^{a_1}_{i_1}(x_1) A^{a_3}_{i_3}(x_3) \rangle_c
\langle A^{a_2}_{i_2}(x_2) \rangle
\nonumber \\
&&+ \langle T A^{a_2}_{i_2}(x_2) A^{a_3}_{i_3}(x_3) \rangle_c
\langle A^{a_1}_{i_1}(x_1) \rangle
+ \langle A^{a_1}_{i_1}(x_1) \rangle
\langle A^{a_2}_{i_2}(x_2) \rangle
\langle A^{a_3}_{i_3}(x_3) \rangle \; , \; \; {\rm etc}.
\label{pappaaaclusterex}
\end{eqnarray}
The expressions for equal-time Green functions are obtained by taking
the well defined equal-time limit which yields the desired operator
ordering in the cluster expansions. We arrive at
\begin{eqnarray*}
\langle A^a_i(\vec{x}) \rangle
=\langle A^a_i(\vec{x}) \rangle_c \; , \; \;
\langle \Pi^a_i(\vec{x}) \rangle
=\langle \Pi^a_i(\vec{x}) \rangle_c \; ,
\end{eqnarray*}
\begin{eqnarray*}
\langle A^{a_1}_{i_1}(\vec{x}_1) A^{a_2}_{i_2}(\vec{x}_2) \rangle
= \langle A^{a_1}_{i_1}(\vec{x}_1) A^{a_2}_{i_2}(\vec{x}_2) \rangle_c
+ \langle A^{a_1}_{i_1}(\vec{x}_1) \rangle
\langle A^{a_2}_{i_2}(\vec{x}_2) \rangle \; ,
\end{eqnarray*}
\begin{eqnarray*}
\langle \Pi^{a_1}_{i_1}(\vec{x}_1) A^{a_2}_{i_2}(\vec{x}_2) \rangle
= \langle \Pi^{a_1}_{i_1}(\vec{x}_1) A^{a_2}_{i_2}(\vec{x}_2) \rangle_c
+ \langle \Pi^{a_1}_{i_1}(\vec{x}_1) \rangle
\langle A^{a_2}_{i_2}(\vec{x}_2) \rangle \; ,
\end{eqnarray*}
\begin{eqnarray*}
\langle \Pi^{a_1}_{i_1}(\vec{x}_1) \Pi^{a_2}_{i_2}(\vec{x}_2) \rangle
= \langle \Pi^{a_1}_{i_1}(\vec{x}_1) \Pi^{a_2}_{i_2}(\vec{x}_2) \rangle_c
+ \langle \Pi^{a_1}_{i_1}(\vec{x}_1) \rangle
\langle \Pi^{a_2}_{i_2}(\vec{x}_2) \rangle \; ,
\end{eqnarray*}
\begin{eqnarray}
\lefteqn{
\langle A^{a_1}_{i_1}(\vec{x}_1) A^{a_2}_{i_2}(\vec{x}_2)
A^{a_3}_{i_3}(\vec{x}_3) \rangle =
\langle A^{a_1}_{i_1}(\vec{x}_1) A^{a_2}_{i_2}(\vec{x}_2)
A^{a_3}_{i_3}(\vec{x}_3) \rangle_c }
\nonumber \\
&&+ \langle A^{a_1}_{i_1}(\vec{x}_1) A^{a_2}_{i_2}(\vec{x}_2) \rangle_c
\langle A^{a_3}_{i_3}(\vec{x}_3) \rangle
+ \langle A^{a_1}_{i_1}(\vec{x}_1) A^{a_3}_{i_3}(\vec{x}_3) \rangle_c
\langle A^{a_2}_{i_2}(\vec{x}_2) \rangle
\nonumber \\
&&+ \langle A^{a_2}_{i_2}(\vec{x}_2) A^{a_3}_{i_3}(\vec{x}_3) \rangle_c
\langle A^{a_1}_{i_1}(\vec{x}_1) \rangle
+ \langle A^{a_1}_{i_1}(\vec{x}_1) \rangle
\langle A^{a_2}_{i_2}(\vec{x}_2) \rangle
\langle A^{a_3}_{i_3}(\vec{x}_3) \rangle \; , \; \; {\rm etc}. \; ,
\label{equaltimeclusterex}
\end{eqnarray}
where all time arguments are now equal and have been suppressed in
our notation. In view of their length the cluster expansions of other
Green functions required for our calculations are not explicitly
given here, but e.g. can be found in \cite{wan95}.

We mention that
instead of (\ref{genfunctionals}), one could equivalently start
with a generating functional that only depends on the source
$J^a_i(x)$ and then obtain Green functions containing conjugate
momenta $\Pi^a_i(x)$ by differentiating with respect to the time
coordinate. An alternative for the derivation of
(\ref{equaltimeclusterex}) without the detour via
time ordered Green functions is the use of an equal-time generating
functional, where due to the absence of the time ordering
operator one has to take care of the nonvanishing commutator
between $A^a_i(\vec{x})$ and $\Pi^a_i(\vec{x})$ by
rewriting the exponential containing the sources $J$ and $\sigma$
by means of the Campbell-Baker-Hausdorff-formula \cite{nac97}.

Inserting the cluster expansions into the hierarchy
equations of type (\ref{greenaeqom}), (\ref{greenpeqom}), etc.,
leads to an infinite coupled hierarchy of equations of motion
for connected equal-time Green functions.
Our approximation scheme now consists in truncating this hierarchy
by neglecting all connected $n$-point functions with $n>N$,
which leads to a closed set of coupled, nonlinear equations
of motion for the connected equal-time Green functions up to
the $N$-point level. These equations are referred to as
$N$-point correlation dynamics.
Since the time derivatives of the full $N$-point
functions couple to the full $N+2$-point functions, all cluster
expansions up to the $N+2$-point level are required for the
derivation of the correlation dynamical equations.

In the truncation scheme with $N=1$ we recover the classical
Hamilton equations for temporal gauge Yang-Mills theory
(the 1-point functions are simply the classical fields).
In the $N=2$ truncation scheme we recover the time dependent
Hartree-Fock-Bogoliubov equations (this truncation scheme corresponds
to a Gaussian trial state and to a Gaussian form of the integral measure;
in the present case this, however, contradicts gauge invariance).
The $N=3$ truncation scheme is the lowest order approximation to take
into account the 3-point vertex if the system is in a (global) colour
singlet configuration.

The highest order approximation investigated in this article
is $N=4$. For the straightforward but very lengthy derivation of the
equations of 2, 3, and 4-point correlation dynamics we have developed
computer algebra codes using the Mathematica programming language,
which are also used to generate (highly optimized) Fortran routines
for the time integration of the equations within a finite set
of plane wave states on a torus.
The resulting equations of motion contain several hundred terms
and are therefore not explicitly displayed in this article. The
reader is, however, referred to our previous article \cite{wan95},
which contains the explicit 4-point equations in coordinate
space in a notation that is compactified using permutation operators.
\section{Hierarchy of Gauss law constraints}
\label{gausshierarchy}
In addition to the equations of motion following from the
Heisenberg equation -- discussed in the preceding section --
the equal-time Green functions have to fulfill the conditions
that can be derived from Gauss law (\ref{gausslaw}), which,
as stated in Section \ref{quantization}, is not automatically
contained in the Hamiltonian dynamics of the theory.
The present section is devoted to the discussion of these
Gauss law conditions.

We first note that from (\ref{gausslaw}) we immediately get
\begin{eqnarray}
O G^a(\vec{x}) |\Psi_{ph}\rangle=0
\label{ogausslaw}
\end{eqnarray}
for any product of field operators $O$ and for any physical
state $|\Psi_{ph}\rangle$. For an arbitrary expectation
value (\ref{traceo}) we thus have
\begin{eqnarray}
\langle O G^a(\vec{x}) \rangle=0 \; ,
\label{ogaussexpectation}
\end{eqnarray}
which in principle already comprises the complete set of
equations for equal-time Green functions that can be derived
from Gauss law. It is important to point out that in
(\ref{ogaussexpectation}) $G^a(\vec{x})$ stands to the right
of all other operators and that the operator ordering cannot
be reversed by invoking hermiticity arguments (the Gauss law
operator, although hermitean in the space of normalizable states,
loses this property in the present case \cite{ros84, hos84, bia84}).

If we replace $O$ in (\ref{ogaussexpectation}) by a product of
$n$ field operators, then (due to the structure of $G^a(\vec{x})$)
we get a relation between a full $n+1$- and a full $n+2$-point
equal-time function. Thus, as in the case of the Heisenberg
equations of motion, we are left with an infinite coupled hierarchy
of equations. In the simplest case, $O=1$, we obtain a relation
between 1- and 2-point functions:
\begin{eqnarray}
\langle G^a(\vec{x}) \rangle =
\partial_{x_i} \langle \Pi^a_i(\vec{x}) \rangle
+ g f^{abc} \langle A^b_i(\vec{x}) \Pi^c_i(\vec{x}) \rangle=0 \; .
\label{stident1}
\end{eqnarray}
On the next level, with $O=A^d_j(\vec{y})$ and $O=\Pi^d_j(\vec{y})$,
we get two different relations between 2- and 3-point functions:
\begin{eqnarray}
\partial_{x_i} \langle A^d_j(\vec{y}) \Pi^a_i(\vec{x}) \rangle
+ g f^{abc} \langle
A^d_j(\vec{y}) A^b_i(\vec{x}) \Pi^c_i(\vec{x}) \rangle=0 \; ,
\label{stident2}
\end{eqnarray}
\begin{eqnarray}
\partial_{x_i} \langle \Pi^d_j(\vec{y}) \Pi^a_i(\vec{x}) \rangle
+ g f^{abc} \langle
\Pi^d_j(\vec{y}) A^b_i(\vec{x}) \Pi^c_i(\vec{x}) \rangle=0 \; .
\label{stident3}
\end{eqnarray}
Again one level higher we get three different relations between
3- and 4-point functions by choosing
$O=A^d_j(\vec{y}) A^e_k(\vec{z})$, $O=\Pi^d_j(\vec{y}) A^e_k(\vec{z})$,
and $O=\Pi^d_j(\vec{y}) \Pi^e_k(\vec{z})$. We note that
$O=A^d_j(\vec{y}) \Pi^e_k(\vec{z})$ is redundant, since it can be
obtained by commutation relations.
\begin{eqnarray}
\partial_{x_i} \langle A^d_j(\vec{y}) A^e_k(\vec{z})
\Pi^a_i(\vec{x}) \rangle + g f^{abc} \langle A^d_j(\vec{y})
A^e_k(\vec{z}) A^b_i(\vec{x}) \Pi^c_i(\vec{x}) \rangle=0 \; ,
\label{stident4}
\end{eqnarray}
\begin{eqnarray}
\partial_{x_i} \langle \Pi^d_j(\vec{y}) A^e_k(\vec{z})
\Pi^a_i(\vec{x}) \rangle + g f^{abc} \langle \Pi^d_j(\vec{y})
A^e_k(\vec{z}) A^b_i(\vec{x}) \Pi^c_i(\vec{x}) \rangle=0 \; ,
\label{stident5}
\end{eqnarray}
\begin{eqnarray}
\partial_{x_i} \langle \Pi^d_j(\vec{y}) \Pi^e_k(\vec{z})
\Pi^a_i(\vec{x}) \rangle + g f^{abc} \langle \Pi^d_j(\vec{y})
\Pi^e_k(\vec{z}) A^b_i(\vec{x}) \Pi^c_i(\vec{x}) \rangle=0 \; .
\label{stident6}
\end{eqnarray}
Further relations are easily obtained for the higher orders.
Within our temporal gauge formulation
(\ref{stident1}), (\ref{stident2}), (\ref{stident3}) and the
corresponding higher order equations are the Slavnov-Taylor
identities for equal-time Green functions and have to be
fulfilled in order to guarantee gauge invariance.

If we consider the (Heisenberg picture) time derivative of
one of the above Slavnov-Taylor identities, we have
\begin{eqnarray}
\partial_t \langle O G^a(\vec{x}) \rangle
= \langle \frac{1}{i} [ O,H] G^a(\vec{x}) \rangle
+ \langle O \frac{1}{i} [G^a(\vec{x}),H] \rangle =0 \; .
\label{slavnovtaylorconserve}
\end{eqnarray}
If $O$ is an $n$-point operator, the first term
in (\ref{slavnovtaylorconserve}) vanishes
due to the Slavnov-Taylor identities that relate the
equal-time Green functions of order $n+1$ and $n+2$,
those of order $n+2$ and $n+3$, and those of order
$n+3$ and $n+4$, since the time derivative of an $n$-point
operator in general couples to $n$-, $n+1$-, and $n+2$-point
operators. The second term vanishes due to
(\ref{gaussconserve}). Thus, in principle, the hierarchy
of Slavnov-Taylor identities
can be fulfilled by an appropriate choice of initial conditions
for the equal-time Heisenberg hierarchy.

This last property, however, only holds if we consider an exact solution
of the hierarchy of equations of motion; it breaks down for solutions
within our approximation scheme.
This comes about for two different reasons:

a) Since for our numerical calculations we consider the
theory on a torus and only take into account a finite number
of plane wave states, identity (\ref{gaussconserve}) is
violated, i.e. if $H$ and $G^a(\vec{x})$ are replaced by
the corresponding finite momentum space representation,
their commutator does not vanish any more.
This is nothing but the well known fact that any simple
momentum cutoff violates gauge invariance.
In Section \ref{propagate} we will discuss how this can be partly
remedied by considering different momentum cutoffs for the 2-
and the 3-point functions.

b) The truncation in terms of connected Green functions is not fully
consistent with the constraint hierarchy following from Gauss law,
so that even for a regularization scheme that preserves gauge invariance,
not all Slavnov-Taylor identities would be conserved in time.

In order to investigate the last point in more detail in the
SU(2) case, we first note that the 2-, 3-, and 4-point correlation
dynamical equations of motion
-- regardless of the finite momentum basis --
conserve the following global colour singlet structure
of connected equal-time Green functions
($O$ denotes fields $A$ and/or conjugate momenta $\Pi$):
\begin{eqnarray}
\langle O^a_i(\vec{x}) \rangle=0 \; ,
\label{onepointcolourstructure}
\end{eqnarray}
\begin{eqnarray}
\langle O^{a_1}_{i_1}(\vec{x}_1) O^{a_2}_{i_2}(\vec{x}_2) \rangle_c
= \delta^{a_1 a_2} \langle \! \langle O_{i_1}(\vec{x}_1)
O_{i_2}(\vec{x}_2) \rangle \! \rangle_c \; ,
\label{twopointcolourstructure}
\end{eqnarray}
\begin{eqnarray}
\langle O^{a_1}_{i_1}(\vec{x}_1) O^{a_2}_{i_2}(\vec{x}_2)
O^{a_3}_{i_3}(\vec{x}_3) \rangle_c
=\varepsilon^{a_1 a_2 a_3} \langle \! \langle O_{i_1}(\vec{x}_1)
O_{i_2}(\vec{x}_2) O_{i_3}(\vec{x}_3) \rangle \! \rangle_c \; ,
\label{threepointcolourstructure}
\end{eqnarray}
where we have $f^{abc}=\varepsilon^{abc}$ for colour SU(2).
Here the double brackets
$\langle \! \langle \cdot \rangle \! \rangle$ denote
a colour independent function, depending only on the spatial
coordinates and the vector indices of the corresponding
field operators. The notation using operators without colour indices
inside the double brackets is to be understood as purely symbolic and
is only supposed to indicate the kind of field (A or $\Pi$) to which
the spatial coordinates and vector indices belong (our notation
should especially not be confused with that of matrix valued fields
in the fundamental representation of SU(2)).
While for the 2-point functions we have
\begin{eqnarray}
&&\langle \! \langle A_{i_1}(\vec{x}_1)
A_{i_2}(\vec{x}_2) \rangle \! \rangle_c=
\langle \! \langle A_{i_2}(\vec{x}_2)
A_{i_1}(\vec{x}_1) \rangle \! \rangle_c \; ,
\nonumber \\
&&\langle \! \langle \Pi_{i_1}(\vec{x}_1)
\Pi_{i_2}(\vec{x}_2) \rangle \! \rangle_c=
\langle \! \langle \Pi_{i_2}(\vec{x}_2)
\Pi_{i_1}(\vec{x}_1) \rangle \! \rangle_c \; ,
\label{colour2pcommute}
\end{eqnarray}
and
\begin{eqnarray}
\langle \! \langle A_{i_1}(\vec{x}_1)
\Pi_{i_2}(\vec{x}_2) \rangle \! \rangle_c=
\langle \! \langle \Pi_{i_2}(\vec{x}_2) A_{i_1}(\vec{x}_1)
\rangle \! \rangle_c
+ i \delta_{i_1 i_2} \delta(\vec{x}_1-\vec{x}_2) \; ,
\label{colour2pnotcommute}
\end{eqnarray}
the colour independent 3-point function, due to the antisymmetry
of the SU(2) structure constants, is antisymmetric with respect to
the exchange of any pair of fields or momenta, i.e. we have
\begin{eqnarray}
\lefteqn{
\langle \! \langle O_{i_1}(\vec{x}_1) O_{i_2}(\vec{x}_2)
O_{i_3}(\vec{x}_3) \rangle \! \rangle}
\nonumber \\
&&=-\langle \! \langle O_{i_2}(\vec{x}_2) O_{i_1}(\vec{x}_1)
O_{i_3}(\vec{x}_3) \rangle \! \rangle =
-\langle \! \langle O_{i_1}(\vec{x}_1) O_{i_3}(\vec{x}_3)
O_{i_2}(\vec{x}_2) \rangle \! \rangle \; .
\label{colour3panticommute}
\end{eqnarray}

The color structure of the connected 4-point functions is more
involved and given in Appendix \ref{fourpointcolourstructure}.
Since our condition for fixing the residual gauge freedom
(\ref{coulombfix}) does not prefer any direction in colour
space, we always have such a global singlet configuration
and can therefore make use of (\ref{onepointcolourstructure}),
(\ref{twopointcolourstructure}), (\ref{threepointcolourstructure}),
and (\ref{fourpointcolourstructure1})-(\ref{fourpointcolourstructure4})
for an explicit reduction of the number of degrees of freedom
appearing in the correlation dynamical equations. This, as well
as the proof of the conservation of the above colour
structure by $n$-point correlation dynamics ($n \leq 4$),
is straightforward but lengthy and
has been carried out by means of computer algebra.

One immediate consequence of (\ref{onepointcolourstructure}) is
that the full 2- and 3-point functions do not have disconnected
parts and therefore equal the connected 2- and 3-point functions.
Furthermore, (\ref{stident1}) is trivially conserved within
correlation dynamics, since for all times we have
\begin{eqnarray}
\langle G^a(\vec{x}) \rangle =
\partial_{x_i} \langle \Pi^a_i(\vec{x}) \rangle
+ g \varepsilon^{abc} \langle A^b_i(\vec{x}) \Pi^c_i(\vec{x}) \rangle
=g \varepsilon^{abc} \delta^{bc} \langle \! \langle
A_i(\vec{x}) \Pi_i(\vec{x}) \rangle \! \rangle_c=0 \; .
\label{stident1conserve}
\end{eqnarray}

Also, due to the global colour singlet structure,
identities (\ref{stident2}) and (\ref{stident3}), relating
full 2- and 3-point functions,
directly become identities for connected Green functions.
The Gauss law constraints of higher order,
relating full $n$- and $n+1$-point functions, in general translate into
identities relating connected $n$- and $n+1$-point functions
and products of all lower order connected equal-time functions
down to the connected 2-point function.
Thus, within an $n$-point truncation scheme, all identities
up to the one relating the full $n-1$- and $n$-point functions
remain unchanged, and the identities relating the full $n$- and
$n+1$-point functions have to be truncated by neglecting the
connected $n+1$-point function; e.g. within a 3-point truncation
scheme one has relations between connected 2- and 3-point
functions and identities relating connected 3-point functions
to products of 2 connected 2-point functions.

Since the conservation of a Gauss law constraint identity,
as stated above, requires the conservation of the identities of the
2 next higher orders, a truncation in terms of connected Green
functions up to a certain order in general leads to a non-conservation
of (\ref{stident2}), (\ref{stident3}), and the higher order constraints.

A subset of conserved identities can nevertheless be obtained if,
instead of multiplying $G^a(\vec{x})$ from the left
with an arbitrary operator product, one considers expectation
values of products of several Gauss law operators. In general,
a product of $n$ Gauss law operators contains products of up to
$2n$ fields or momenta, and its expectation value is thus conserved
within a $2n$-point truncation scheme as long as
$[G^a(\vec{x}),H]=0$ \cite{wan294}. The conservation
of these identities is of course destroyed as soon as we work
within a finite momentum basis.

It is worthwhile to mention, that within the framework of correlation
dynamics for nonrelativistic fermion systems there exists a
constraint, that is violated in a similar way by the $n$-point
truncation scheme as the Gauss law in the present case of
correlation dynamics for a nonabelian gauge theory. This constraint
is the requirement that the system has good particle number,
i.e. if $N_{Fermi}=\sum_{\alpha} a^\dagger_{\alpha} a_{\alpha}$
is the number operator and $H_{Fermi}$ is the Hamiltonian,
with $[H_{Fermi},N_{Fermi}]=0$,
and $|\Psi_{Fermi}\rangle$ is the state
describing the system in Fock space, we impose the constraint
\begin{eqnarray}
\left(N_{Fermi}-\langle N_{Fermi} \rangle \right)
|\Psi_{Fermi}\rangle =0 \; ,
\label{goodnumber}
\end{eqnarray}
from which we directly obtain
\begin{eqnarray}
\langle O \left( N_{Fermi} -\langle N_{Fermi}\rangle\right)\rangle=0
\label{tracerel}
\end{eqnarray}
for any operator $O$. By inserting any possible product of
creation and annihilation operators for $O$ we obtain a constraint
hierarchy similar to the hierarchy of Slavnov-Taylor identities
as obtained above from Gauss law. This constraint hierarchy is
nothing but the set of trace relations relating the $n$-body
($2n$-point) and the $n+1$-body ($2n+2$-point) density matrices
of a pure $N$-particle wavefunction, which is well known to be
violated within the correlation dynamics approach.
Just as in the case of Gauss law a subset of conserved identities
within a $2n$-point truncation can be obtained by considering
the expectation values of $N_{Fermi}^k$, $k \leq n$ (e.g.
$\langle N_{Fermi}^2 \rangle$ is conserved within 2-body (4-point)
correlation dynamics).

However, there is still a major qualitative difference
between the Gauss law in Yang-Mills theory and the trace relations
in fermionic many-body theory. While the Gauss law (\ref{gausslaw})
is induced by a local gauge symmetry, condition (\ref{goodnumber})
is induced by a global symmetry.
Therefore particle number conservation is not violated on the
operator level, if the theory is considered within a finite single
particle basis.

While the violation of the trace relations
was found to be only weak and without significant impact
on the time evolution of the system at least for
the application of correlation dynamics to light nuclei
\cite{pet94}, it is not a priori clear if this will also be
the case with the violation of Gauss law in temporal gauge
Yang-Mills theory and has to be studied explicitly.
\section{Numerical investigations for SU(2) in 2+1 dimensions}
\label{propagate}
For our numerical calculations we consider SU(2) and expand the fields
and conjugate momenta into plane wave states on a torus in $2+1$
spacetime dimensions according to
\begin{eqnarray}
A^a_i(\vec{x})=\sum_{\vec{k}} \frac{1}{L} e^{i\vec{k}\vec{x}}
A^a_i(\vec{k}) \; , \; \;
\Pi^a_i(\vec{x})=\sum_{\vec{k}} \frac{1}{L} e^{i\vec{k}\vec{x}}
\Pi^a_i(\vec{k}) \; ,
\label{basisexpand}
\end{eqnarray}
where $L$ is the size of the 2-dimensional box (of volume $L^2$) and
the momentum vectors $\vec{k}$ are chosen according to the periodicity
condition
\begin{eqnarray}
\vec{k}=\frac{2\pi}{L} \vec{n} \; , \; \;
\vec{n} \in Z \! \! \! Z^2 \; .
\label{kdef}
\end{eqnarray}
The equal-time Green functions and the correlation dynamical
equations of motion are expanded accordingly.

In this section we investigate the behaviour of the system in
the weak coupling limit by choosing certain initial conditions and
propagating the connected equal-time Green functions according to the
correlation dynamical equations of motion in different approximation
schemes.

The weak coupling limit is considered, since the properties of the
regularized system with a given number of basis states are uniquely
determined by the dimensionless quantity $g L^{1/2}$
(recall that in $2+1$ dimensions $g^2$ has the dimension of energy);
the limit $gL^{1/2}\to 0$ is therefore equivalent to the small
volume limit. This should in turn make sure that the different numbers
of basis states considered suffice to allow for an investigation of
the asymptotic behaviour as the ultraviolet cutoff is taken to infinity.

In order to investigate the nonabelian aspects we, furthermore, start
with perturbative initial conditions that are close to the
configuration given by the equal-time limit of (\ref{mypropagator})
at $t=t_0$. The advantage is that with (\ref{mypropagator}),
at $t=t_0$ only the coupling terms contribute in the equations of motion,
while the free (abelian) terms vanish (up to the explicit time dependence
already present in the abelian ground state solution).
Otherwise the time evolution in
the weak coupling limit is usually dominated by the (uninteresting)
propagation of a free field. The equal-time limit of
(\ref{mypropagator}) at $t=t_0$ leads to
\begin{eqnarray}
\langle A^a_i(\vec{k}) A^b_j(\vec{q}) \rangle_c
=\delta^{ab} \delta_{\vec{k}+\vec{q},\vec{0}}
\left( \delta_{ij}-\frac{k_i k_j}{\vec{k}^2} \right)
\frac{1}{2|\vec{k}|} \; ,
\label{aaccmini}
\end{eqnarray}
\begin{eqnarray}
\langle \Pi^a_i(\vec{k}) \Pi^b_j(\vec{q}) \rangle_c
=\delta^{ab} \delta_{\vec{k}+\vec{q},\vec{0}}
\left( \delta_{ij}-\frac{k_i k_j}{\vec{k}^2} \right)
\frac{|\vec{k}|}{2} \; ,
\label{ppccmini}
\end{eqnarray}
\begin{eqnarray}
\langle A^a_i(\vec{k}) \Pi^b_j(\vec{q}) \rangle_c
=\delta^{ab} \delta_{\vec{k}+\vec{q},\vec{0}}
\left( \delta_{ij}-\frac{k_i k_j}{\vec{k}^2} \right)
\frac{i}{2} \; \; \; {\rm and}
\label{apccmini}
\end{eqnarray}
\begin{eqnarray}
\langle \Pi^a_i(\vec{k}) A^b_j(\vec{q}) \rangle_c
=\delta^{ab} \delta_{\vec{k}+\vec{q},\vec{0}}
\left( \delta_{ij}+\frac{k_i k_j}{\vec{k}^2} \right)
\frac{-i}{2} \; ;
\label{paccmini}
\end{eqnarray}
all higher order connected Green functions vanish.
Note that while $\langle A A \rangle_c$, $\langle \Pi \Pi \rangle_c$,
and $\langle A \Pi \rangle_c$ are purely transverse
due to the gauge fixing condition (\ref{coulombfix})
and the abelian limit of Gauss law (\ref{gausslaw}),
this is no longer the case for $\langle \Pi A \rangle_c$.

While in principle the size $L$ of the box introduces an infrared
scale into the theory, (\ref{aaccmini}) -- due to the periodic
boundary conditions -- still contains an infrared
singularity which has to be regularized in some way.
The task of finding gauge
invariant infrared regulators for Yang-Mills theories is nontrivial
\cite{bla96} and the resulting theory would in general be
hard to handle in the framework of correlation dynamics.
We, therefore, resort to two different ways of handling this problem
without having to introduce an additional mass scale by hand:
we either leave out the zero momentum mode (which is not a gauge
invariant prescription) or selfconsistently generate a mass for
the initial condition by solving the quantum mechanical problem
for the zero mode, only, which is described by the Hamiltonian
\begin{eqnarray}
H=\frac{1}{2} \Pi^a_i \Pi^a_i + \frac{g^2}{4 L^2} \varepsilon^{abc}
\varepsilon^{ade} A^b_i A^c_j A^d_i A^e_j \; ,
\label{zeromodehamiltonian}
\end{eqnarray}
with $A^a_i \equiv A^a_i(\vec{k}=\vec{0})$,
$\Pi^a_i \equiv \Pi^a_i(\vec{k}=\vec{0})$ and
$[A^a_i,\Pi^b_j]=i \delta^{ab} \delta_{ij}$.
The Gaussian ansatz
\begin{eqnarray}
\Psi(A)=\prod_{i,a} \left( \frac{2 \alpha}{\pi} \right)^{1/4}
e^{-\alpha (A^a_i)^2}
\label{zeromodeansatz}
\end{eqnarray}
yields the configuration of lowest energy within the
2-point truncation scheme for
$\alpha=\frac{1}{2} \left( \frac{g}{L} \right)^{2/3}$
and hence an effective mass
\begin{eqnarray}
m_{\rm eff}= 2 \alpha = \left( \frac{g}{L} \right)^{2/3} \; .
\label{effectivemass}
\end{eqnarray}
Thus instead of (\ref{aaccmini}) and (\ref{ppccmini}) we use
($\omega(\vec{k})=\sqrt{\vec{k}^2+m_{\rm eff}^2}$)
\begin{eqnarray}
\langle A^a_i(\vec{k}) A^b_j(\vec{q}) \rangle_c
=\delta^{ab} \delta_{\vec{k}+\vec{q},\vec{0}}
\left( \delta_{ij}-\frac{k_i k_j}{\vec{k}^2} \right)
\frac{1}{2\omega(\vec{k})} \; ,
\label{aameffini}
\end{eqnarray}
\begin{eqnarray}
\langle \Pi^a_i(\vec{k}) \Pi^b_j(\vec{q}) \rangle_c
=\delta^{ab} \delta_{\vec{k}+\vec{q},\vec{0}}
\left( \delta_{ij}-\frac{k_i k_j}{\vec{k}^2} \right)
\frac{\omega(\vec{k})}{2} \; ,
\label{ppmeffini}
\end{eqnarray}
together with (\ref{apccmini}) and (\ref{paccmini}) as initial
conditions and include $\vec{k}=\vec{0}$, such that at $t=t_0$ the
free propagation terms and the terms describing the coupling to
the zero mode via the 4-point vertex cancel out.

Before we begin our further investigation we note that, according to our
experience, the solutions of the equations of $n$-point correlation
dynamics with $n \geq 3$ (for any theory we have investigated so far,
and also in the present case) become unstable at a certain point in
time, when the absolute values of the equal-time functions suddenly
start to rise steeply as a function of time. This does not seem to
be a numerical effect, since the corresponding solutions remain
unchanged if we use higher machine accuracies and smaller time step sizes.
In the following we therefore focus on the time intervals where the
solutions are still stable.
\subsection{Infrared properties and gauge invariance}
\label{unmodi}
We now i) either leave out the zero momentum mode and initialize
(\ref{aaccmini})-(\ref{paccmini}) or ii) include the zero
momentum mode and initialize (\ref{aameffini}), (\ref{ppmeffini}),
(\ref{apccmini}), and (\ref{paccmini}) at $t=0$ and then numerically
integrate the correlation dynamical equations, which conserve the
total energy in time.

Fig.~\ref{bild1} shows the resulting electric field energy
($Re E_{el}L$, $Im E_{el}L$)
\begin{eqnarray}
E_{el}=\langle \hat{E}_{el} \rangle=
\frac{1}{2} \int d^2x \langle \Pi^a_i(\vec{x}) \Pi^a_i(\vec{x}) \rangle
\label{eel}
\end{eqnarray}
as a function of time, where we have shifted the value at $t=0$
to zero. Since $\hat{E}_{el}$ is a gauge invariant
operator, which is hermitean in the space of physical states,
$E_{el}$ should be a real quantity. However, the figures
in the right column show that within all approximation schemes
(first row: 2-point; second row: 3-point; third row: 4-point) we
obtain an imaginary part that is roughly as large as the real part.
These oscillations of $E_{el}$ in the complex plane are due to a
violation of gauge invariance, leading to a non-cancellation of the
terms in (\ref{partialintegrate}) that differentiate between
the Heisenberg and the Schr\"odinger picture in the propagation of
gauge dependent quantities. While it is clear that within the 2-point
truncation scheme (first row) gauge invariance must be violated due to
the neglect of the 3-point vertex, our calculation shows that
within the 3- and 4-point truncation schemes (second and third row),
there is no significant improvement in this respect, which is partly
due to a non-gauge invariant UV cutoff.
One might argue that our initial conditions could also be a
problem since they fulfill Gauss law only in the abelian limit,
i.e. while at $t=0$ the Gauss law identities relating the 2- and
3-point functions are fulfilled, those relating the full 3- and 4-point
functions are not. However, we have checked that by enforcing these
identities at $t=0$ (cf. the discussion in context with Fig.
\ref{bild6}), we obtain almost no visible effect on our results.

Some results concerning the infrared convergence are worth mentioning:
Fig.~\ref{bild1} shows that within 2- and 4-point correlation dynamics
(first and third row) the inclusion of the $\vec{k}=\vec{0}$ mode
(considering the 9 lowest lying momentum states) leads to a significant
reduction of the amplitudes of the oscillation as compared to a neglect
of the $\vec{k}=\vec{0}$ mode (considering the remaining 8 lowest
lying momentum states); in the 4-point case the amplitudes are reduced
by almost one order of magnitude.
In contrast to that, within 3-point correlation
dynamics the inclusion of the zero momentum mode only leads to a
significant increase of fast oscillations with comparably small
amplitudes, mainly in the real part of $E_{el}$,
with frequencies given by the dynamics of the free field.
The overall shape of the curves with respect to the slow, large
amplitude oscillations -- that violate gauge invariance -- is almost
unaffected by the inclusion or neglect of $\vec{k}=\vec{0}$.
This can be viewed as an indication of infrared convergence within the
3-point truncation scheme, whereas there is no infrared convergence on
the 2- and 4-point level.

In Fig.~\ref{bild2} we display the transverse and longitudinal
contributions to the real part of $E_{el}$ for the same calculation as in
Fig.~\ref{bild1}; the left column shows the results with the
$\vec{k}=\vec{0}$ mode not included, the right column contains those with
the $\vec{k}=\vec{0}$ mode. Within the 2-point truncation scheme
(first row) we see that without the zero mode, the oscillations
almost entirely (up to small deviations within linewidth)
show up in the longitudinal sector of the theory. The inclusion
of the zero mode then leads to oscillations in the transverse sector
as well. This, however, is merely due to the fact that with our
prescription for the longitudinal projector at $\vec{k}=\vec{0}$, the
zero mode is by definition completely transverse; the observed transverse
oscillations are in fact entirely those of the zero mode.
Within 3-point correlation dynamics the inclusion of the zero momentum
mode leaves the longitudinal sector almost unaffected and only leads to
an increase of fast oscillations already mentioned in the discussion
of Fig.~\ref{bild1}. These oscillations apparently only take place in
the transverse sector; a coupling between the transverse and the
longitudinal sector only appears by generating an amplitude modulation
for the fast, transverse oscillations; the transverse modes are of
almost no effect on the longitudinal modes.
Within 4-point correlation dynamics we observe a similar behaviour
as in 2-point correlation dynamics, i.e.\ slow oscillations of the zero
mode appearing in the transverse sector, with an additional increase of
fast transverse oscillations as observed in the 3-point truncation scheme.

The violation of gauge invariance in the present approximation schemes
can be seen directly by looking at the violation of the Gauss law
identities. The upper part of Fig.~\ref{bild3} shows the quantity
$\frac{1}{3} \sum_{\vec{k},i} \left| \langle A^a_i(\vec{k})
G^a(-\vec{k}) \rangle \right|$, which is a measure for the violation
of the corresponding identity between 2-point functions of the
form $\langle A \Pi \rangle$ and 3-point functions of the form
$\langle A A \Pi \rangle$ (cf. eq. (\ref{stident2})).
The calculations are the same as in Figs. \ref{bild1} and \ref{bild2}
with the zero momentum mode included.
Up to about $t/L=100$, where the 3- and the 4-point approximation are
still stable, all truncation schemes lead to a violation of roughly the
same order of magnitude. In contrast to that, the quantity
$\frac{1}{6} \sum_{\vec{k}_1,\vec{k}_2,\vec{k}_3,i,j}
\delta_{\vec{k}_1+\vec{k}_2+\vec{k}_3} \left| \varepsilon^{abc}
\langle A^a_i(\vec{k}_1) A^b_j(\vec{k}_2) G^c(\vec{k}_3) \rangle
\right|$ displayed in the lower part of Fig.~\ref{bild3},
measuring the violation of the corresponding identity between 3-point
functions of the form $\langle A A \Pi \rangle$ and full 4-point functions
of the form $\langle A A A \Pi \rangle$ (cf. eq. (\ref{stident4})),
becomes significantly larger with increasing order of the truncation
scheme. Both results clearly indicate that the problem with gauge
invariance cannot be cured by simply going to higher orders of
correlation dynamics.
\subsection{Scaling behaviour and UV convergence}
\label{scaluv}
In order to gain a better understanding of the behaviour of the theory
within correlation dynamics, we now turn to an investigation of the system
for different values of the dimensionless coupling $g L^{1/2}$ and for
different numbers of plane wave states in the single particle basis.

In the left column of Fig.~\ref{bild4}
we again display the real part of the electric field energy as a function
of time, using the 21 lowest lying momentum states
($\vec{k}=\vec{0}$ included), for 3 different values of $g L^{1/2}$.
In each of the 3 approximation schemes, the curves for the different
values of the coupling apparently (up to the fast transverse oscillations
in 3- and 4-point correlation dynamics, second and third row) have the
same shape and are only squeezed in the direction of the time axis and
stretched in the direction of the amplitude axis. This indicates that
in the longitudinal (plus the zero mode) sector the theory exhibits a
scaling behaviour with respect to the coupling. In fact, the corresponding
frequencies (and amplitudes of dimension energy) behave like
$\omega \propto g^{2/3}$ as shown in the right column of
Fig.~\ref{bild4} by rescaling all curves according to the scaling law
\begin{eqnarray}
E_{el}(g_2,t)=\left( \frac{g_2}{g_1} \right)^{2/3}
E_{el}(g_1,\left( \frac{g_2}{g_1} \right)^{2/3} t) \; ,
\label{g23scalinglaw}
\end{eqnarray}
which is a property of any system with a purely quartic coupling.
This can be seen as follows: consider a Hamiltonian of the form
\begin{eqnarray}
H=\frac{1}{2} \Pi^2 + \frac{g^2}{4} A^4 \; ,
\label{purequartichamiltonian}
\end{eqnarray}
where this is meant as a purely formal notation with any indices or
spacetime dependencies suppressed. One can now perform a canonical
transformation according to
\begin{eqnarray}
A=g^{-1/3} A' \; \; , \; \; \; \; \Pi=g^{1/3} \Pi'
\; \; , \; \; \; \; [A, \Pi]=[A', \Pi'] \;  ,
\label{canontrafo}
\end{eqnarray}
which leads to
\begin{eqnarray}
H=g^{2/3} \left( \frac{1}{2} \Pi'^2 + \frac{1}{4} A'^4 \right)
\label{rescaledpurequartham}
\end{eqnarray}
and thus to a $g^{2/3}$ scaling of all eigenvalues of the Hamiltonian.

In case of 2-point correlation dynamics for Yang-Mills theory,
a $g^{2/3}$ scaling can indeed be expected a priori for the
longitudinal sector of the theory (if it decouples from the
transverse sector) since the 3-point vertex is
not taken into account and the 2-point terms $\propto A^2$ contain
a transverse projection operator. The fact that the $g^{2/3}$ scaling
survives in 3- and 4-point correlation dynamics despite the inclusion
of the 3-point vertex is, however, in line with the result obtained by
L\"uscher \cite{lue83}, that all eigenvalues of the (purely discrete)
spectrum of Yang-Mills theory on a torus can be expanded in a power
series of $\lambda=g^{2/3}$.

In order to investigate the asymptotic behaviour of the system as
the UV cutoff is taken to infinity, the left column of Fig.~\ref{bild5}
again displays the real part of the electric field energy as a function
of time for different numbers of momentum states that are taken into
account. In the first row we have plotted the corresponding result
within the 2-point truncation scheme and with the zero momentum mode
excluded. All curves apparently have the same shape and are only squeezed
or stretched with respect to each other along the two axes.
In the right column we show the curves to obey the relation
\begin{eqnarray}
E_{el}(n_2,t)=\left( \frac{n_2}{n_1} \right)^{4/3}
E_{el}(n_1,\left( \frac{n_2}{n_1} \right)^{1/3} t)
\label{basisscalinglaw}
\end{eqnarray}
by rescaling them, where $n_i$ denotes the number of basis
states taken into account; as in the case of Fig.~\ref{bild4} the
rescaled curves are all on top of each other.
This shows that the frequencies and the amplitudes of dimension
energy diverge as the UV cutoff is removed -- which is an unwanted
behaviour since Yang-Mills theory in $2+1$ spacetime dimensions
is supposed to be finite.

In the second row, the same results are plotted within the 2-point
truncation scheme and with the $\vec{k}=\vec{0}$ mode now included.
While the frequencies still scale like $\omega \propto n^{1/3}$ as in
(\ref{basisscalinglaw}), the $n^{4/3}$ scaling of the amplitude of
$E_{el}$ is destroyed by the inclusion of the zero mode.

In the third row the corresponding results are shown in the 3-point
approximation with the $\vec{k}=\vec{0}$ mode included; we have chosen
only to display the longitudinal part in order to get rid of the fast
oscillations in the transverse sector. From Figs. \ref{bild1} and
\ref{bild2} we already know that the results are essentially identical
in this case if the zero mode is excluded. The rescaling of the curves
works reasonably well in the first half of the displayed time interval;
in the second half the application of (\ref{basisscalinglaw}) while still
``going into the right direction'' does not lead to a really impressive
agreement of the rescaled curves. Furthermore, within the accuracy
reached here, it only makes a small difference if in
(\ref{basisscalinglaw}) the zero mode is counted or not.

Relation (\ref{basisscalinglaw}) can be a priori expected within the
2-point approximation since there the longitudinal sector (if it
decouples from the transverse sector) essentially behaves like $\Phi^4$
theory with a purely quartic coupling, i.e. like a system described
by the (momentum space) Hamiltonian
\begin{eqnarray}
H=\frac{1}{2} \sum_{\vec{k}} \Pi(\vec{k}) \Pi(-\vec{k})
+\frac{g^2}{4 V} \sum_{\vec{k}_1 \vec{k}_2 \vec{k}_3 \vec{k}_4}
\delta_{\vec{k}_1+\vec{k}_2+\vec{k}_3+\vec{k}_4,\vec{0}}
\Phi(\vec{k}_1) \Phi(\vec{k}_2) \Phi(\vec{k}_3) \Phi(\vec{k}_4) \; ,
\label{phi4hamilton}
\end{eqnarray}
where $V$ denotes the quantization volume in any spacetime dimensionality.
Computing the expectation value of (\ref{phi4hamilton}) within a
Hartree-Fock-Bogoliubov ansatz with effective single particle energies
$\Omega(\vec{k})$ then yields the energy functional
\begin{eqnarray}
\langle H \rangle_{\Omega} \equiv E[\Omega]
=\frac{1}{4} \sum_{\vec{k}} \Omega(\vec{k}) +
\frac{3 g^2}{16 V} \left( \sum_{\vec{k}} \frac{1}{\Omega(\vec{k})}
\right)^2 \; ,
\label{efunctional}
\end{eqnarray}
which is minimized by
\begin{eqnarray}
\Omega(\vec{q})^2=\frac{3 g^2}{2 V} \sum_{\vec{k}}
\frac{1}{\Omega(\vec{k})} \; .
\label{efunctionalminimize}
\end{eqnarray}
Eq. (\ref{efunctionalminimize}) directly implies that $\Omega(\vec{q})$
is independent of $\vec{q}$ and therefore leads to
\begin{eqnarray}
\Omega(\vec{q}) \equiv \Omega = g^{2/3}
\left( \sum_{\vec{k}} \right)^{1/3} \left( \frac{3}{2 V} \right)^{1/3},
\label{phi4frequencyscaling}
\end{eqnarray}
which is the frequency scaling contained in (\ref{basisscalinglaw})
and in (\ref{g23scalinglaw}). With
\begin{eqnarray}
E_{el}=\frac{1}{4} \sum_{\vec{k}} \Omega(\vec{k})
= \frac{1}{4} g^{2/3} \left( \sum_{\vec{k}} \right)^{4/3}
\left( \frac{3}{2 V} \right)^{1/3}
\label{phi4amplitudescaling}
\end{eqnarray}
we also recover the corresponding dependencies for the amplitudes.

The result that (\ref{basisscalinglaw}) remains approximately valid
in the 3-point truncation scheme indicates that, in contrast to
the infrared behaviour, the cancellations required to guarantee that
the theory is UV convergent do {\it not} take place.
This is again a consequence of the violation of gauge invariance.

Furthermore, we note that within the 3-point approximation, in the
equations of motion for the 3-point functions, the contribution of the
terms coupling to a product of a 2- and a 3-point function via the 4-point
vertex can to a good approximation be neglected.
\subsection{Restoration of Gauss law}
\label{implemgauss}
In Sections \ref{unmodi} and \ref{scaluv} we observed a violation
of Gauss law, resulting in unphysical imaginary parts for
gauge invariant hermitean quantities and in the non-cancellation
of UV divergences.

We therefore now try to improve our approximation scheme by partly
replacing our dynamical equations of motion by the constraint
equations arising from Gauss law (cf. Section \ref{gausshierarchy}).
There is of course no unique way to do this;
we implement Gauss law by means of the replacement
\begin{eqnarray}
i k_i \Pi^a_i(\vec{k}) \to - \frac{g}{L} \varepsilon^{abc}
\sum_{\vec{q}_1 \vec{q}_2} \delta_{\vec{k},\vec{q}_1+\vec{q}_2}
A^b_i(\vec{q}_1) \Pi^c_i(\vec{q}_2) \; ,
\label{gaussreplacement}
\end{eqnarray}
i.e. in an equal-time function that contains a longitudinal
field momentum, the latter is commuted through to the right and
the resulting $n$-point function is replaced by a sum over $n+1$-point
functions according to (\ref{gaussreplacement}) (since we only use
this prescription to replace 2- and 3-point functions, which possess no
disconnected parts, we do not have to make the distiction between
full and connected equal-time functions on the l.h.s. of
(\ref{gaussreplacement})). This is done throughout
the time integration and the remaining functions are propagated as before
using the correlation dynamical equations of motion.
Of course (\ref{gaussreplacement}) cannot be applied for
$\vec{k}=\vec{0}$; the Gauss law for the zero mode, however, turns out
to be fulfilled automatically within our investigation.

We apply this method within the 3-point truncation scheme, where
we investigate the implications of a) either enforcing the Gauss law
identities relating the 2- and the 3-point functions (\ref{stident2}),
(\ref{stident3}) or b) enforcing the identities relating the 3- and
the 4-point functions (\ref{stident4})-(\ref{stident6}) as well. In the
latter case, the full 4-point function is given by products of two
2-point functions and we thus have a coupled system of equations for
2- and 3-point functions, which we solve by iteration.

We mention that whenever there is a Green function containing
more than one longitudinal field momentum, there is an ambiguity in the
application of (\ref{gaussreplacement}). In case of the 2-point/3-point
identities, i.e. having a 2-point function containing two longitudinal
momenta, this does not create any problem, since due to the symmetries
of the system both ways for applying (\ref{gaussreplacement}) lead to
the same result. In case of the 3-point/4-point identities, this is not
the case and we choose to symmetrize between the various possibilities;
the stationary solution of our iteration scheme can then not exactly
fulfill the corresponding identities. However, the remaining
violation of the 3-point/4-point identities turns out to be very small
compared to the original violation.

Fig.~\ref{bild6} shows the electric field energy $E_{el}$ and the
magnetic field energy
\begin{eqnarray}
E_{mag} \equiv \langle \hat{E}_{mag} \rangle
=\frac{1}{4} \int d^2 x
\langle F^a_{ij}(\vec{x}) F^a_{ij}(\vec{x}) \rangle
\label{emag}
\end{eqnarray}
as a function of time obtained within the scheme above for 21
momentum states ($\vec{k}=\vec{0}$ included).
While the calculation that only implements the 2-point/3-point
identity (upper part) becomes unstable at about $t/L=40$,
the main effect of the additional inclusion of the 3-point/4-point
identity (lower part) is a stabilization of the system.
In both calculations the real part of $E_{mag}$ on average decreases
with time, while the real part of $E_{el}$ on average stays constant
(left column); both exhibit additional fast oscillations with frequencies
that are essentially given by the dynamics of the free field.
Since the total energy is given by $E_{tot}=E_{el}+E_{mag}$, this
result implies that the price we have to pay for our restoration
of Gauss law is the violation of energy conservation; the reason for
this is the replacement of the 2-point function containing two
longitudinal momenta, which contributes to $E_{el}$, by means of
(\ref{gaussreplacement}).

Energy conservation can in principle be restored by instead using
the corresponding 2-point/3-point identity to replace one of the
3-point functions (these do not show up in the total energy).
Any corresponding prescription would, however, seem highly arbitrary,
since there is a momentum summation and a contraction over
vector indices on the r.h.s. of (\ref{gaussreplacement}).
Moreover, our attempts in that direction have lead to
equations of motion that become unstable much earlier than the original
ones and that still exhibit large imaginary parts for the electric
and magnetic field energies.

The imaginary parts of $E_{el}$ and $E_{mag}$ (right column)
both stay very small up to about $t/L=10$, and then fast oscillations
start to build up; if these are subtracted, the imaginary parts
stay very small, which might be seen as an improvement.
We note in passing that when neglecting the zero momentum mode,
the fast free field oscillations almost completely vanish.

In Fig.~\ref{bild7} the violation of the 3-point/4-point Gauss law
identities is compared for three different approximation schemes with
the same parameters as in Fig.~\ref{bild6}: i) without implementing
Gauss law, ii) implementing the 2-point/3-point identities and iii)
implementing the 3-point/4-point identities as well.
In the left picture we have plotted the quantity
$\frac{1}{6} \sum_{\vec{k}_1,\vec{k}_2,\vec{k}_3,i,j}
\delta_{\vec{k}_1+\vec{k}_2+\vec{k}_3} \left| \varepsilon^{abc}
\langle A^a_i(\vec{k}_1) A^b_j(\vec{k}_2) G^c(\vec{k}_3) \rangle
\right|$. Since there is no ambiguity with more than one longitudinal
momentum, there is no such violation if the 3-point/4-point
identities are implemented. While the strength of the violation
oscillates without the implementation of Gauss law, it simply continues
to rise if only the 2-point/3-point identity is implemented until
the system becomes unstable at about $t/L=40$.
The same behaviour can be seen for the quantities
$\frac{1}{6} \sum_{\vec{k}_1,\vec{k}_2,\vec{k}_3,i,j}
\delta_{\vec{k}_1+\vec{k}_2+\vec{k}_3} \left| \varepsilon^{abc}
\langle \Pi^a_i(\vec{k}_1) A^b_j(\vec{k}_2) G^c(\vec{k}_3) \rangle
\right|$ (middle picture) and\\
$\frac{1}{6} \sum_{\vec{k}_1,\vec{k}_2,\vec{k}_3,i,j}
\delta_{\vec{k}_1+\vec{k}_2+\vec{k}_3} \left| \varepsilon^{abc}
\langle \Pi^a_i(\vec{k}_1) \Pi^b_j(\vec{k}_2) G^c(\vec{k}_3) \rangle
\right|$ (right picture), where the time interval has been chosen
larger. One can see that without the implementation of Gauss law
the equations of motion remain stable up to about $t/L=100$.
In the middle and the right picture, one can additionally observe the
small violation arising in the scheme implementing the 3-point/4-point
identities due to the above mentioned ambiguity in the application
of (\ref{gaussreplacement}); the corresponding system of equations
becomes unstable at about $t/L=70$.

In Fig.~\ref{bild8} we compare the time evolution of the real
part of the magnetic field energy obtained with the implementation
of 2-point/3-point and 3-point/4-point identities for
different values of the coupling (upper part). After rescaling
both axes (as in Fig.~\ref{bild4}), all curves are on top of each
other (lower part), indicating the validity of the $g^{2/3}$ scaling
law (\ref{g23scalinglaw}). The same result is obtained if only the
2-point/3-point identity is implemented.

Fig.~\ref{bild9} shows the corresponding comparison for fixed coupling
and different numbers of momentum states taken into account (upper
part). After a rescaling of both axes (as in Fig.~\ref{bild5}), all
curves are approximately on top of each other (lower part),
indicating that the $(\sum_{\vec{k}})^{1/3}$ scaling law
(\ref{basisscalinglaw}) holds as well. This result can also be
reproduced within the scheme only implementing the 2-point/3-point
identity. The ultraviolet behaviour is thus unchanged, the
restoration of Gauss law has not helped to obtain the necessary
cancellations and to achieve UV convergence.

The behaviour observed in Figs.~\ref{bild6}-\ref{bild9} can again be
reproduced by considering the simplified model for the 2-point dynamics
of the longitudinal sector of the theory that is obtained from the
$\Phi^4$-like Hamiltonian (\ref{phi4hamilton}).
Within the 2-point truncation scheme the Gauss law identities
(\ref{stident1}) and (\ref{stident2}) reduce to the abelian version
of Gauss law, since the 3-point function is neglected; in our scheme
given by (\ref{gaussreplacement}) this corresponds to setting to zero
all 2-point functions with longitudinal momenta commuted through to
the right. In order to investigate the resulting behaviour within the
$\Phi^4$-like model, we thus implement the conditions
\begin{eqnarray}
\langle \Phi(\vec{k}) \Pi(-\vec{k}) \rangle=\langle \Pi(\vec{k})
\Pi(-\vec{k}) \rangle = 0 \; , \; \;
\langle \Pi(\vec{k}) \Phi(-\vec{k}) \rangle=-i
\label{phi4gausslaw}
\end{eqnarray}
throughout the time evolution. The only remaining equation of motion
then is
\begin{eqnarray}
&& \partial_t \langle \Phi(\vec{k}) \Phi(-\vec{k}) \rangle
=\langle \Pi(\vec{k}) \Phi(-\vec{k}) \rangle
+\langle \Phi(\vec{k}) \Pi(-\vec{k}) \rangle=-i
\nonumber \\
&& \Rightarrow \langle \Phi(\vec{k}) \Phi(-\vec{k}) \rangle = -i t \; .
\label{phi4eom}
\end{eqnarray}
If we split up the expectation value of (\ref{phi4hamilton}) into a part
$E_{el}$ corresponding to the electric field energy and a part $E_{mag}$
corresponding to the magnetic field energy in the longitudinal sector of
Yang-Mills theory according to
\begin{eqnarray}
&& E_{el}=\frac{1}{2} \sum_{\vec{k}}
\langle \Pi(\vec{k}) \Pi(-\vec{k}) \rangle \; ,
\nonumber \\
&& E_{mag}=\frac{g^2}{4 V}
\sum_{\vec{k}_1 \vec{k}_2 \vec{k}_3 \vec{k}_4}
\delta_{\vec{k}_1+\vec{k}_2+\vec{k}_3+\vec{k}_4,\vec{0}}
\langle \Phi(\vec{k}_1) \Phi(\vec{k}_2)
\Phi(\vec{k}_3) \Phi(\vec{k}_4) \rangle \; ,
\label{phi4esplitup}
\end{eqnarray}
the insertion of (\ref{phi4gausslaw}) and (\ref{phi4eom}) within the
2-point truncation scheme yields
\begin{eqnarray}
E_{el}=0 \; , \; \; E_{mag}=- \frac{3 g^2 t^2}{4 V}
\left(\sum_{\vec{k}}\right)^2 \; .
\label{phi4esolution}
\end{eqnarray}
Eq. (\ref{phi4esolution}) is a particularly simple solution that fulfills
the $g^{2/3}$ scaling law (\ref{g23scalinglaw}) and the $(\sum_{k})^{1/3}$
scaling law (\ref{basisscalinglaw}), and that moreover yields no imaginary
parts for $E_{el}$ and $E_{mag}$. Indeed, even though the 3-point functions
are included, the magnetic field energies displayed in Fig.~\ref{bild6}
on average fulfill $E_{mag} \propto t^2$, while the electric field energies
on average remain constant.

In order to demonstrate the effect of the inclusion of the 3-point
function, Fig.~\ref{bild10} shows the real part of the total energy
obtained within the 2-point truncation scheme with the abelian Gauss
law implemented in comparison with the total energy obtained within
the 3-point truncation scheme with the 2-point/3-point and the
3-point/4-point Gauss law identities implemented. In case of the
2-point approximation, the $\vec{k}=\vec{0}$ mode is excluded,
since otherwise the time evolution would be dominated by oscillations
due to the zero mode dynamics. The inclusion of the 3-point function
leads to a significant reduction of the violation of energy
conservation; it, however, fails to lead to a complete cancellation
of the corresponding terms.
\subsection{An extended momentum basis}
\label{modmombas}
After having investigated the consequences of a restoration of the Gauss
law identities, we now turn to a different method of improving 3-point
correlation dynamics, which consists in altering the momentum basis.

Within a {\it finite} momentum basis the violation of gauge invariance
already shows up on the operator level, which means that the
commutator of the Gauss law and the Hamiltonian becomes nonzero.
This is due to the fact that for a number of terms containing the
projection operator in momentum space
$\sum_{\vec{k}} |\vec{k} \rangle \langle \vec{k}|$
the completeness relation cannot be applied,
since the sum of two or more momenta can lead out of the
momentum space taken into account.

By writing down the time derivative of one of the Gauss law identities
relating 2- and 3-point functions, one can see that for these particular
identities this problem can be cured by taking into account 3-point
functions with one momentum outside the momentum space allowed for the
2-point functions. We, therefore, increase the allowed momentum space for
the 3-point functions correspondingly. The remaining violation of the
2-point/3-point identities, which is then due to the 3-point
truncation scheme, i.e. to the violation of higher order identities,
is expected to be of a smaller effect on the time evolution than the
original violation.

We note in passing that it turns out not to have a
significant effect if we also allow 3-point functions with
2 or 3 momenta larger than the 2-point cutoff,
which is connected to the fact that the time
derivatives of these 3-point functions then do not receive any
contributions from the 3-point vertex term, which is the dominant
nonabelian contribution (cf. the remark at the end of Section
\ref{unmodi}).

We can also combine the theory with the increased momentum basis
for the 3-point functions with the implementation of Gauss law
identities discussed in Section \ref{implemgauss}.

Fig.~\ref{bild11} shows the electric field energy $E_{el}$
(upper part) and the magnetic field energy $E_{mag}$ (lower part)
as a function of time obtained within the above method by taking
into account the 9 lowest lying momenta for the 2-point functions,
which then (by summing up two of the momenta in a 3-point function)
leads to an allowed space of 25 momenta for the third momentum
(the sum of all three momenta has to vanish due to translation
invariance). We compare the results obtained without an additional
implementation of Gauss law identities and with the additional
implementation of either the 2-point/3-point identity only or the
3-point/4-point identity as well, as in Section \ref{implemgauss}.
For both the real part (left column) and the imaginary part (right
column) of $E_{mag}$ we see that the results without Gauss law
identities and those using the 2-point/3-point identity are almost
identical. Up to about $t/L=10$ the same holds for $E_{el}$.
While for both approximations the system of equations becomes unstable
at about $t/L=25$, the additional implementation of the 3-point/4-point
identities stabilizes the system up to about $t/L=80$.

Within all three approximations, on average, there is a large decrease in
the real part of $E_{el}$ and a comparably small decrease in
the real part of $E_{mag}$ which implies that energy conservation is
violated. The situation is similar to that of Section \ref{implemgauss},
only with $E_{el}$ and $E_{mag}$ exchanged.

The imaginary parts of $E_{el}$ and $E_{mag}$ stay very small until
about $t/L=10$. In case of no Gauss law constraints or only the
2-point/3-point identity implemented,
the imaginary part of $E_{el}$ then starts to
build up fast (free field dynamics) oscillations with small amplitudes
and in addition a slow large amplitude oscillation. If the
3-point/4-point identity is included, the slow oscillation vanishes.
In the imaginary part of $E_{mag}$, the fast oscillations build up
comparably large amplitudes, which are significantly reduced by the
inclusion of the 3-point/4-point identity.

In Fig.~\ref{bild12} we demonstrate that the time evolution of the real
part of the electric field energy again follows the $g^{2/3}$ scaling
of frequencies and amplitudes; the upper picture shows the corresponding
curves within the approximation scheme implementing the 2-point/3-point
and the 3-point/4-point Gauss law identities for different couplings.
The lower picture shows that the same curves with the axes rescaled
according to (\ref{g23scalinglaw}) (as in Figs. \ref{bild4} and
\ref{bild8}) agree very well.
The same result can be obtained within the other two
approximations investigated in Fig.~\ref{bild11}.

We note that at present we are not able to investigate the UV properties
of this approximation since we cannot take into account enough basis
states for the 3-point functions.

Fig.~\ref{bild13} compares the violation of Gauss law, measured by the
same quantities as in Fig.~\ref{bild3}, for all three approximation
schemes discussed in this section and the unmodified 3-point correlation
dynamics discussed in Sections \ref{unmodi} and \ref{scaluv}.
In the beginning, the
inclusion of higher momenta for the 3-point functions leads to a
significantly smaller violation of the 2-point/3-point identity
(upper picture) than
in the unmodified case. The strength of the violation decreases, if
more Gauss law identities are included, and there is a nonzero violation
(becoming large as the system becomes unstable) even if the 2-point/3-point
identity is implemented, which is due to the fact that the Gauss law
for the zero mode, which is not enforced, is no longer fulfilled
automatically as it is in case of Section \ref{implemgauss}.

The 3-point/4-point identity, if not enforced, exhibits a steeply rising
violation (lower picture), which is much stronger than
in the completely unmodified dynamics (Sections \ref{unmodi} and
\ref{scaluv}). If the 3-point/4-point identity is enforced, there is only
a very small violation due to the zero mode. In view of these dramatic
differences, it is remarkable how insensitive the results shown
in Fig.~\ref{bild11} are to the inclusion of Gauss law identities
especially in the early stages of the propagation.
\section{Adiabatic switching}
\label{adiabaticswitching}
In Yang-Mills theory, as in any field theory, it is crucial
to have a good approximation for the ground state of the theory.
In this section, we therefore aim at obtaining an approximate
solution for the equal-time Green functions in the ground state of
SU(2) Yang-Mills theory on a torus in $2+1$ spacetime dimensions
within the framework of the equations of motion for connected
equal-time Green functions as discussed in Section \ref{eomcgf}.

In our previous applications of $n$-point correlation dynamics to
the nuclear many-body problem \cite{pet94, pfi94, hae195} and
to $\Phi^4$-theory in $0+1$, $1+1$ and $2+1$ spacetime dimensions
\cite{hae295, pet96, pet97}, we found that in the manifold
of stationary solutions the total energy of the system was in
general not bounded from below, so that without additional boundary
conditions the equations of motion cannot be used for a determination
of ground state Green functions by a variational calculation.

In the present case of equations of motion for Yang-Mills theory
defined within the Heisenberg picture, the situation is even more
complex, since in an equilibrium state only expectation values
of gauge-invariant operators are stationary, and thus the task
of finding the manifold of equilibrium solutions is significantly more
involved than in one of the above non-gauge theories.

In our previous investigations
\cite{pfi94, hae195, hae295, pet96, pet97},
in order to avoid the difficulties associated with a variational
calculation, we successfully resorted to the method of adiabatically
switching on either the coupling or the residual interaction,
starting from the free solution or a stationary
Hartree-Fock-Bogoliubov solution as initial condition, respectively.
This method also has the advantage of requiring a significantly
lower numerical effort than a variational calculation.

However, in the following we will demonstrate that,
at least within our approximation scheme,
the method of adiabatically generating a ground state
configuration encounters severe problems in the present case
of $2+1$-dimensional SU(2) Yang-Mills theory on a torus.
In the $2+1$-dimensional continuum,
without the presence of any regulators,
it is obvious that the adiabatic method cannot work, since
the coupling is the only dimensionful parameter and therefore
the system must exhibit a discontinuous transition to the abelian
limit. Even on a torus the theory can be shown to
have such a discontinuous transition -- in the framework of an
implementation of a purely spatial axial gauge condition --
due to the different numbers of independent zero modes for the
Gauss law constraint in the abelian and the nonabelian case
\cite{len94}. Nevertheless, one could hope that in the presence of an
additional infrared regulator (such as leaving out the zero momentum
mode), this behaviour might change.
Moreover, since the lowest energy state in the whole Hilbert space
should automatically fulfill Gauss law \cite{fey81},
one could hope that in a possible adiabatic limit no
additional difficulties arise by the fact
that the propagation with a time dependent coupling, due to the
induced explicit time dependence of the Gauss law operator, does
not conserve gauge invariance.

We proceed by parameterizing the coupling time dependently according
to $g(t) L^{1/2} = \alpha t$ and start from the perturbative initial
condition given by (\ref{aaccmini})-(\ref{paccmini}). We exclude the
zero mode, since at $t=0$ we have $g(t=0)=0$ and therefore the
effective zero mode mass (\ref{effectivemass}) vanishes, and we are left
with the original infrared singularity of the free photon propagator.
If the adiabatic method would work, then in the limit $\alpha L \to 0$
the system should propagate along an adiabatic trajectory.

In order to investigate this we have plotted the real part of the total
energy as a function of the coupling for different values of $\alpha$
in Fig.~\ref{bild14}.
In all three approximation schemes -- unmodified 3-point correlation
dynamics (first row), implementation of the 2-point/3-point Gauss law
identity (second row), and implementation of Gauss law up to the
3-point/4-point identities (third row) -- the curves, instead of
converging towards asymptotic values, apparently exhibit a scaling
behaviour similar to the coupling or UV-cutoff scaling discussed in
Section \ref{propagate} (left column). The underlying scaling
relation can be shown to be
\begin{eqnarray}
E_{tot}(\alpha_2,g)=\left( \frac{\alpha_2}{\alpha_1}
\right)^{2/5} E_{tot}(\alpha_1,\left( \frac{\alpha_1}{\alpha_2}
\right)^{3/5} g)
\label{a25scalinglaw}
\end{eqnarray}
by rescaling the curves accordingly (right column);
the rescaled curves agree very well.

Fig.~\ref{bild15} shows the same behaviour within the various
approximation schemes taking into account an increased set of basis
states for the 3-point function (cf. Section \ref{modmombas}).

Within (unmodified) 2- and 4-point correlation
dynamics, the validity of (\ref{a25scalinglaw}) can be seen as well,
which is, however, not explicitly demonstrated here.

All of the above results imply that the method of adiabatic
switching does not work in the present case.

In order to get an explanation for this behaviour we consider
the simple example of a quantum mechanical system described by the
(purely quartic) time dependent Hamiltonian
\begin{eqnarray}
H_\alpha(x,p,t)=\frac{p^2}{2}+(\alpha t)^2 \frac{x^4}{4} \; .
\label{adihamilton}
\end{eqnarray}
Suppose we have found a solution of the time-dependent Schr\"odinger
equation for $\alpha=1$:
\begin{eqnarray}
\left[ i \partial_t - H_1(x,p,t) \right] \Psi_1(x,t)=0 \; ,
\label{a1solution}
\end{eqnarray}
and that we try to construct a corresponding solution for $\alpha \ne 1$
using the ansatz
\begin{eqnarray}
\Psi_\alpha(x,t)=\Psi_1(x',\gamma t)
\label{adiansatz}
\end{eqnarray}
with $x'$ defined by the canonical transformation
\begin{eqnarray}
x=\beta^{-1/3} x' \; \; , \; \; \; \; p=\beta^{1/3} p' \; .
\label{adicanontrafo}
\end{eqnarray}
Inserting (\ref{adiansatz}) into the time dependent Schr\"odinger
equations yields
\begin{eqnarray}
&& \left[ i \partial_t - H_\alpha(x,p,t) \right] \Psi_\alpha(x,t)
\nonumber \\
&& = \left[ i\gamma \frac{\partial}{\partial \gamma t}
- \beta^{2/3} \left\{ \frac{p'^2}{2}
+ \left( \frac{\alpha}{\beta} \right)^2 t^2 \frac{x'^4}{4} \right\}
\right] \Psi_1(x',\gamma t) \stackrel{!}{=} 0
\nonumber \\
&& \Rightarrow \beta^{2/3} \left\{ \frac{p'^2}{2}
+ \left( \frac{\alpha}{\beta} \right)^2 t^2 \frac{x'^4}{4} \right\}
\stackrel{!}{=} \gamma H_1(x',p',\gamma t)
\nonumber \\
&& \Rightarrow \gamma=\beta^{2/3} \; , \; \; \gamma=\frac{\alpha}{\beta}
\Rightarrow \gamma=\alpha^{2/5} \; ,
\label{adifinalresult}
\end{eqnarray}
which directly implies a behaviour as in (\ref{a25scalinglaw}).
If the initial condition is chosen independent of $\alpha$, we
furthermore have
\begin{eqnarray}
\Psi_\alpha(x,t=0)=\Psi_1(x,t=0)=\Psi_1(\beta^{1/3} x,0)
\Rightarrow \Psi_\alpha(x,0)=const
\label{adiincon}
\end{eqnarray}
and thus our derivation only applies for the plane wave of momentum
zero as initial condition, which is exactly what we have when starting
with the unperturbed ground state. For plane waves with
momenta $\ne 0$ our consideration does not rule out (\ref{a25scalinglaw}),
since the wavefunctions can always pick up phases that cancel out in
expectation values.

It is a priori clear that for the above example the method of adiabatic
switching can never work, because a non-normalizable scattering state
(e.g., a plane wave) cannot be continuously transformed into a normalizable
bound state of a quartic potential. For this simple example, it is easy to
circumvent this problem by simply introducing a mass (harmonic
oscillator) term, which is switched off adiabatically as the quartic
coupling is switched on. In the case of Yang-Mills theory, however, the
situation is different: both states, the unperturbed one and the
perturbed one, are non-normalizable due to their invariance under
abelian or nonabelian gauge transformations, respectively. Therefore,
the introduction of a mass term for the longitudinal Yang-Mills fields
is no solution for the present problem since then the initial state
would be a normalizable wave functional.
\section{Summary}
\label{summary}
The main results of this work concerning the investigation
of the weak coupling (small volume) limit of SU(2) Yang-Mills theory
on a torus in $2+1$ spacetime dimensions can be summarized as follows:
\begin{itemize}
\item{
The violation of Gauss law within 2-, 3-, and 4-point correlation dynamics
leads to large imaginary parts for the electric and magnetic field
energies, which is clearly an unphysical result, whereas the total
energy is conserved.}
\item{
The frequencies of the slow oscillations
in the longitudinal sector of the theory scale like $g^{2/3}$,
as for a pure quartic oscillator.}
\item{
Within 2- and 3-point correlation dynamics we also find the frequencies
to scale like $\left( \sum_{\vec{k}} \right)^{1/3}$, as in a
$\Phi^4$-like theory described by a Hamiltonian of the form
$H=\int d^dx \left(\frac{1}{2} \Pi^2 + \frac{g^2}{4} \Phi^4 \right)$.
This result indicates that the cancellations necessary to ensure the
ultraviolet finiteness of the theory do not occur on the 3-point level.
This can also be attributed to the violation of gauge invariance.}
\item{
Within the 3-point truncation scheme we find that the behaviour of the
low frequency oscillations in the longitudinal sector is unaffected
by the inclusion or exclusion of the zero momentum mode, which is
an indication of infrared convergence.}
\item{
The modification of the 3-point truncation scheme by restoration
of Gauss law identities of different orders does not produce any change
in the coupling or ultraviolet scaling behaviour of the system;
it, however, leads to a reduction of the unphysical imaginary parts of
the electric and magnetic field energies, but also to the violation of
energy conservation.}
\item{
By taking into account an increased momentum basis for the 3-point
functions we obtain similar results as for the restoration of Gauss
law identities alone; the additional implementation of Gauss law
identities, while stabilizing the system, does not lead to a significant
change in the behaviour of the electric and magnetic field energies
especially in the early stages of the propagation.}
\item{
The adiabatic generation of a perturbed ground state from
the free (abelian) ground state turns out to be impossible due
to a scaling behaviour similar to that of a pure quartic oscillator.}
\end{itemize}
We thus conclude that standard many-body techniques such as the
connected Green function approach in lower order (up to the 4-point
level), which are successful for the nuclear many-body problem
and for $\Phi^4$-theory, do not allow to compute the low-energy response
of Yang-Mills systems, since either gauge invariance or the conservation
of total energy is violated and since the UV divergences do not
cancel out.
This is mainly due to the fact that, for all approximation schemes
studied in this work, the necessary cancellations between the
3-point vertex terms and the Hartree-Fock-Bogoliubov contribution to
the 4-point-vertex terms (the dressed tadpole) do not occur.
\begin{appendix}
\section{A simple quantum mechanical example}
\label{simpleex}
In this appendix we discuss a simple example, taken from
ordinary quantum mechanics in $1+1$ spacetime dimensions,
in order to present the essential features of the approach of
Section \ref{quantization} to gauge fixing in the canonical formalism
without having to deal with the complicated structure of
Yang-Mills theory.
The same example has been used in \cite{gir86} to
illustrate a different regularization procedure, which has
the box normalization of plane wave states as quantum mechanical
analogue. Both regularization schemes lead to the same result.

We consider the Hamiltonian of a free nonrelativistic
particle of unit mass ($m=1$):
\begin{eqnarray}
\hat{H}=\frac{\hat{p}^2}{2}  \; , \; \;
H^x \Psi(x) \equiv \langle x | \hat{H} |\Psi\rangle
=-\frac{\partial^2_x}{2} \Psi(x)
\label{exhamiltonian}
\end{eqnarray}
(in order to avoid confusion,
throughout this appendix we denote operators by $\hat{\cdot}$,
their coordinate space representations are in
analogy to Section \ref{quantization} denoted by $\cdot^x$).
The eigenstates of (\ref{exhamiltonian}) are the non-normalizable
plane wave states $|p\rangle$ with
\begin{eqnarray}
\langle x|p \rangle=N e^{ipx} \; .
\label{explanewave}
\end{eqnarray}

We can now view $\hat{p}$ as the analogue of the Gauss law operator
(\ref{gausslaw}), i.e. we have
\begin{eqnarray}
[\hat{p},\hat{H}]=0 \; .
\label{excommute}
\end{eqnarray}
The ''gauge transformations'' generated by $\hat{p}$ are the
translations
\begin{eqnarray}
&&\hat{U}(\Theta)=e^{i\Theta\hat{p}} \; , \; \;
U^x(\Theta)\Psi(x)=\Psi(x+\Theta) \; ,
\nonumber \\
&&\hat{U}^{-1}(\Theta)\hat{x}\hat{U}(\Theta)=\hat{x}-\Theta \; , \; \;
\hat{U}^{-1}(\Theta)\hat{p}\hat{U}(\Theta)=\hat{p} \; .
\label{exgaugetraf}
\end{eqnarray}
If we define a space of ''physical'' states by the requirement
of gauge (translation) invariance, the only state to survive is the
plane wave with momentum zero, i.e.
\begin{eqnarray}
\widehat{p}|\Psi_{ph}\rangle=0
\Leftrightarrow |\Psi_{ph}\rangle=|p=0\rangle
\; \; .
\label{exphysstate}
\end{eqnarray}
For the other plane wave states the momentum eigenvalue $p$ can be
regarded as the analogue to a static background charge distribution.

Let $\widehat{O}_{inv}$ be a translation invariant operator, i.e.
$[\widehat{O}_{inv},\widehat{p}]=0$. In order to obtain the expectation
value of $\widehat{O}_{inv}$ in the physical state we first introduce
the integral measure corresponding to (\ref{invariantintmeasure})
by dividing out the volume of the translation group,
\begin{eqnarray}
&&d\mu(x)=(\int d\Theta )^{-1} dx \; ,
\nonumber \\
&&\langle \Psi_{ph}| \widehat{O}_{inv} |\Psi_{ph}\rangle
=\int d\mu(x) \Psi^*_{ph}(x) O^x_{inv} \Psi_{ph}(x)
=(\int d\Theta)^{-1} \int dx \; O^x_{inv} 1 \; ,
\label{exinvariantintmeasure}
\end{eqnarray}
where the constant $N$ is determined to be 1 by the requirement
$\langle 1 \rangle=1$.
However, $d\mu(x)$ still leads to ill defined (infinite) expectation
values of non-translation invariant operators, such as $\widehat{x}^2$.
In analogy to the derivation of (\ref{intmeasure}) we thus write
\begin{eqnarray}
&&1\equiv\Delta(x) \int d\Theta \; \delta\left(f(x-\Theta)\right) \; ,
\nonumber \\
&&\int d\mu(x) \Psi^*_{ph}(x) O^x_{inv} \Psi_{ph}(x)
\nonumber \\
&&=\int d\Theta \int d\mu(x-\Theta)
\Delta(x-\Theta) \delta\left(f(x-\Theta)\right)
\Psi^*_{ph}(x-\Theta) O^{x-\Theta}_{inv} \Psi_{ph}(x-\Theta)
\nonumber \\
&&=\int dx \; \Delta(x) \delta\left(f(x)\right) O^x_{inv} 1
\label{exfadpoptrick}
\end{eqnarray}
and define $d\nu(x) =dx \Delta(x) \delta\left(f(x)\right)$, which in
case of a function $f(x)$ with only one simple root at $x=x_0$ becomes
\begin{eqnarray}
d\nu(x)=dx \; \delta(x-x_0) \; .
\label{exintmeasure}
\end{eqnarray}
Instead of evaluating the spatial density corresponding
to a translation invariant quantity by integrating over space
and dividing by the volume, one thus can equivalently just pick the
value at one specific point.

The non-hermiticity of the Hamiltonian (\ref{exhamiltonian}), when
acting after a non-translation invariant operator within the
regularization given by (\ref{exintmeasure}), can easily
be discussed for the example of the operator $\widehat{x}^2$:
\begin{eqnarray}
\lefteqn{
\langle \Psi_{ph} | \widehat{H} \widehat{x}^2 |\Psi_{ph}\rangle
=\int d\nu (x) \Psi^*_{ph}(x) H^x x^2 \Psi_{ph}(x) }
\nonumber \\
&&=\int dx \; \delta(x-x_0) \left( -\frac{1}{2} \partial_x^2 \right) x^2
=\int dx \; \delta(x-x_0) \left( -1 \right)=-1
\nonumber \\
&&\neq \langle \widehat H \Psi_{ph} | \widehat{x}^2 |\Psi_{ph}\rangle
=\int d\nu(x) \left( -\frac{1}{2} \partial_x^2 \Psi^*_{ph}(x) \right)
x^2 \Psi_{ph}(x) = 0 \; .
\label{exnonhermi}
\end{eqnarray}
Furthermore, we can give a simple example to illustrate
that only if $\hat{x}(t=0)-x_0=\hat{x}-x_0$
appears to the left of all other operators (that it does not
commute with), the gauge fixing condition $x-x_0=0$ leads to
vanishing expectation values:
\begin{eqnarray}
\lefteqn{
\langle \Psi_{ph} | (\hat{x}-x_0) \hat{p} \hat{x}|\Psi_{ph}\rangle
=\int dx \; \delta(x-x_0) (x-x_0) \frac{1}{i}=0}
\nonumber \\
&&\neq \langle \Psi_{ph}| \hat{p} (\hat{x}-x_0) \hat{x} |\Psi_{ph}\rangle
=\int dx \; \delta(x-x_0) \frac{1}{i} (2x-x_0)=\frac{1}{i} x_0 \; .
\label{gaugefixvanish}
\end{eqnarray}
\section{Conserved colour structure of 4-point functions}
\label{fourpointcolourstructure}
In this appendix we present the colour structure of the
various connected four-point functions
corresponding to a global colour singlet configuration,
which is conserved in time by the equations of motion of 4-point
correlation dynamics. We have
\begin{eqnarray}
\lefteqn{
\langle A^{a_1}_{i_1}(\vec{x}_1) A^{a_2}_{i_2}(\vec{x}_2)
A^{a_3}_{i_3}(\vec{x}_3) A^{a_4}_{i_4}(\vec{x}_4) \rangle_c}
\nonumber \\
&&= \delta^{a_1 a_2} \delta^{a_3 a_4} (1-\delta^{a_1 a_3})
\langle \! \langle A_{i_1}(\vec{x}_1) A_{i_2}(\vec{x}_2)
A_{i_3}(\vec{x}_3) A_{i_4}(\vec{x}_4)
\rangle \! \rangle^{aabb}_c
\nonumber \\
&&+\delta^{a_1 a_3} \delta^{a_2 a_4} (1-\delta^{a_1 a_2})
\langle \! \langle A_{i_1}(\vec{x}_1) A_{i_3}(\vec{x}_3)
A_{i_2}(\vec{x}_2) A_{i_4}(\vec{x}_4)
\rangle \! \rangle^{aabb}_c
\nonumber \\
&&+ \delta^{a_1 a_4} \delta^{a_2 a_3} (1-\delta^{a_1 a_2})
\langle \! \langle A_{i_1}(\vec{x}_1) A_{i_4}(\vec{x}_4)
A_{i_2}(\vec{x}_2) A_{i_3}(\vec{x}_3)
\rangle \! \rangle^{aabb}_c
\nonumber \\
&&+ \delta^{a_1 a_2} \delta^{a_2 a_3} \delta^{a_3 a_4}
\langle \! \langle A_{i_1}(\vec{x}_1) A_{i_2}(\vec{x}_2)
A_{i_3}(\vec{x}_3) A_{i_4}(\vec{x}_4)
\rangle \! \rangle^{aaaa}_c \; ,
\label{fourpointcolourstructure1}
\end{eqnarray}
\begin{eqnarray}
\lefteqn{
\langle \Pi^{a_1}_{i_1}(\vec{x}_1) A^{a_2}_{i_2}(\vec{x}_2)
A^{a_3}_{i_3}(\vec{x}_3) A^{a_4}_{i_4}(\vec{x}_4) \rangle_c}
\nonumber \\
&&= \delta^{a_1 a_2} \delta^{a_3 a_4} (1-\delta^{a_1 a_3})
\langle \! \langle \Pi_{i_1}(\vec{x}_1) A_{i_2}(\vec{x}_2)
A_{i_3}(\vec{x}_3) A_{i_4}(\vec{x}_4)
\rangle \! \rangle^{aabb}_c
\nonumber \\
&&+\delta^{a_1 a_3} \delta^{a_2 a_4} (1-\delta^{a_1 a_2})
\langle \! \langle \Pi_{i_1}(\vec{x}_1) A_{i_3}(\vec{x}_3)
A_{i_2}(\vec{x}_2) A_{i_4}(\vec{x}_4)
\rangle \! \rangle^{aabb}_c
\nonumber \\
&&+ \delta^{a_1 a_4} \delta^{a_2 a_3} (1-\delta^{a_1 a_2})
\langle \! \langle \Pi_{i_1}(\vec{x}_1) A_{i_4}(\vec{x}_4)
A_{i_2}(\vec{x}_2) A_{i_3}(\vec{x}_3)
\rangle \! \rangle^{aabb}_c
\nonumber \\
&&+ \delta^{a_1 a_2} \delta^{a_2 a_3} \delta^{a_3 a_4}
\langle \! \langle \Pi_{i_1}(\vec{x}_1) A_{i_2}(\vec{x}_2)
A_{i_3}(\vec{x}_3) A_{i_4}(\vec{x}_4)
\rangle \! \rangle^{aaaa}_c \; ,
\label{fourpointcolourstructure2}
\end{eqnarray}
\begin{eqnarray}
\lefteqn{
\langle \Pi^{a_1}_{i_1}(\vec{x}_1) \Pi^{a_2}_{i_2}(\vec{x}_2)
A^{a_3}_{i_3}(\vec{x}_3) A^{a_4}_{i_4}(\vec{x}_4) \rangle_c}
\nonumber \\
&&= \delta^{a_1 a_2} \delta^{a_3 a_4} (1-\delta^{a_1 a_3})
\langle \! \langle \Pi_{i_1}(\vec{x}_1) \Pi_{i_2}(\vec{x}_2)
A_{i_3}(\vec{x}_3) A_{i_4}(\vec{x}_4)
\rangle \! \rangle^{aabb}_c
\nonumber \\
&&+\delta^{a_1 a_3} \delta^{a_2 a_4} (1-\delta^{a_1 a_2})
\langle \! \langle \Pi_{i_1}(\vec{x}_1) \Pi_{i_2}(\vec{x}_2)
A_{i_3}(\vec{x}_3) A_{i_4}(\vec{x}_4)
\rangle \! \rangle^{abab}_c
\nonumber \\
&&+ \delta^{a_1 a_4} \delta^{a_2 a_3} (1-\delta^{a_1 a_2})
\langle \! \langle \Pi_{i_1}(\vec{x}_1) \Pi_{i_2}(\vec{x}_2)
A_{i_4}(\vec{x}_4) A_{i_3}(\vec{x}_3)
\rangle \! \rangle^{abab}_c
\nonumber \\
&&+ \delta^{a_1 a_2} \delta^{a_2 a_3} \delta^{a_3 a_4}
\langle \! \langle \Pi_{i_1}(\vec{x}_1) \Pi_{i_2}(\vec{x}_2)
A_{i_3}(\vec{x}_3) A_{i_4}(\vec{x}_4)
\rangle \! \rangle^{aaaa}_c \; ,
\label{fourpointcolourstructure3}
\end{eqnarray}
\begin{eqnarray}
\lefteqn{
\langle \Pi^{a_1}_{i_1}(\vec{x}_1) \Pi^{a_2}_{i_2}(\vec{x}_2)
\Pi^{a_3}_{i_3}(\vec{x}_3) A^{a_4}_{i_4}(\vec{x}_4) \rangle_c}
\nonumber \\
&&= \delta^{a_1 a_2} \delta^{a_3 a_4} (1-\delta^{a_1 a_3})
\langle \! \langle \Pi_{i_1}(\vec{x}_1) \Pi_{i_2}(\vec{x}_2)
\Pi_{i_3}(\vec{x}_3) A_{i_4}(\vec{x}_4)
\rangle \! \rangle^{aabb}_c
\nonumber \\
&&+\delta^{a_1 a_3} \delta^{a_2 a_4} (1-\delta^{a_1 a_2})
\langle \! \langle \Pi_{i_1}(\vec{x}_1) \Pi_{i_3}(\vec{x}_3)
\Pi_{i_2}(\vec{x}_2) A_{i_4}(\vec{x}_4)
\rangle \! \rangle^{aabb}_c
\nonumber \\
&&+ \delta^{a_1 a_4} \delta^{a_2 a_3} (1-\delta^{a_1 a_2})
\langle \! \langle \Pi_{i_2}(\vec{x}_2) \Pi_{i_3}(\vec{x}_3)
\Pi_{i_1}(\vec{x}_1) A_{i_4}(\vec{x}_4)
\rangle \! \rangle^{aabb}_c
\nonumber \\
&&+ \delta^{a_1 a_2} \delta^{a_2 a_3} \delta^{a_3 a_4}
\langle \! \langle \Pi_{i_1}(\vec{x}_1) \Pi_{i_2}(\vec{x}_2)
\Pi_{i_3}(\vec{x}_3) A_{i_4}(\vec{x}_4)
\rangle \! \rangle^{aaaa}_c \; ,
\label{fourpointcolourstructure4}
\end{eqnarray}
and $\langle \Pi \; \Pi \; \Pi \; \Pi \rangle_c$ has the
same structure as $\langle A \; A \; A \; A \rangle_c$.
As for the 2- and 3-point functions the double brackets
$\langle \! \langle \cdot \rangle \! \rangle_c$ denote a function
that does not depend on colour degrees of freedom. The superscripts
$\cdot^{aaaa}$, $\cdot^{aabb}$, and $\cdot^{abab}$ refer to the
configuration of colour indices on the l.h.s. of the
equation for the corresponding function to contribute, where
the order of the colour indices refers to the order in which
the corresponding operators appear in
$\langle \! \langle O \; O \; O \; O \rangle \! \rangle_c$
on the r.h.s. (the notation becomes self-explaining by
looking at the corresponding Kronecker-deltas in front of
the colour independent functions).

In the case of a colour-singlet configuration, as investigated in
the present article, one thus has 11 different 4-point functions
that depend only on the spatial coordinates and on the
vector indices of the corresponding field operators $A_i$ and momenta
$\Pi_i$. While the original, colour dependent connected equal-time
4-point functions are invariant under any permutation of operators
(note that $A^a_i(\vec{x})$ and $\Pi^b_j(\vec{y})$ commute in
all connected Green functions of order $\geq 3$), the corresponding
colour independent functions are invariant under a simultaneous
exchange of two of the fields and the corresponding colour
superscripts.
\end{appendix}
\newpage
\newpage
{\large \bf Figure Captions}
\newcounter{figno}
\begin{list}
{\underline{Fig.\arabic{figno}}:}
{\usecounter{figno}\setlength{\rightmargin}{\leftmargin}}
\item
\label{bild1}
Electric field energy $E_{el} = \frac{1}{2} \sum_{\vec{k}} \langle
\Pi^a_i(\vec{k}) \Pi^a_i(-\vec{k}) \rangle$ as a function of time
in 2-point (row 1), 3-point (row 2), and 4-point (row 3) approximation
with 8 momentum states ($\vec{k}=\vec{0}$ mode not included)
and 9 momentum states ($\vec{k}=\vec{0}$ mode included);
left column: $Re E_{el}$, right column: $Im E_{el}$
\item
\label{bild2}
Real part of the transverse and longitudinal parts of the
electric field energy,
$E^{tr}_{el}= \frac{1}{2} \sum_{\vec{k}}
\left( \delta_{ij}-l_{ij}(\vec{k}) \right)
\langle \Pi^a_i(\vec{k}) \Pi^a_j(-\vec{k}) \rangle$,
$E^{lo}_{el}= \frac{1}{2} \sum_{\vec{k}} l_{ij}(\vec{k})
\langle \Pi^a_i(\vec{k}) \Pi^a_j(-\vec{k}) \rangle$,
$l_{ij}(\vec{k})=k_i k_j/\vec{k}^2 (\vec{k} \ne \vec{0})$,
$l_{ij}(\vec{k})=0 (\vec{k} = \vec{0})$,
as a function of time in 2-point (row 1), 3-point (row 2), and
4-point (row 3) approximation; left column: 8 momentum states
($\vec{k}=\vec{0}$ mode not included), right column: 9 momentum states
($\vec{k}=\vec{0}$ mode included)
\item
\label{bild3}
Violation of Gauss law identities measured by
$\frac{1}{3} \sum_{\vec{k},i} \left| \langle A^a_i(\vec{k})
G^a(-\vec{k}) \rangle \right|$ (upper part)
and $\frac{1}{6} \sum_{\vec{k}_1,\vec{k}_2,\vec{k}_3,i,j}
\delta_{\vec{k}_1+\vec{k}_2+\vec{k}_3} \left| \varepsilon^{abc}
\langle A^a_i(\vec{k}_1) A^b_j(\vec{k}_2) G^c(\vec{k}_3) \rangle
\right|$ (lower part) as a function of time in 2-point, 3-point,
and 4-point approximation with 9 momentum states
\item
\label{bild4}
Real part of the electric field energy as a function of
time in 2-point (row 1), 3-point (row 2), and 4-point (row 3)
approximation with 21 momentum states for different couplings
(left column); curves rescaled onto $g L^{1/2}=0.01$ via the assumption
$E_{el}(g_2,t)=\left( \frac{g_2}{g_1} \right)^{2/3}
E_{el}(g_1,\left( \frac{g_2}{g_1} \right)^{2/3} t)$ (right column)
\item
\label{bild5}
Real part of the electric field energy as a function of
time in 2-point approximation without the $\vec{k}=\vec{0}$ mode
(row 1), 2-point approximation including the $\vec{k}=\vec{0}$ mode
(row 2), and in 3-point approximation including the $\vec{k}=\vec{0}$
mode (only transverse part of $E_{el}$, row 3)
for different numbers $n$ of momentum states (left column);
curves rescaled onto lowest number of states via the assumption
$E_{el}(n_2,t)=\left( \frac{n_2}{n_1} \right)^{4/3}
E_{el}(n_1,\left( \frac{n_2}{n_1} \right)^{1/3} t)$ (right column)
\item
\label{bild6}
Electric and magnetic field energy as a function of time in 3-point
approximation with 21 states using the Gauss law identities
relating 2- and 3-point functions (upper part), and using the Gauss
law identities relating 2- and 3-point functions and 3- and 4-point
functions (lower part); left column: real part, right column:
imaginary part
\item
\label{bild7}
Violation of Gauss law identities measured by\\
$\frac{1}{6} \sum_{\vec{k}_1,\vec{k}_2,\vec{k}_3,i,j}
\delta_{\vec{k}_1+\vec{k}_2+\vec{k}_3} \left| \varepsilon^{abc}
\langle A^a_i(\vec{k}_1) A^b_j(\vec{k}_2) G^c(\vec{k}_3) \rangle
\right|$ (left part),\\
$\frac{1}{6} \sum_{\vec{k}_1,\vec{k}_2,\vec{k}_3,i,j}
\delta_{\vec{k}_1+\vec{k}_2+\vec{k}_3} \left| \varepsilon^{abc}
\langle \Pi^a_i(\vec{k}_1) A^b_j(\vec{k}_2) G^c(\vec{k}_3) \rangle
\right|$ (middle part),\\
and $\frac{1}{6} \sum_{\vec{k}_1,\vec{k}_2,\vec{k}_3,i,j}
\delta_{\vec{k}_1+\vec{k}_2+\vec{k}_3} \left| \varepsilon^{abc}
\langle \Pi^a_i(\vec{k}_1) \Pi^b_j(\vec{k}_2) G^c(\vec{k}_3) \rangle
\right|$ (right part) as a function of time in 3-point approximation
with 21 momentum states i) without using any Gauss law identities,
ii) using the Gauss law identities relating 2- and 3-point functions,
and iii) using the Gauss law identities relating 2- and 3-point functions
and 3- and 4-point functions
\item
\label{bild8}
Real part of the magnetic field energy as a function of time in 3-point
approximation with 21 states using the Gauss law identities relating
2- and 3-point functions and 3- and 4-point functions for different
couplings (upper part); curves rescaled onto $g L^{1/2}=0.01$ via the
assumption $E_{mag}(g_2,t)=\left( \frac{g_2}{g_1} \right)^{2/3}
E_{mag}(g_1,\left( \frac{g_2}{g_1} \right)^{2/3} t)$ (lower part)
\item
\label{bild9}
Real part of the magnetic field energy as a function of time in 3-point
approximation using the Gauss law identities relating 2- and 3-point
functions and 3- and 4-point functions for different numbers $n$ of
momentum states (upper part);
curves rescaled onto 9 states via the assumption
$E_{mag}(n_2,t)=\left( \frac{n_2}{n_1} \right)^{4/3}
E_{mag}(n_1,\left( \frac{n_2}{n_1} \right)^{1/3} t)$ (lower part)
\item
\label{bild10}
Real part of the total energy as a function of time in 2-point
approximation with 20 states ($\vec{k}=\vec{0}$ mode not included)
using the abelian Gauss law identities and in 3-point approximation
with 21 states ($\vec{k}=\vec{0}$ mode included) using the Gauss law
identities relating 2- and 3-point functions and 3- and 4-point functions
\item
\label{bild11}
Electric field energy (upper part) and magnetic field energy
(lower part) as a function of time in 3-point approximation
with 9 momentum states for the 2-point functions and 25 momentum
states for the 3-point functions i) without using any Gauss law
identities, ii) using the Gauss law identities relating 2- and
3-point functions, and iii) using the Gauss law identities relating
2- and 3-point functions and 3- and 4-point functions;
left column: real part, right column: imaginary part
\item
\label{bild12}
Real part of the electric field energy as a function of time in 3-point
approximation with 9 states for the 2-point functions and 25 states for
the 3-point functions using the Gauss law identities relating 2- and
3-point functions and 3- and 4-point functions for different
couplings (upper part); curves rescaled onto $g L^{1/2}=0.01$ via the
assumption $E_{mag}(g_2,t)=\left( \frac{g_2}{g_1} \right)^{2/3}
E_{mag}(g_1,\left( \frac{g_2}{g_1} \right)^{2/3} t)$ (lower part)
\item
\label{bild13}
Violation of Gauss law identities measured by
$\frac{1}{3} \sum_{\vec{k},i} \left| \langle A^a_i(\vec{k})
G^a(-\vec{k}) \rangle \right|$ (upper part)
and $\frac{1}{6} \sum_{\vec{k}_1,\vec{k}_2,\vec{k}_3,i,j}
\delta_{\vec{k}_1+\vec{k}_2+\vec{k}_3} \left| \varepsilon^{abc}
\langle A^a_i(\vec{k}_1) A^b_j(\vec{k}_2) G^c(\vec{k}_3) \rangle
\right|$ (lower part) as a function of time in 3-point approximation
with 9 states and in 3-point approximation with 9 states for the
2-point functions and 25 states for the 3-point functions i) without
using any Gauss law identities, ii) using the Gauss law identities
relating 2- and 3-point functions, and iii) using the Gauss law
identities relating 2- and 3-point functions
and 3- and 4-point functions
\item
\label{bild14}
Real part of the total energy as a function of the time dependent
coupling in 3-point approximation (row 1), in 3-point approximation
using the Gauss law identities relating 2- and 3-point functions
(row 2), and in 3-point approximation using the Gauss law identities
relating 2- and 3-point functions and 3- and 4-point functions (row 3)
with 20 states ($\vec{k}=\vec{0}$ mode not included) (left column);
curves rescaled onto $\alpha L=0.01$ via the
assumption $E_{tot}(\alpha_2,g)=\left( \frac{\alpha_2}{\alpha_1}
\right)^{2/5} E_{tot}(\alpha_1,\left( \frac{\alpha_1}{\alpha_2}
\right)^{3/5} g)$ (right column)
\item
\label{bild15}
Real part of the total energy as a function of the time dependent
coupling in 3-point approximation (row 1), in 3-point approximation
using the Gauss law identities relating 2- and 3-point functions (row 2),
and in 3-point approximation using the Gauss law identities relating
2- and 3-point functions and 3- and 4-point functions (row 3)
with 8 states for the 2-point functions and 24 states for the 3-point
functions ($\vec{k}=\vec{0}$ mode not included) (left column);
curves rescaled onto $\alpha L=0.01$ via the assumption
$E_{tot}(\alpha_2,g)=\left( \frac{\alpha_2}{\alpha_1}
\right)^{2/5} E_{tot}(\alpha_1,\left( \frac{\alpha_1}{\alpha_2}
\right)^{3/5} g)$ (right column)
\end{list}
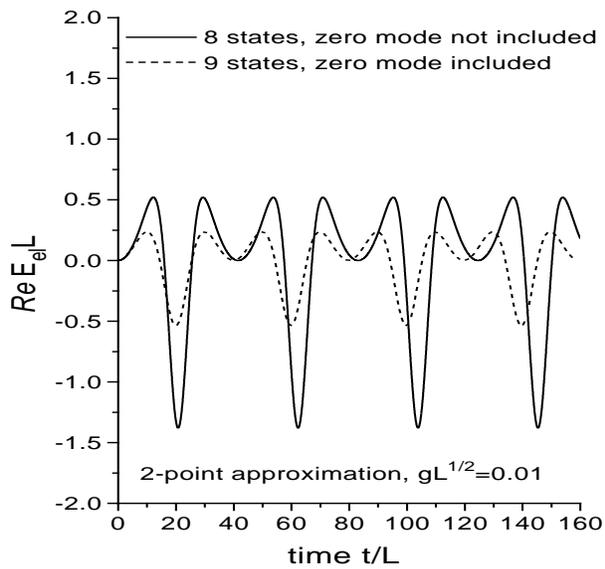
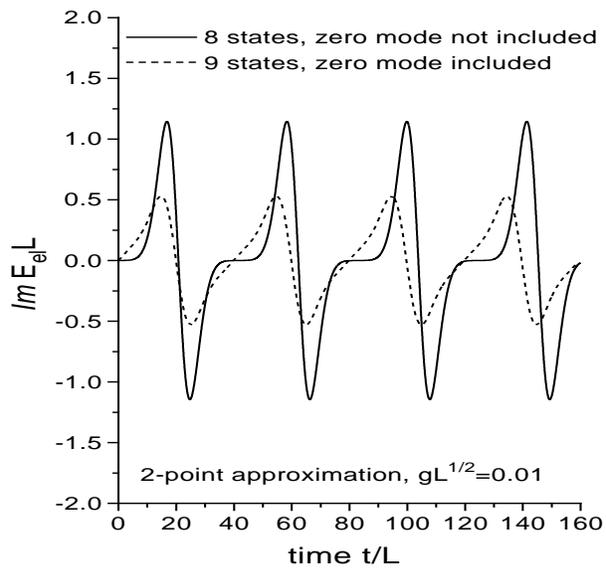
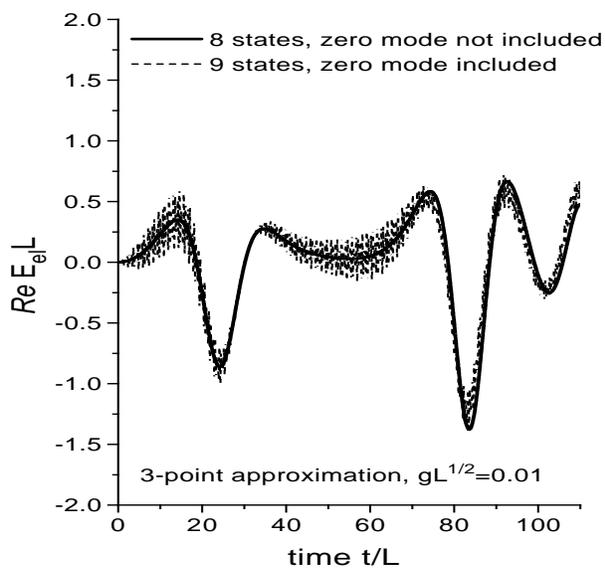
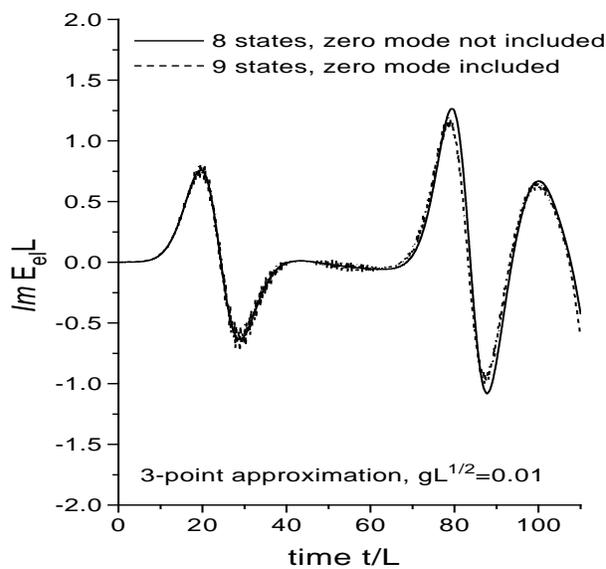
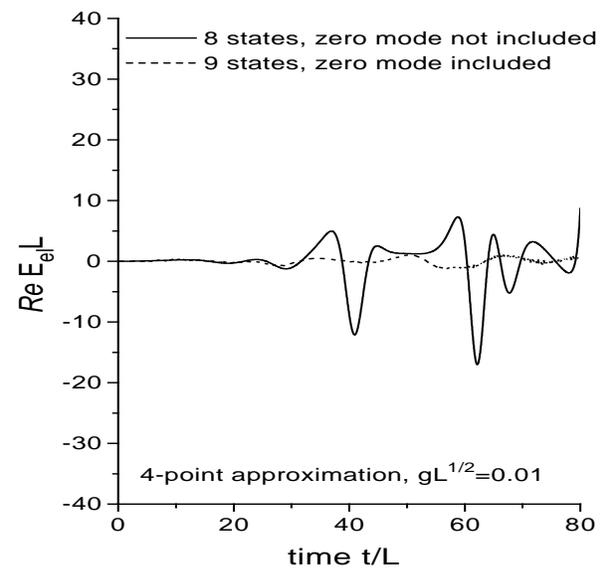
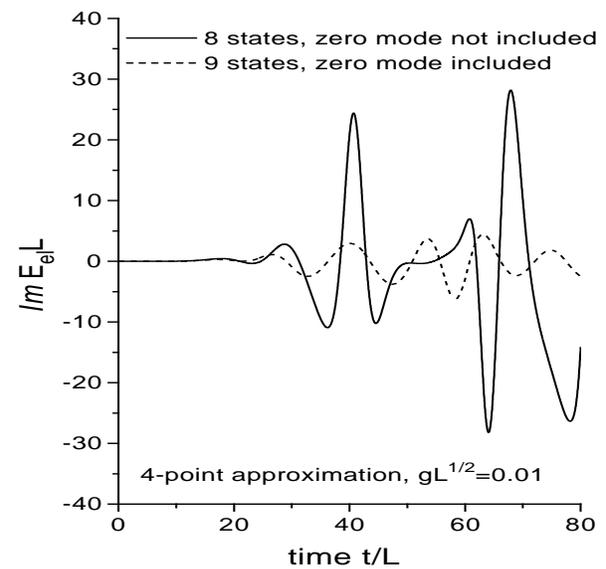
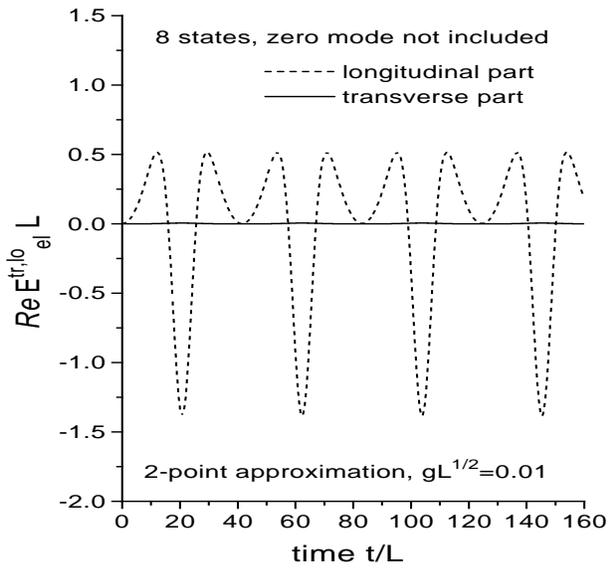
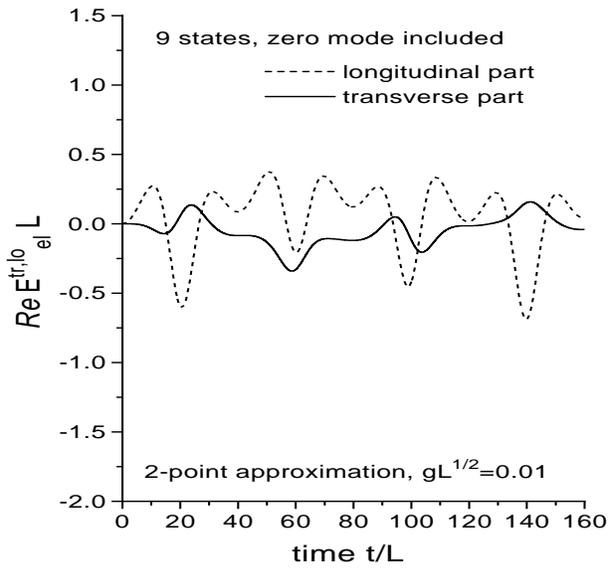
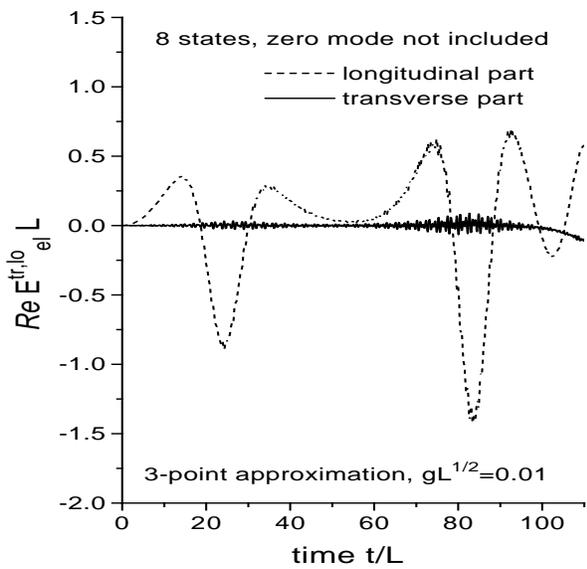
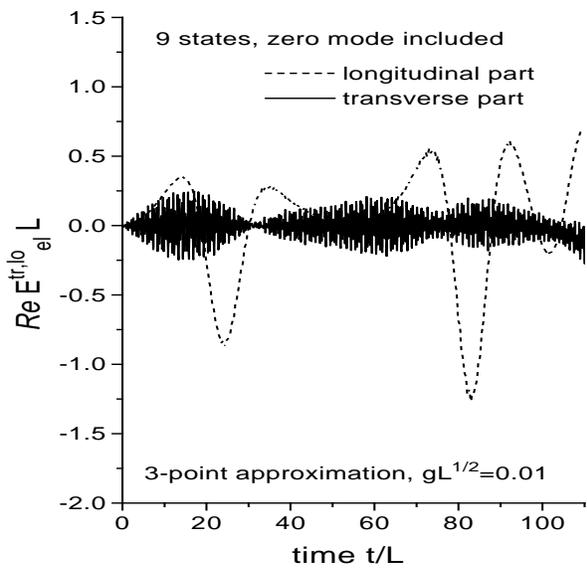
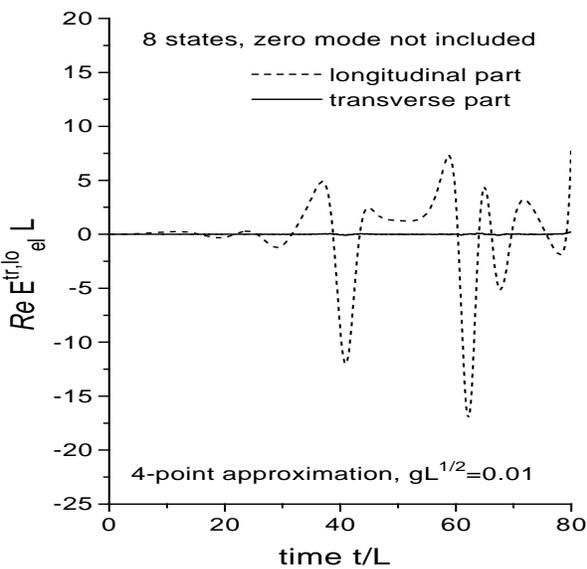
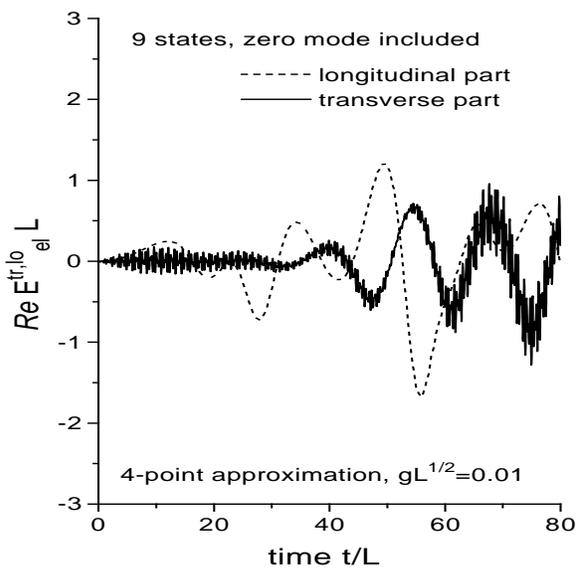
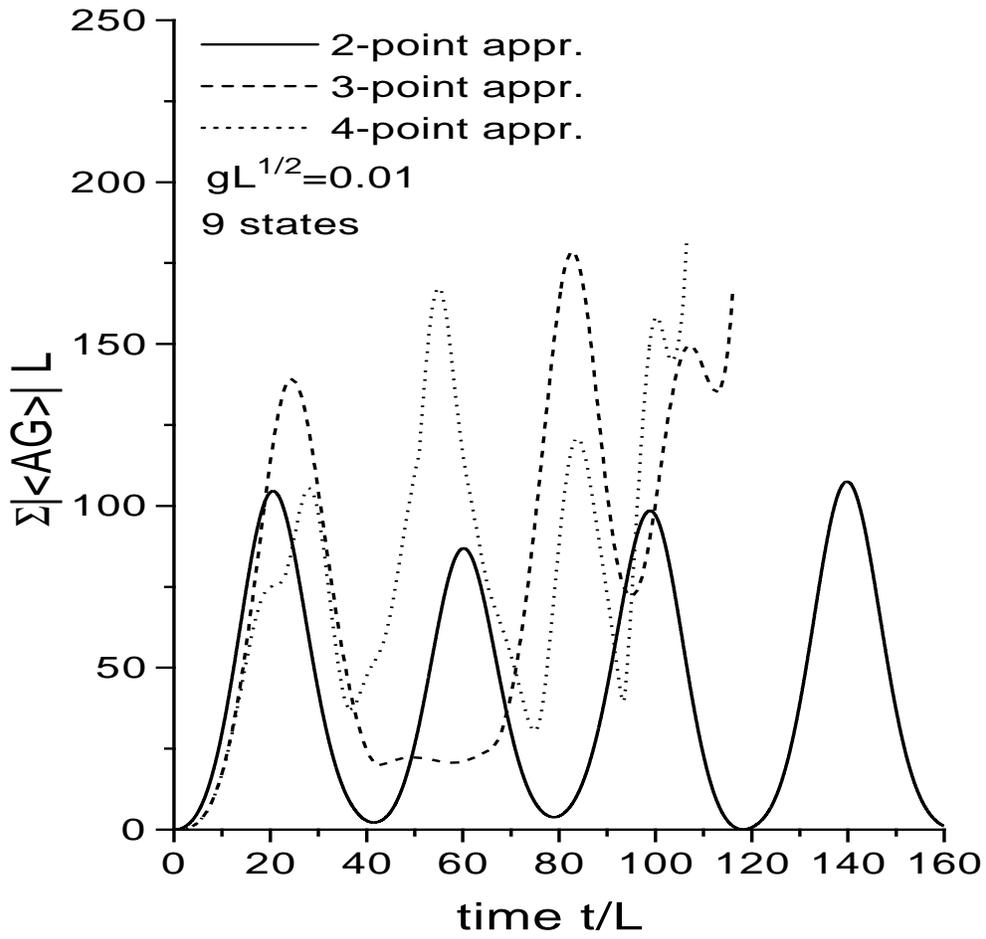
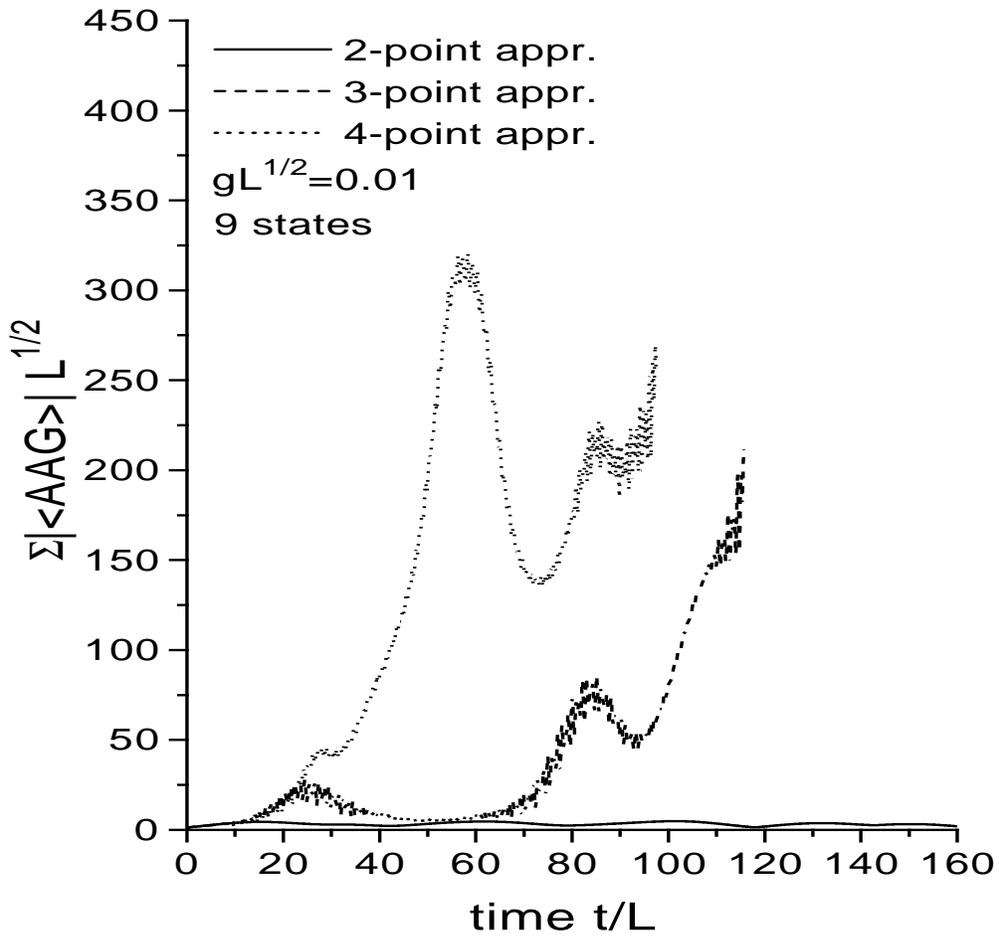
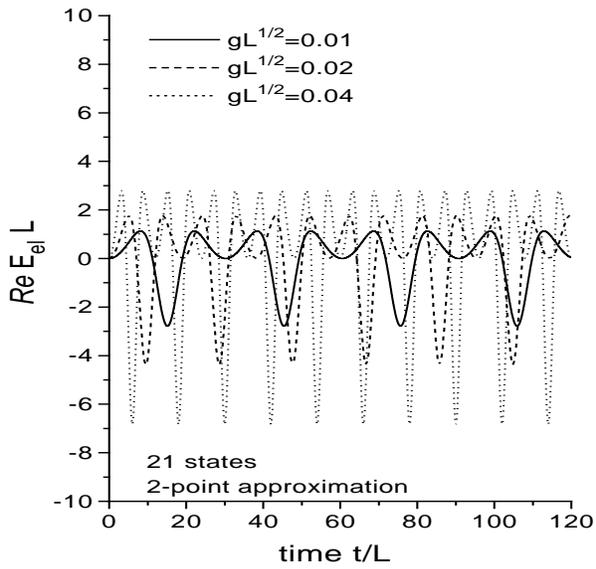
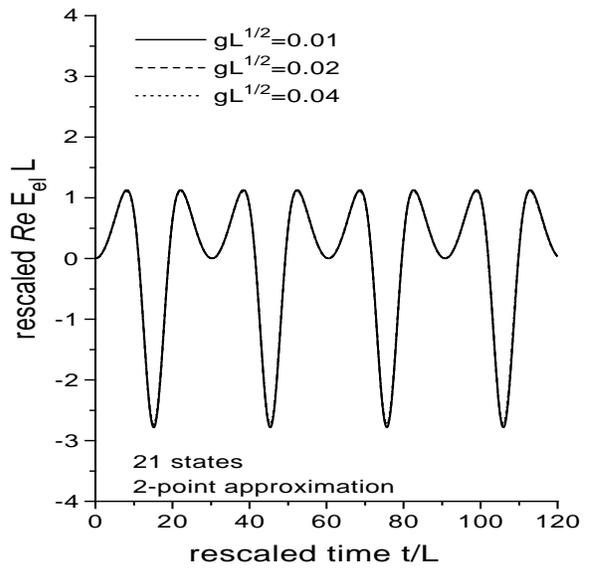
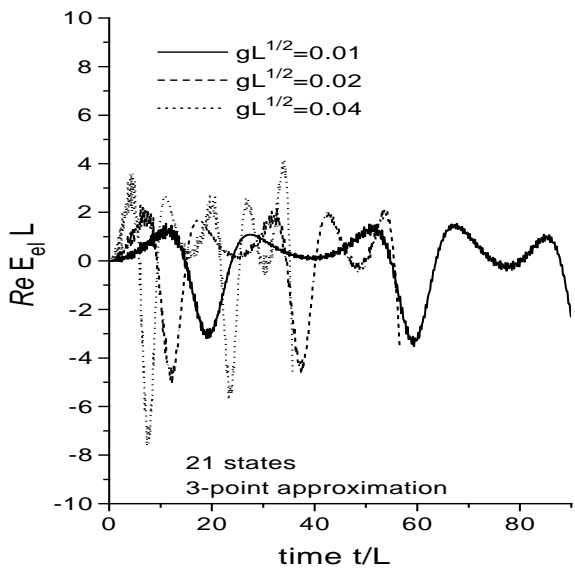
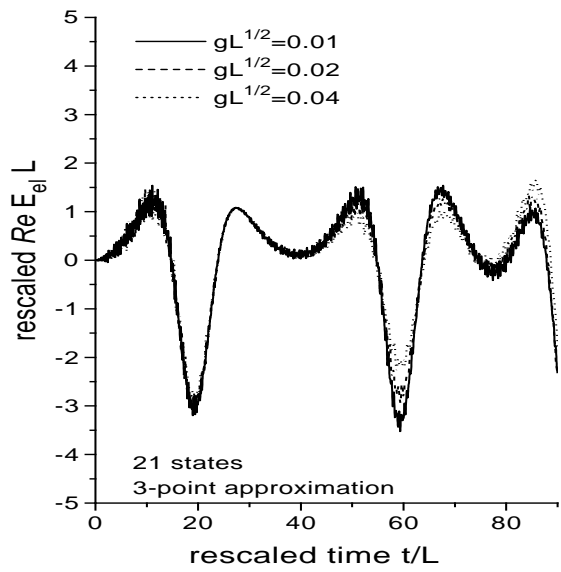
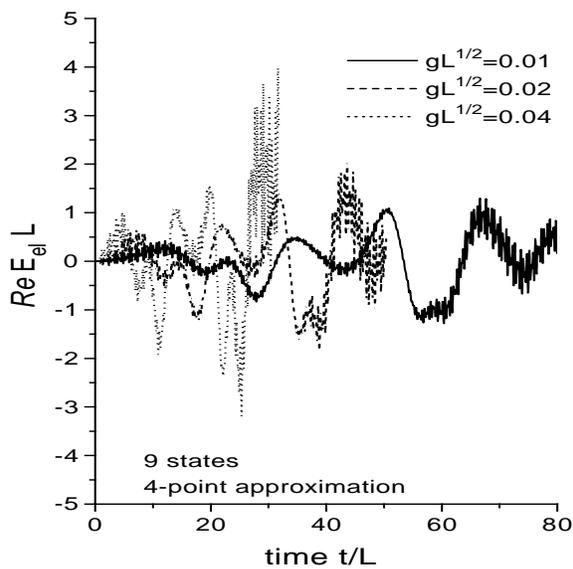
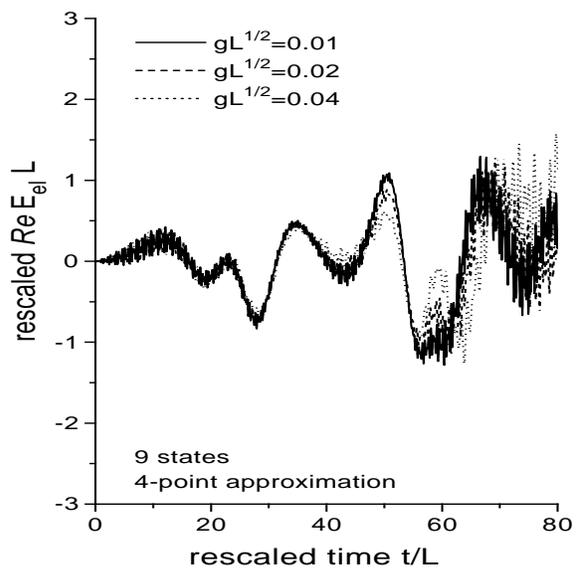
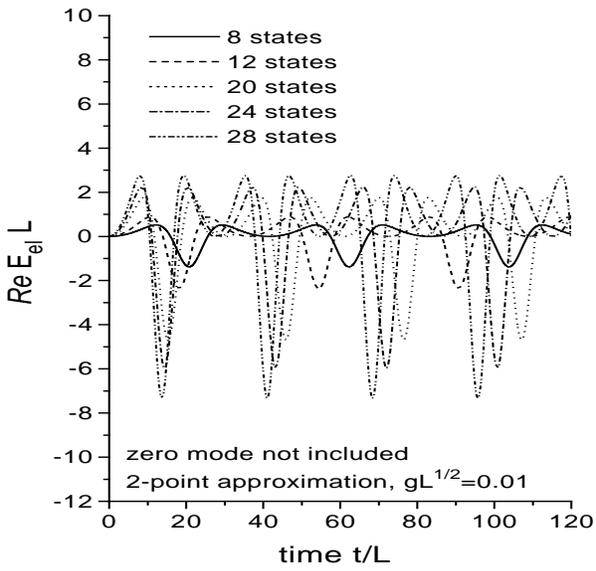
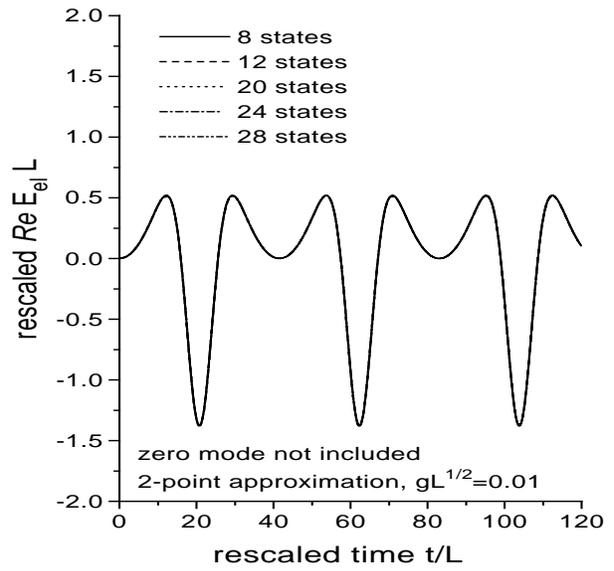
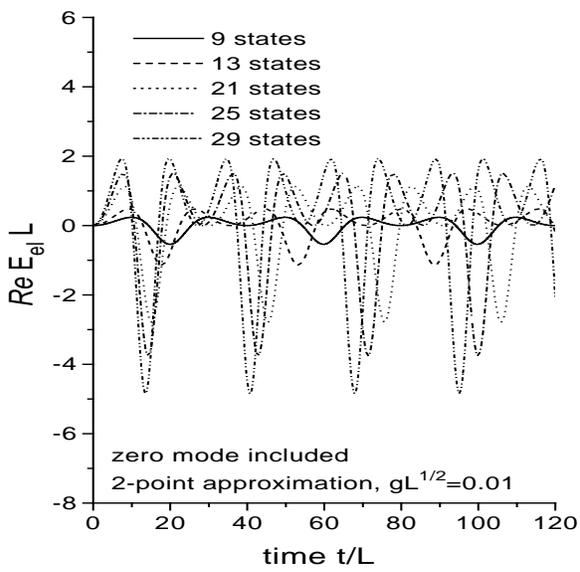
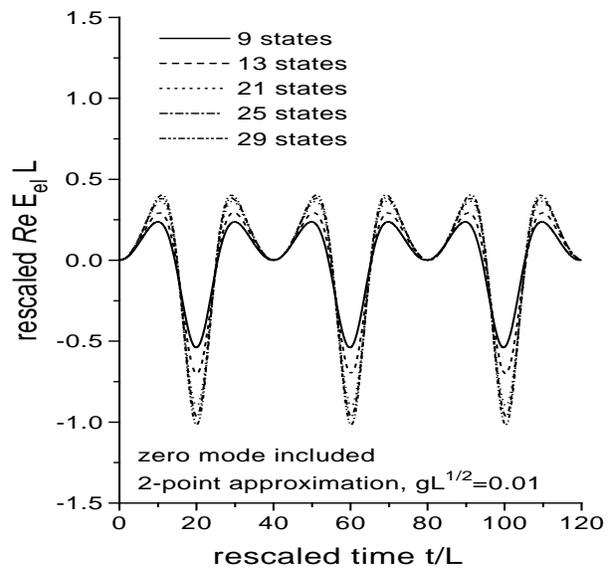
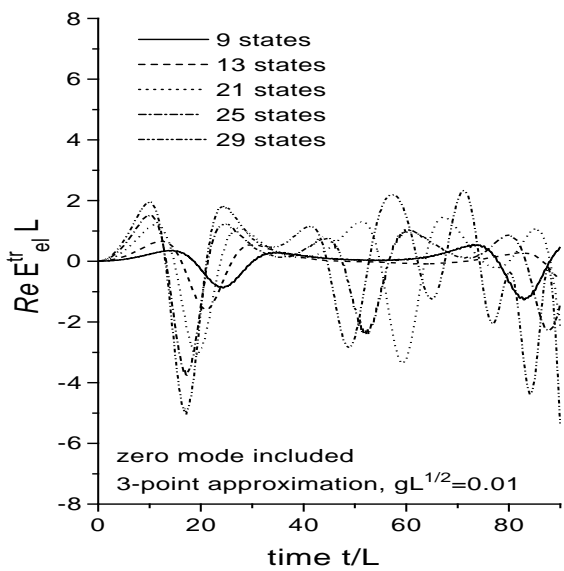
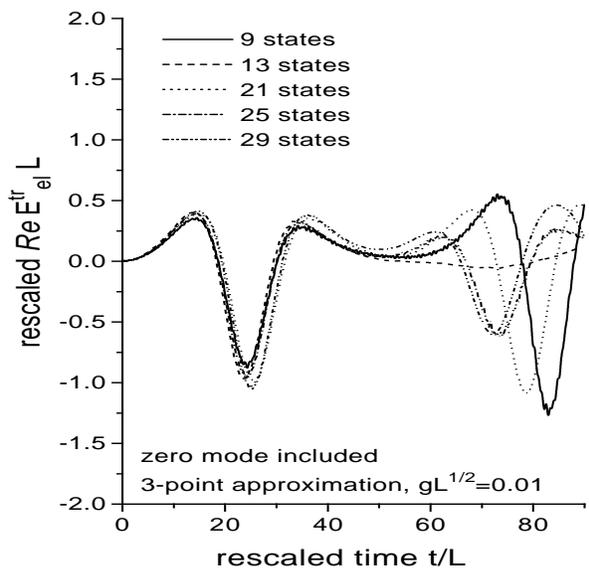
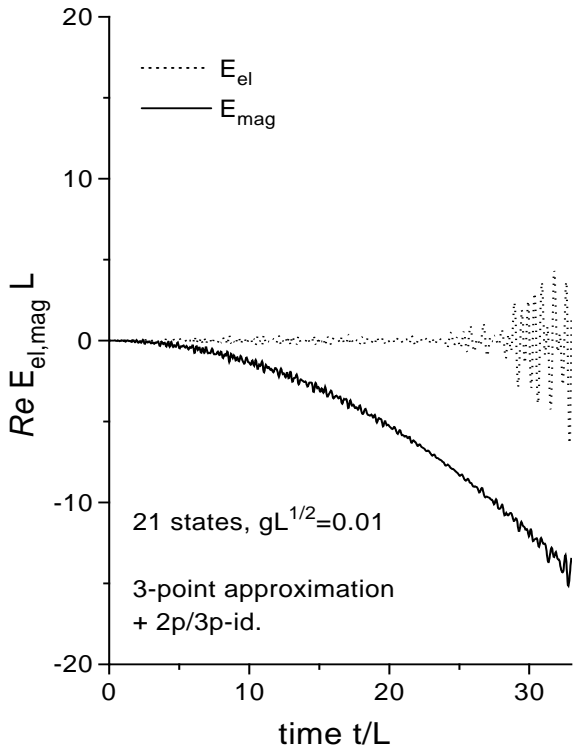
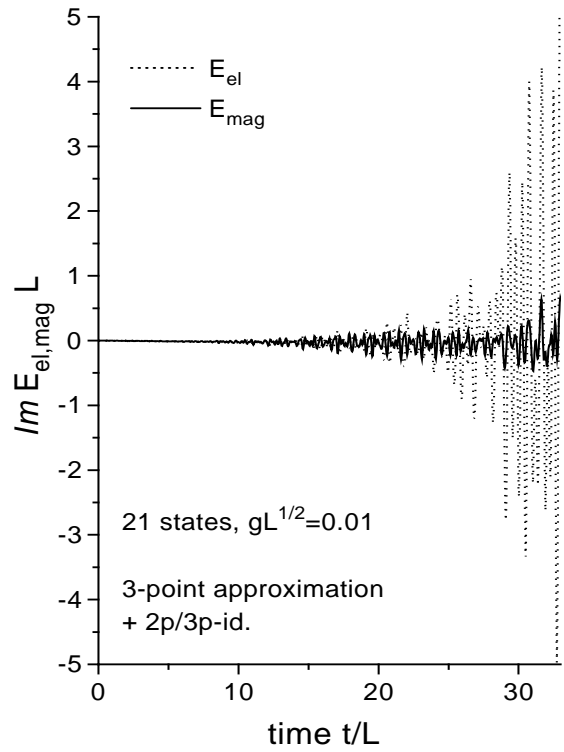
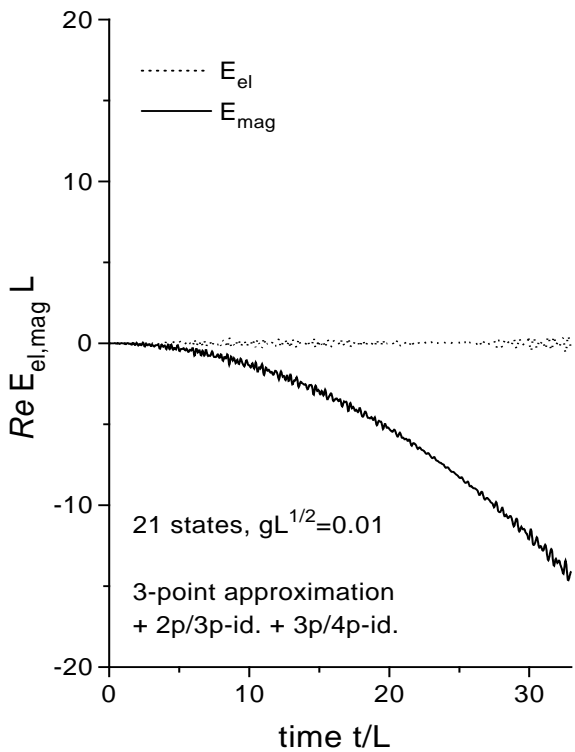
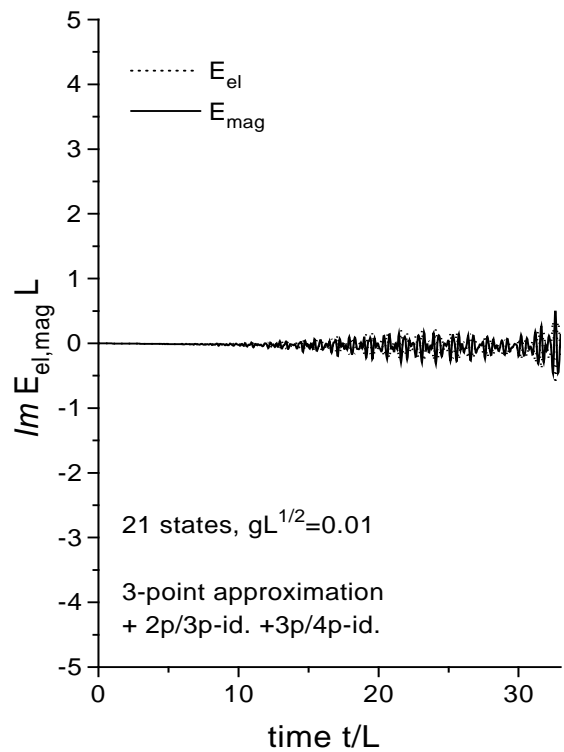
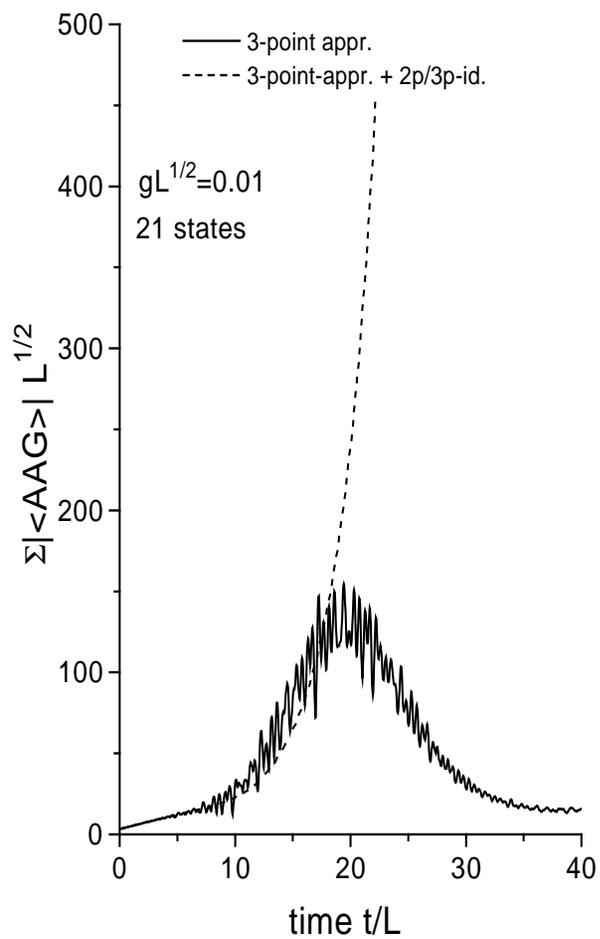
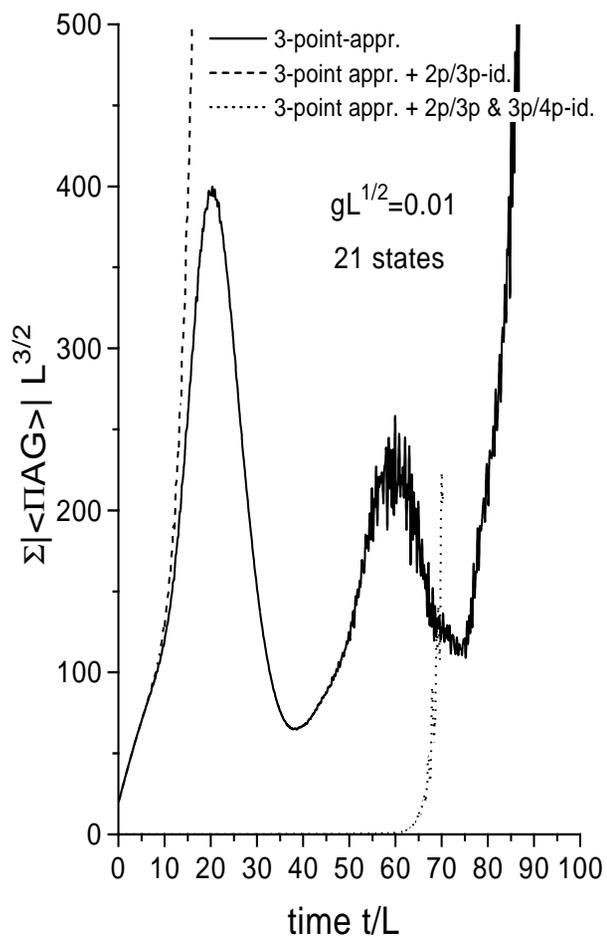
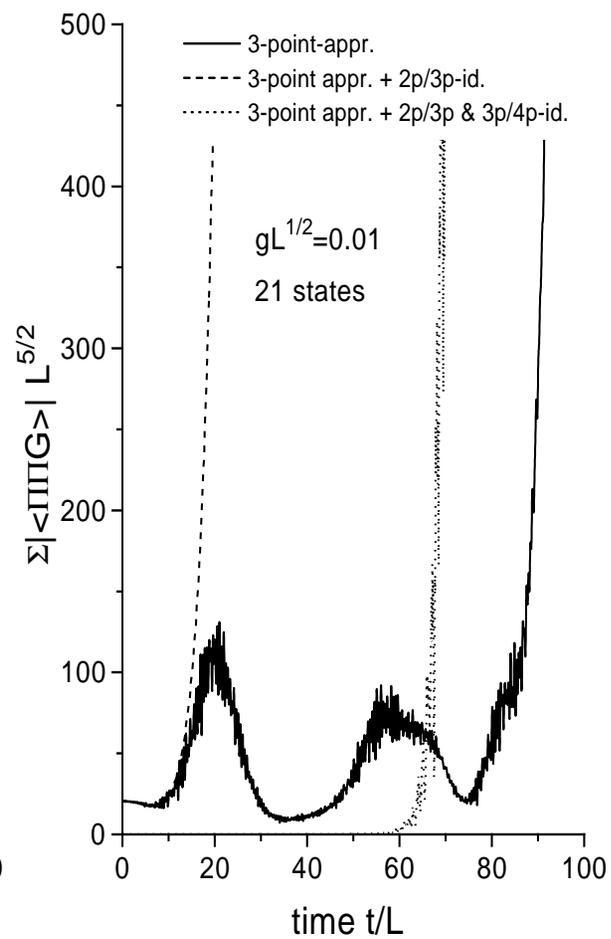
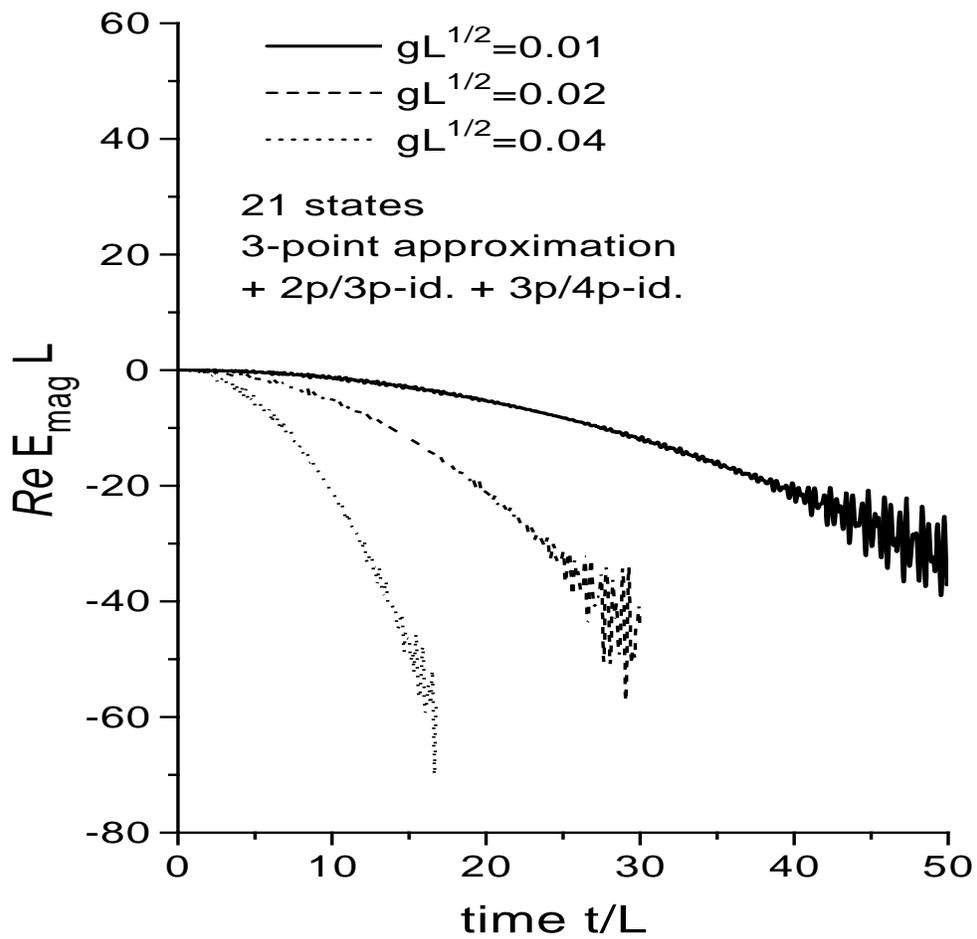
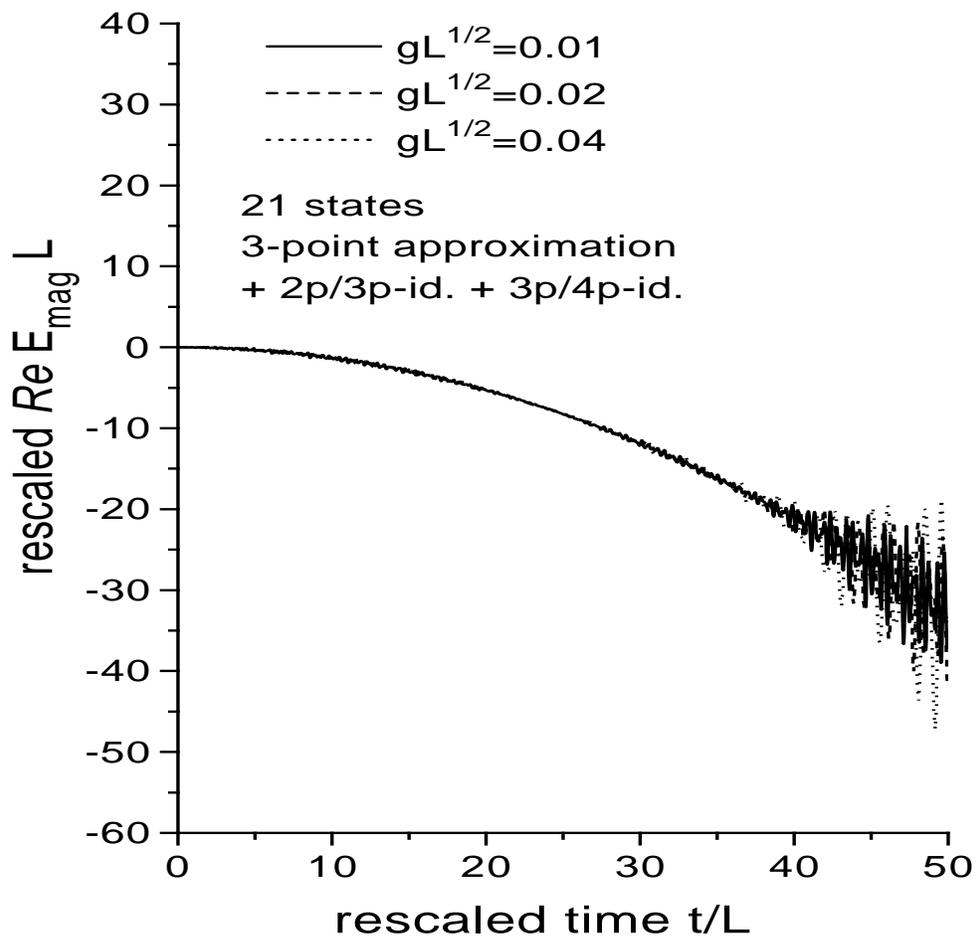
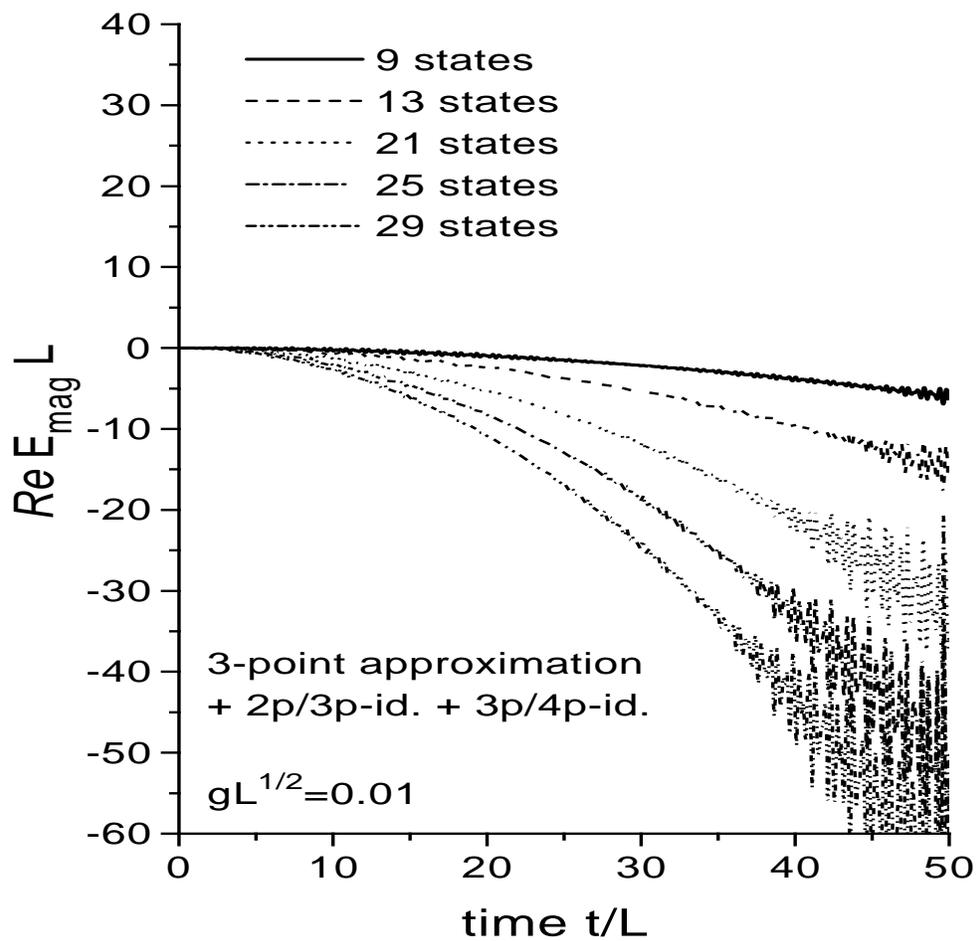
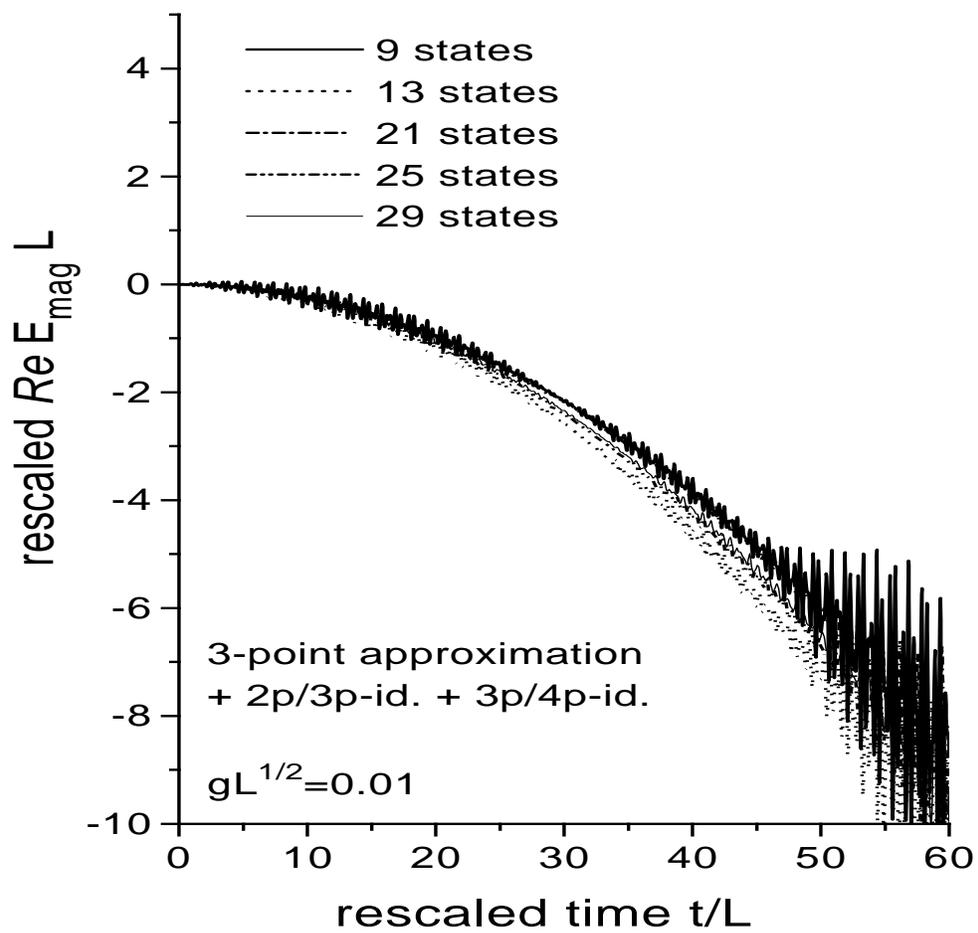
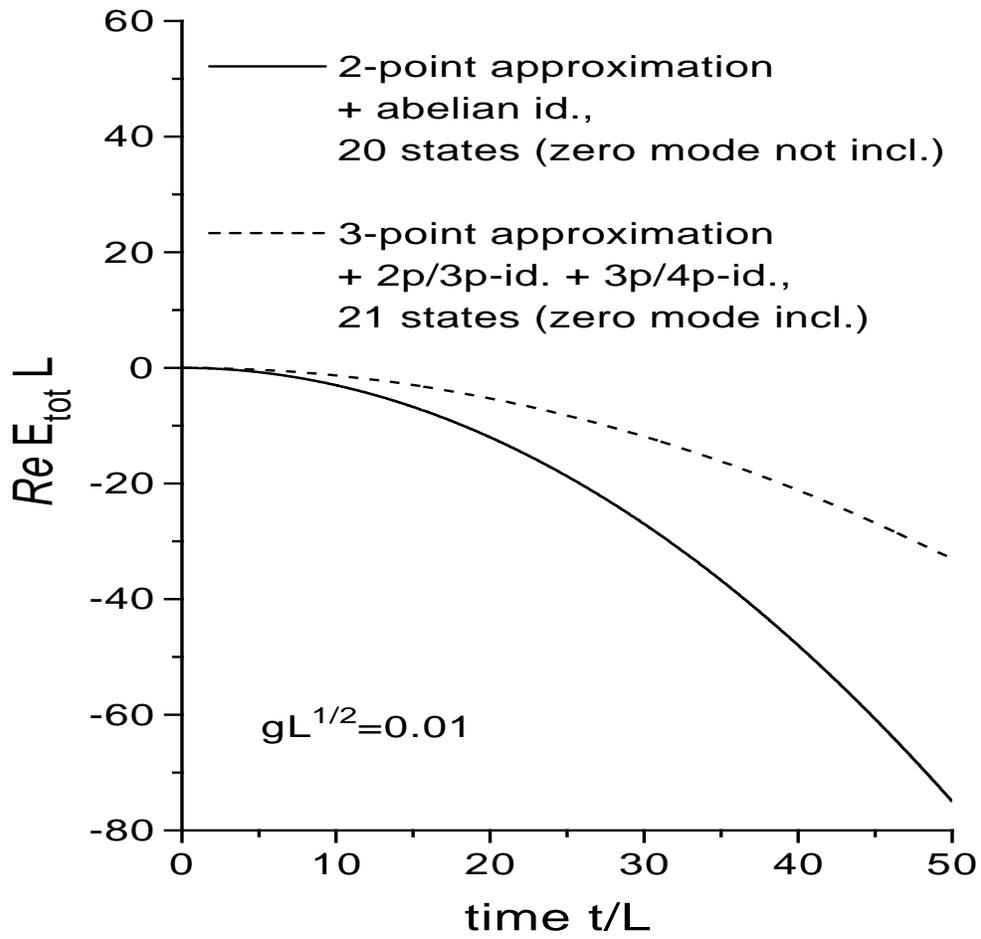
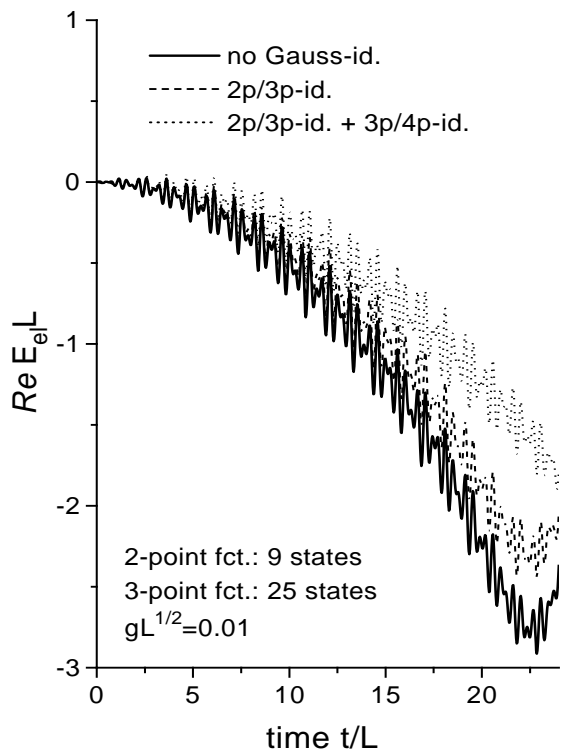
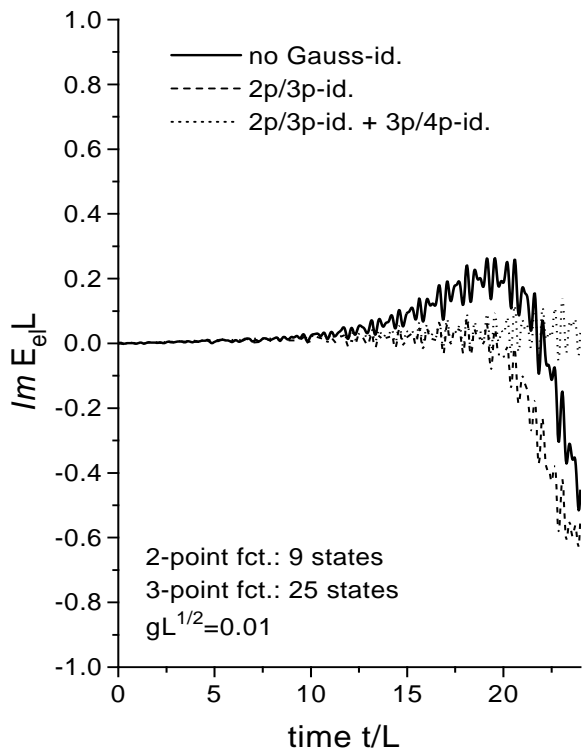
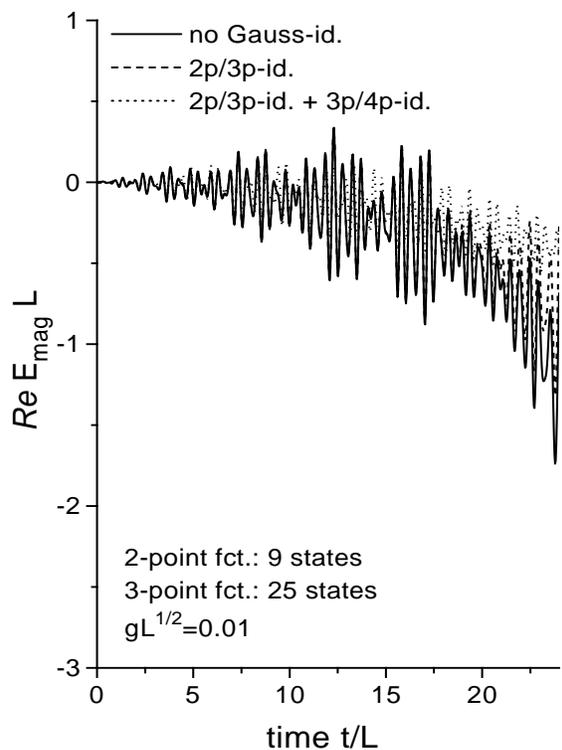
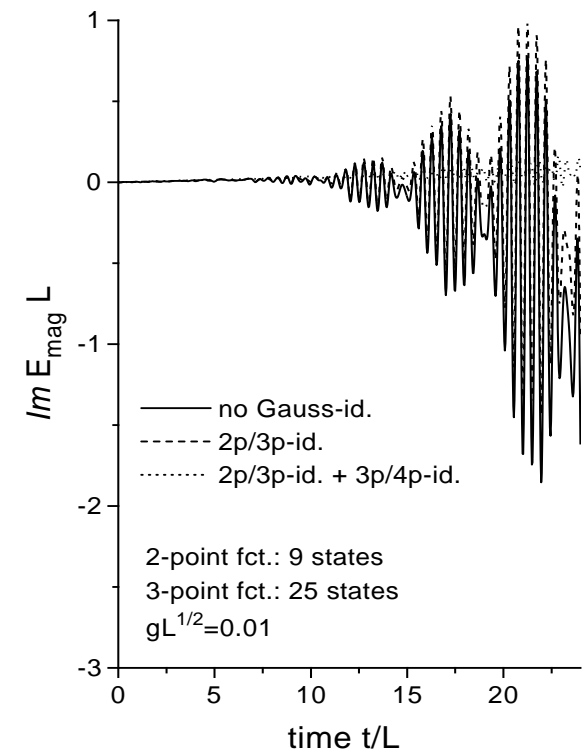
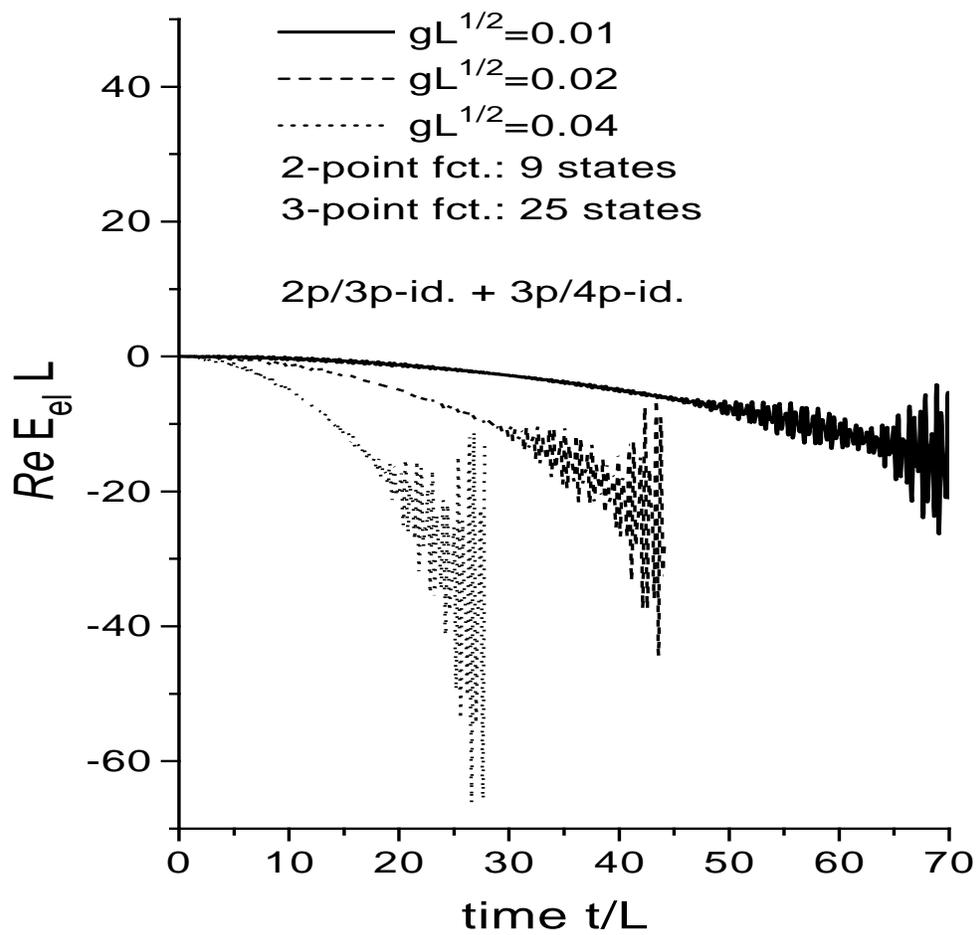
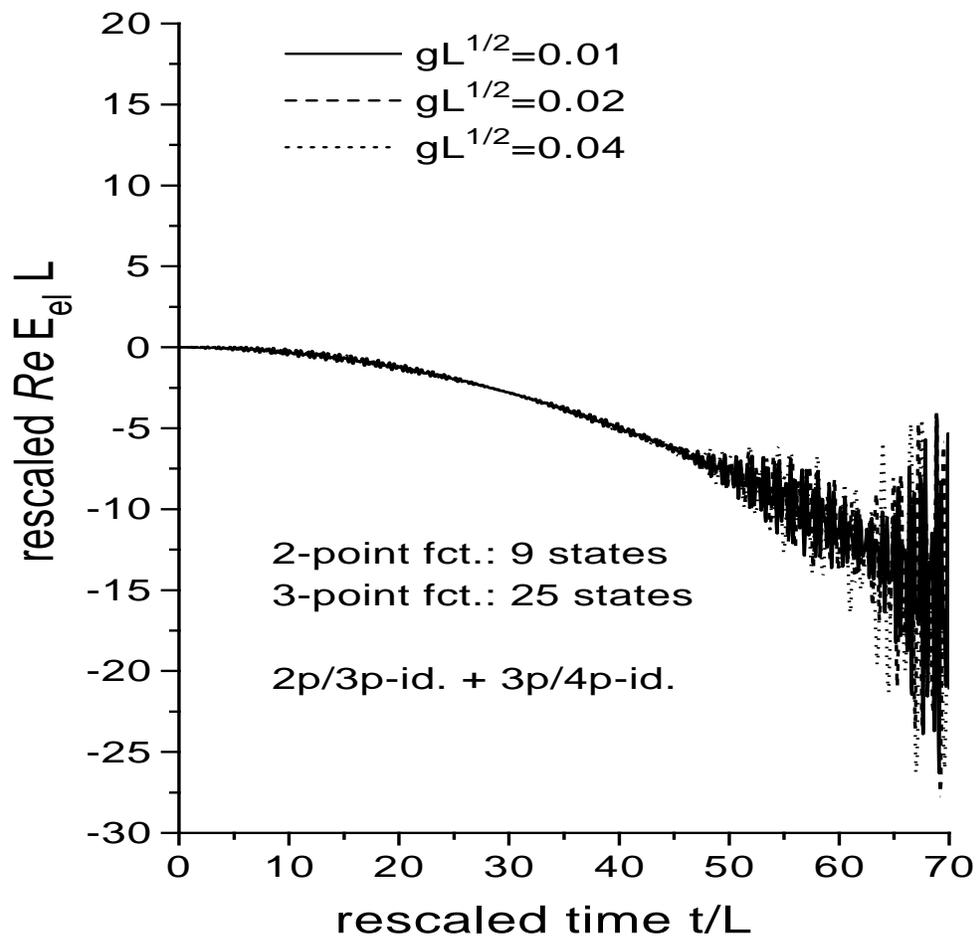
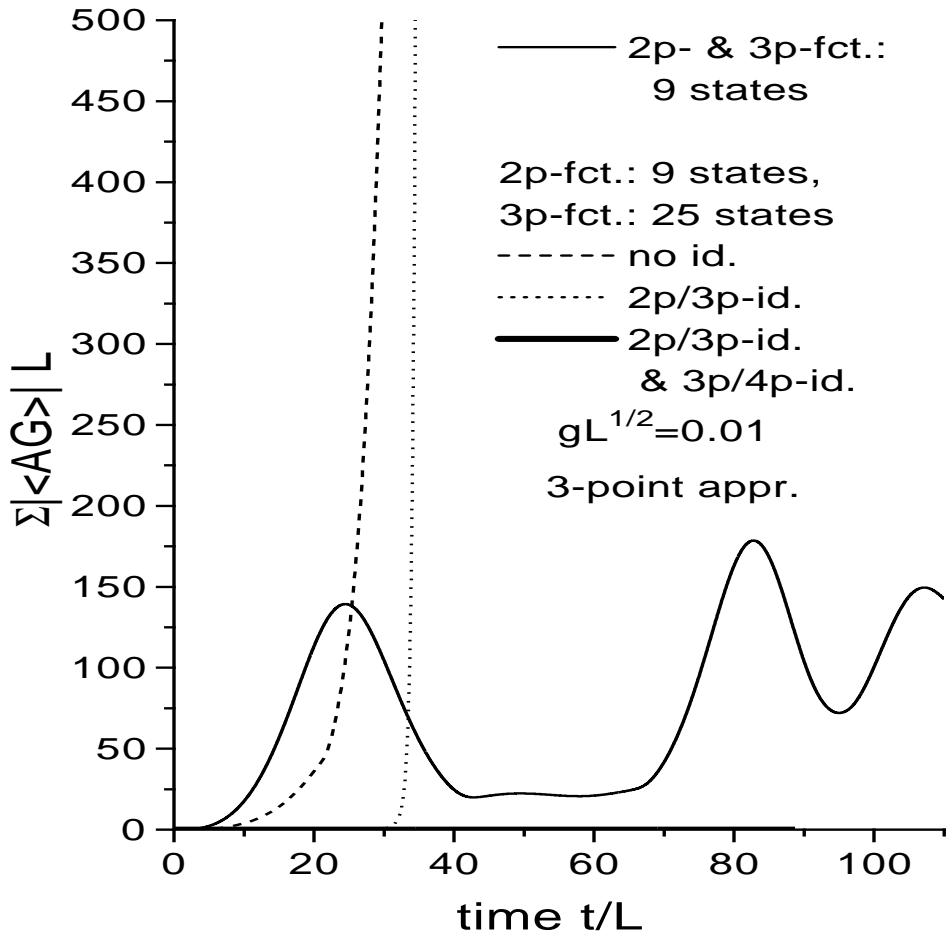
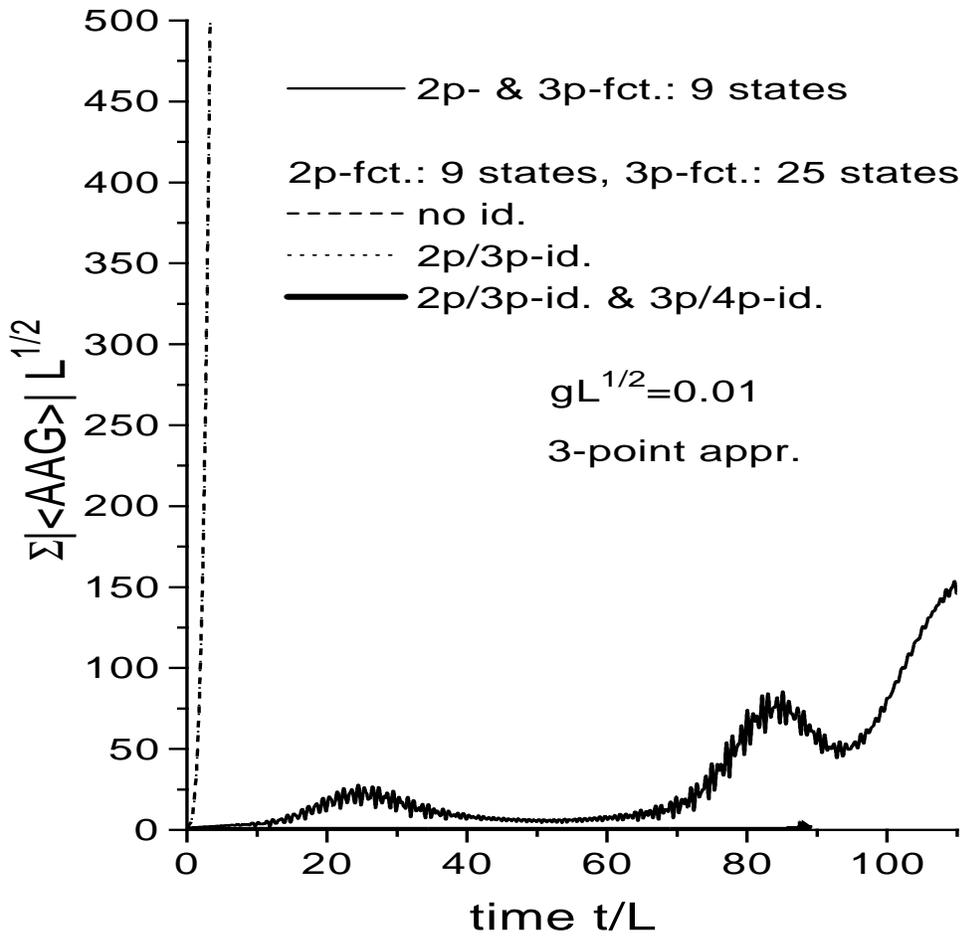
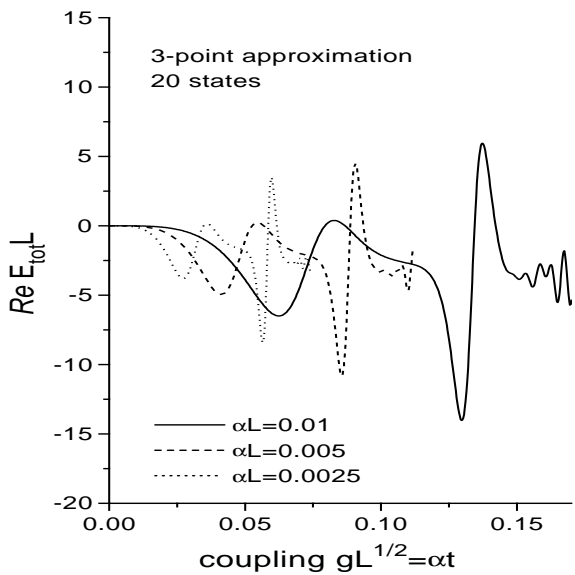
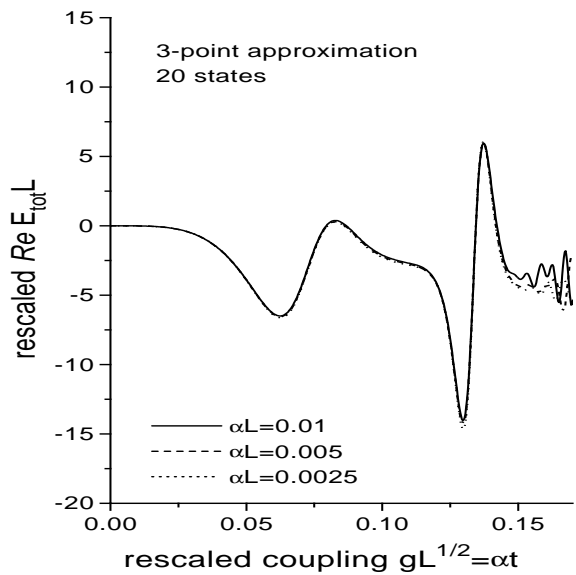
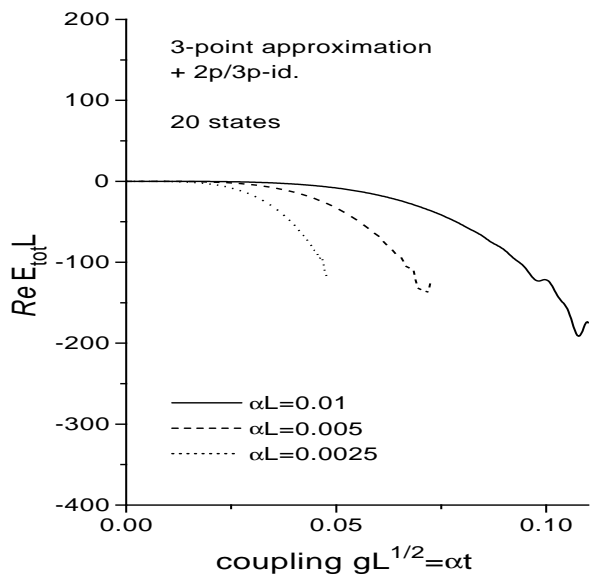
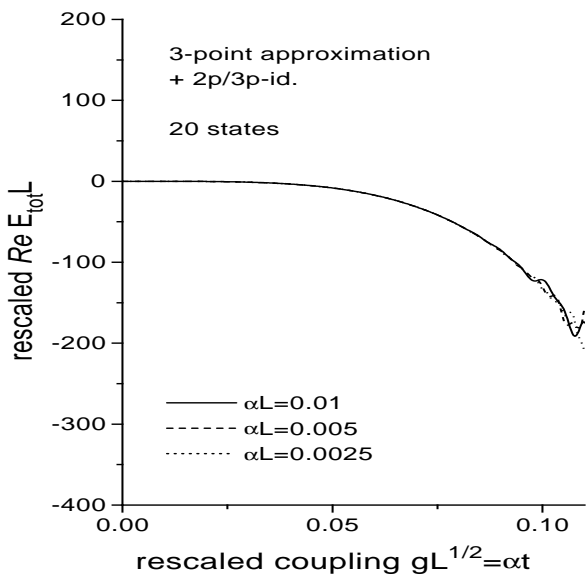
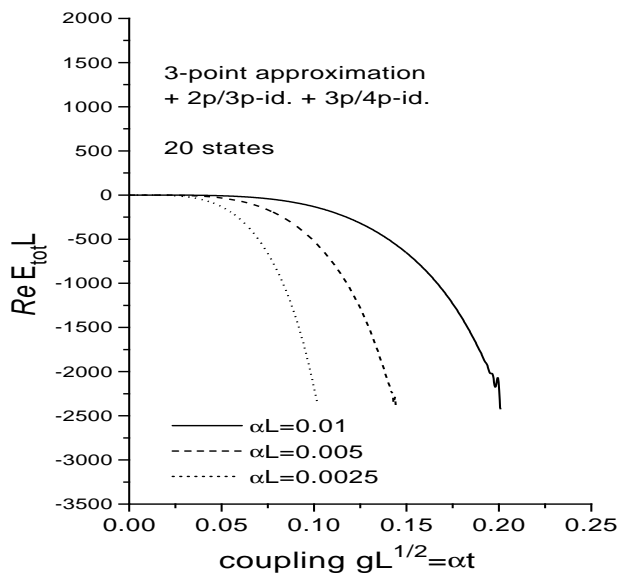
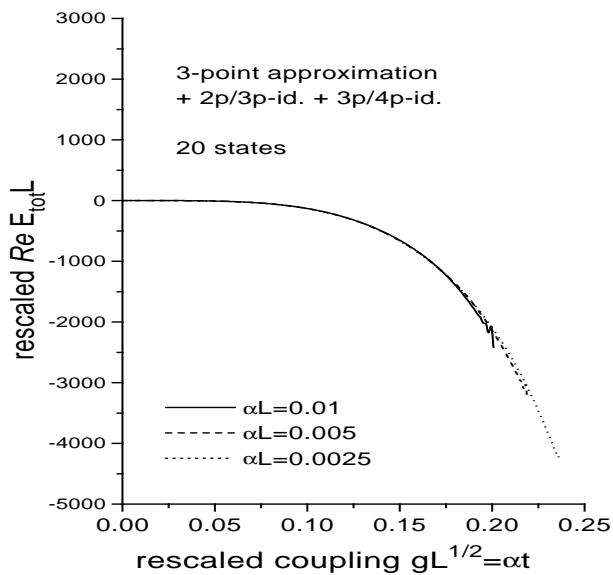
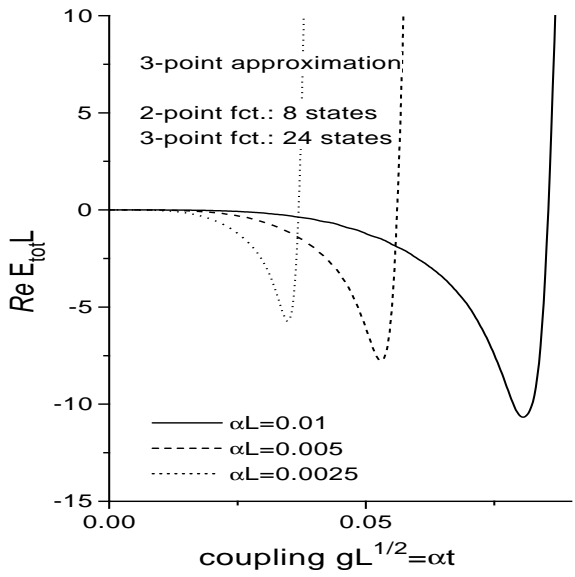
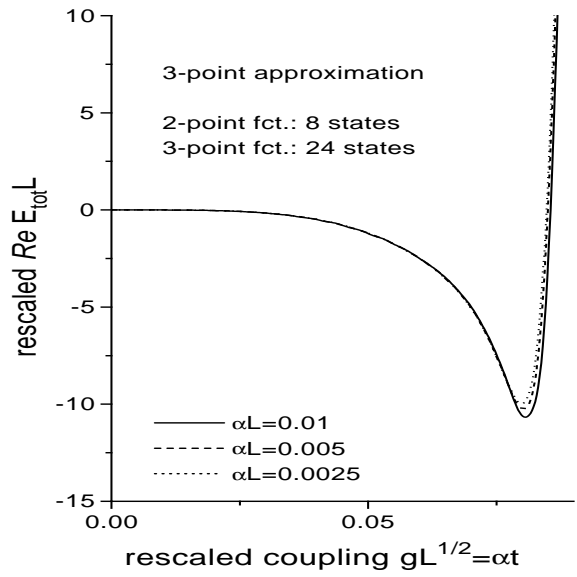
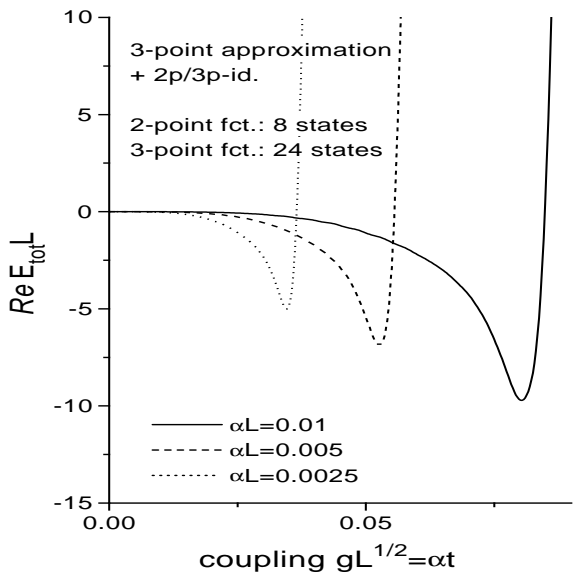
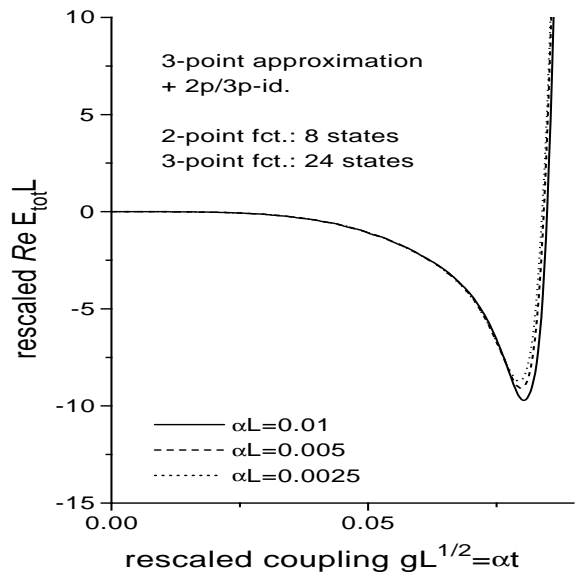
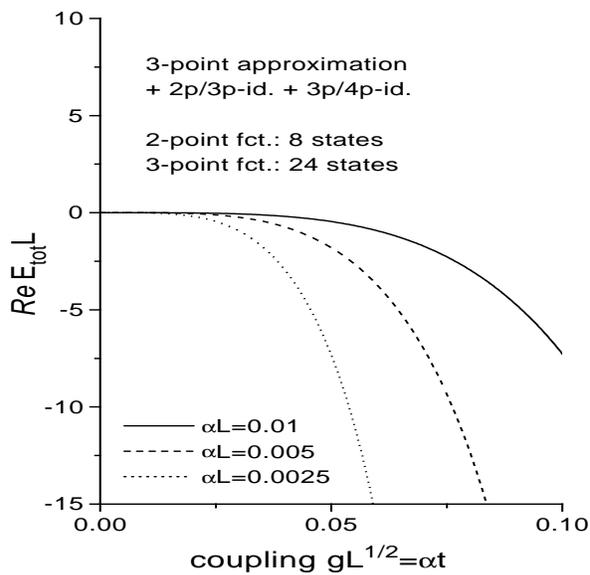
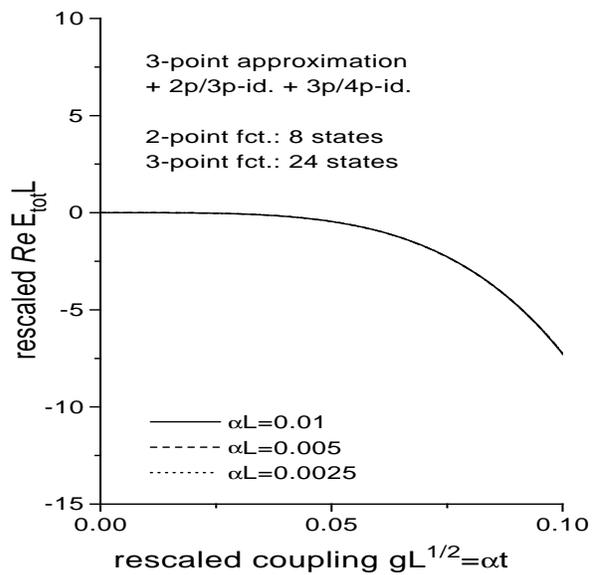
\end{document}